\RequirePackage{fix-cm}
\documentclass[smallextended]{svjour3}       % onecolumn (second format)
\smartqed  % flush right qed marks, e.g. at end of proof
\usepackage{amsmath}
\usepackage{booktabs}
\usepackage{graphicx}
\usepackage{epstopdf, epsfig}
\usepackage{dcolumn}% Align table columns on decimal point
\usepackage{bm}% bold math
\usepackage{ stmaryrd } % for jumps
\usepackage{color}
\usepackage{float}
\usepackage[caption = false]{subfig}
\usepackage{empheq} % to box equations
\usepackage[round]{natbib}
\usepackage[colorlinks=true,citecolor=blue,linkcolor=blue]{hyperref}
\begin{document}

\title{Higher order homogenized boundary conditions for flows over rough and porous surfaces
\thanks{This work was supported by SSF, the Swedish Foundation for Strategic Research (Future Leaders grant FFL15:0001). Y. S. acknowledges funding from the European Union's Horizon 2020 research and innovation programme under the Marie Sklodowska-Curie grant agreement number 708281, and DST-SERB Ramanujan fellowship (SB/S2/RJN-037/2018). U.L. and S.B. acknowledge funding from Swedish Research Council (INTERFACE center and grant nr. VR-2014-5680). S.P. acknowledges funding from the Swiss National
Science foundation (project nr. P2ELP2 181788).}
}
% Grants or other notes about the article that should go on the front
% page should be placed within the \thanks{} command in the title
% (and the %-sign in front of \thanks{} should be deleted)
%
% General acknowledgments should be placed at the end of the article.

%\subtitle{Macroscopic model for flows over porous media}
\titlerunning{A macroscopic model for flows over porous media}        % if too long for running head

\author{Y. Sudhakar \and Ugis L\={a}cis \and 
Simon Pasche \and Shervin Bagheri
}


\institute{Y. Sudhakar \at
              Linne FLOW centre, KTH Mechanics, Royal Institute of Technology, Sweden \\
              School of Mechanical Sciences, Indian Institute of Technology Goa, India\\
              \email{sudhakar@iitgoa.ac.in}           %  \\
           \and
           U. L\={a}cis, S. Pasche, and S. Bagheri \at
			Linne FLOW centre, KTH Mechanics, Royal Institute of Technology, Sweden
}

\date{Received: date / Accepted: date}

\maketitle

\begin{abstract}
We derive a homogenized macroscopic model for fluid flows over ordered homogeneous porous surfaces. The unconfined free-flow is described by the Navier-Stokes equation, and the Darcy equation governs the seepage flow within the porous domain. Boundary conditions that accurately capture mass and momentum transport at the contact surface between these two  domains are derived using the multiscale homogenization technique.  In addition to obtaining the generalized version of the widely used Beavers-Joseph slip condition for tangential velocities, the present work provides an accurate formulation for the transpiration velocity and pressure jump at fluid-porous interfaces; these two conditions are essential for handling two- and three-dimensional flows over porous media. All the constitutive parameters appearing in the interface conditions are computed by solving a set of Stokes problems on a much smaller computational domain, making the formulations free of empirical parameters. The tensorial form of the proposed interface conditions makes it possible to handle flows over isotropic, orthotropic, and anisotropic media. A subset of interface conditions, derived for porous media, can be used to model flows over rough walls. The accuracy of the proposed macroscopic model is numerically quantified for flows over porous and rough walls by comparing the results from our homogenized model with those obtained from geometry-resolved microscopic simulations.
\end{abstract}
 %========================================================================================
\section{Introduction}
\label{intro}
Boundary conditions at the interface between a porous medium and the adjacent free-fluid region is a classical problem in fluid mechanics. Despite numerous analytical~\citep{richardson1971,saffman1971,jones1973,chandesris2006,jamet2009,lacis2016,nakshatrala2019}, numerical~\citep{larson1986,larson1987,carraro2013, suga2016, suga2017}, and experimental \citep{beavers1967,taylor1971,gupte1997,goharzadeh2005,agelinchaab2006,arthur2009,morad2009,carotenuto2011,terzis2019} efforts to shed light on the nature of such conditions, widely accepted formulations to accurately capture transport phenomena across fluid-porous interfaces are not available yet. By employing the principle of virtual power, \cite{nakshatrala2019} have provided fluid-porous interface conditions from which a variety of models can be obtained.  The methodology employed in their work is different compared to classical approaches through homogenization and volume averaging. However, detailed validation of their proposed virtual power concept is yet to be carried out. In the present work, we address this problem by deriving a macroscopic description of two- and three-dimensional flows over porous  surfaces using multiscale homogenization approach. In addition, we will show that a reduced form of the formulations can be used to accurately model macroscopic flows over rough surfaces.

By macroscopic description of flows over porous media, we mean all of the following objectives: (1)~Obtaining a macroscopic description that can model flows \textit{through} porous media, and compute the associated constitutive parameters, (2)~Deriving coupling conditions at the fluid-porous interface, and (3)~Compute constitutive parameters of the interface.

Item (1) of the above list is relatively matured in its development, and various models have been put forward to describe flows through porous media~\citep{LAGE19981}. In the present work, we make use of Darcy's law, which describes a linear relation between creeping velocity inside the porous medium ($\textbf{u}^-$) and the pressure gradient~($\partial p^-/\partial \textbf{x}$):
\begin{equation}
\textbf{u}^-=-\frac{\mathcal{\textbf{K}}^i}{\mu}\cdot \frac{\partial p^-}{\partial \textbf{x}},
\label{eqn:darcy}
\end{equation}
where $\mu$ is the dynamic viscosity of the fluid and $\mathcal{\textbf{K}}^i$ is the permeability tensor of the medium, which can be computed for any ordered homogeneous porous medium by solving a Stokes problem on a periodic unit cell~\citep{meibook}. The present work is focused on the items~(2) and (3) i.e., derivation of interface conditions, and the computation of associated constitutive parameters.

In their classical work, \cite{beavers1967} conducted experiments on Poiseuille flow over a porous block to understand the nature of fluid-porous interface conditions. They observed that the viscous shear from the free-fluid region penetrates into the porous medium, and as a result, the fluid velocity changes from Darcy value in the interior of the porous medium ($u_1^-$) into a larger slip value at the interface ($u_1$). This boundary layer was modeled by the following linear relation between the slip velocity and the shear rate at the interface,
\begin{equation}
u_1-u_1^-=\frac{\sqrt{K^i}}{\alpha}\frac{\partial u_1}{\partial x_2},
\label{eqn:bj}
\end{equation}
where subscripts 1 and 2 denote quantities in the direction parallel and normal to the interface, respectively. Moreover, $K^i$ is the permeability of the isotropic porous medium and $\alpha$ is a constitutive parameter that describes the structure of fluid-porous interface. The presence of such shear-driven slip condition is verified by various studies~\citep{agelinchaab2006,arthur2009,carotenuto2011}, and it is the most widely used condition in the simulation of flows over porous media.

Due to the significant practical importance of understanding the flow behavior over porous media, numerous works have attempted to arrive at more accurate and more general fluid-porous interface conditions. Based on the nature of approximations, these works can be classified as either one- or two-domain approach.

The main advantage of the one-domain approach~\citep{vafai1990,basu1999,goyeau2003,breugem2006,chen2014} is that the flow in the free-fluid and the porous domains are governed by the same form of conservation equations. This is made possible by introducing a heterogeneous transition zone between these two regions. The properties of the flow as well as porous media (porosity, permeability, and apparent viscosity) are assumed to vary rapidly -- but continuously -- within this transition zone to match the values between the two regions. The thickness of the transition zone as well as the functional form of the spatial variation of macroscopic properties within this zone are essential for the complete mathematical description of the flow field. Although many studies have aimed at understanding the nature of the transition layer~\citep{gupte1997,goharzadeh2005,morad2009,khalili2014}, to the best of our knowledge, even for a simple unidirectional channel flow, there are no universally accepted guidelines to specify the properties of this layer. The inability to relate transition layer characteristics to the microstructural details of the porous medium results in an empirical formulation, and hence one-domain approaches require  support from physical experiments.

In the two-domain approach, the free-fluid and the porous domains are separated by a sharp interface. Flows within the porous domain are modeled using the Darcy (or other forms such as the Darcy-Brinkmann or Forchheimer) equation; the Navier-Stokes equations describe the free-fluid region. Across the interface, properties may vary continuously or exhibit jumps~\citep{ochoatapia1995a,ochoatapia1995b,chandesris2006,duman2008}. The crucial factor that dictates the accuracy of two-domain approaches is the \textit{correct} specification of interface conditions that captures transport phenomena across the interface.
%The main challenge is to prescribe coupling conditions at the interface that corresponds to microscale velocity and stress continuity.
Since the experiments of \cite{beavers1967}, several researchers have attempted to derive improved interface conditions by employing either %volume averaging, 
boundary layer matching~\citep{mikelic2001} or homogenization methods~\citep{lacis2016}. The striking feature of this class of methods is the possibility of using microscale geometrical details of the porous medium for computing all coefficients appearing in the interface conditions. This leads to a non-empirical formulation. However, majority of the studies are limited to unidirectional flow over porous layers and are focused on predicting accurate interfacial tangential velocity alone. In the present work, we provide  coupling conditions for tangential and transpiration velocities, as well as on the pressure. Two  works that are relevant to the present formulations are \cite{carraro2013} and \cite{lacis2016}. In \cite{carraro2013}, a pressure jump condition across the fluid-porous interface is proposed, which is active only for anisotropic porous layers. They validated their formulation with microscopic simulations of unidirectional channel flows. \cite{lacis2016} studied two-dimensional flow over (isotropic) porous medium and showed the accuracy of predicting slip and transpiration velocity variation over the interface. Although both studies provide a non-empirical framework, transpiration velocity was absent in the former, while pressure condition was not investigated in the latter.

The present work falls into the category of two-domain approaches. We provide an accurate non-empirical macroscopic description of three-dimensional flow over isotropic as well as anisotropic porous media by employing the multiscale homogenization technique. While previous works~\citep{mikelic2012,lacis2016} have focused on obtaining only leading order conditions at the fluid-porous interface, the present work provides higher-order terms that are essential to represent relevant physical features as discussed in \cite{lacisjfm1}. Moreover, higher order terms are essential to accurately capture transpiration velocity and pressure jump across fluid-porous interfaces. This will be demonstrated in section~\ref{sec:results}.

Similar to porous media, wall roughness also induces a shear-dependent slip velocity~\citep{luchini1991,miksis1994,sarkar1996,stroock2002, kamrin2010,luchini2013,bolanos2017}. For a unidirectional flow, it can be deduced by simply dropping the creeping velocity term from Beavers-Joseph condition given in equation~\eqref{eqn:bj}, yielding
\begin{equation}
u_1=\frac{\sqrt{K^i}}{\alpha}\frac{\partial u_1}{\partial x_2}.
\end{equation}
While the above equation describes a leading order interface condition, few studies have focused on deriving higher order conditions also \citep{achdou1998,bottaro2020}.
The goal of this paper is to rigorously derive  interface conditions for flows over porous media. We will show that a reduced form of these fluid-porous conditions can be used to accurately model flow over rough surfaces. The main features of the derived results are:
\vspace{-0.2cm}\begin{itemize}
\item The proposed method is free of empirical parameters. All the coefficients introduced in the formulation are computed by solving a small set of inexpensive microscale problems.
\item It is applicable to isotropic, anisotropic, and layered porous media that have different microscale geometry at the boundary from that of the interior.
\item A subset of the derived conditions is directly applicable to accurately model the interaction of fluid flows with rough surfaces.
\end{itemize}

%\subsection{Similarities between flows over rough and porous walls}
%Flow over patterned surfaces and porous media exhibit several common characteristics. At low permeabilities, the main effect of porous media is the slip flow at fluid-porous interfaces~\cite{hahn2002}; this feature is the same for walls with smaller roughness scales~\cite{luchini1991}. Moreover, when porosity is increased (or equivalently for a surface patterned with large roughness elements), both classes of flows suffer Kelvin-Helmholtz instabilities~\cite{breugem2006,garcia2011}. Owing to similarities in physics, it is no surprise that the mathematical form of macroscopic description of porous and rough walls also shares commonalities. For example, the interface conditions on flow over rough walls also involves shear-driven slip velocity~\cite{miksis1994,sarkar1996,stroock2002, kamrin2010,luchini2013,bolanos2017}. For a unidirectional flow, it can be deduced by simply dropping the creeping velocity term from Beavers-Joseph condition~(equation~\eqref{eqn:bj})
%\begin{equation}
%u^f=\frac{\sqrt{K^i}}{\alpha}\frac{\partial u^f}{\partial z}
%\end{equation}
%thus enabling the quantification of similar qualitative flow features observed for flows over rough and porous walls. In this paper, we rigorously derive conditions at fluid-porous interface, and same as the aforementioned discussions, we will show that a reduced form of the derived formulations can be used to accurately model flow over rough surfaces.

The paper is organized as follows. The macroscopic model for flows over porous and rough walls is described in section~\ref{sec:main}. For simplicity, \S\ref{sec:main} considers a flat interface aligned along one of the coordinate axes. The complete derivation of the interface conditions using the multiscale homogenization technique is presented in section~\ref{sec:homog}. \S\ref{sec:coeffcompute} elaborates on the auxiliary Stokes problems that are to be solved to compute the constitutive coefficients arising in the interface conditions. In \S\ref{sec:results}, comparison between the proposed macroscopic model and geometry resolved microscopic simulations is presented to demonstrate the accuracy of the present model for rough and porous walls. Finally, conclusions are provided in section~\ref{sec:conc}.
 %========================================================================================
 \section{Main results}
 \label{sec:main}
The main result of the present work is the derivation of accurate interface conditions at fluid-porous contact surfaces. A reduced set of these conditions are explained here for a 2D flow over porous- and rough-surfaces for clarity. The full derivation of more generalized conditions is detailed in section \ref{sec:homog}. It
 will be demonstrated later in section \ref{sec:results} that the reduced conditions introduced here can be used to model macroscopic flows over rough/patterned surfaces.

\begin{figure}
\centering
\includegraphics[trim = 0cm 0cm 0cm 0cm, clip, height=4cm]{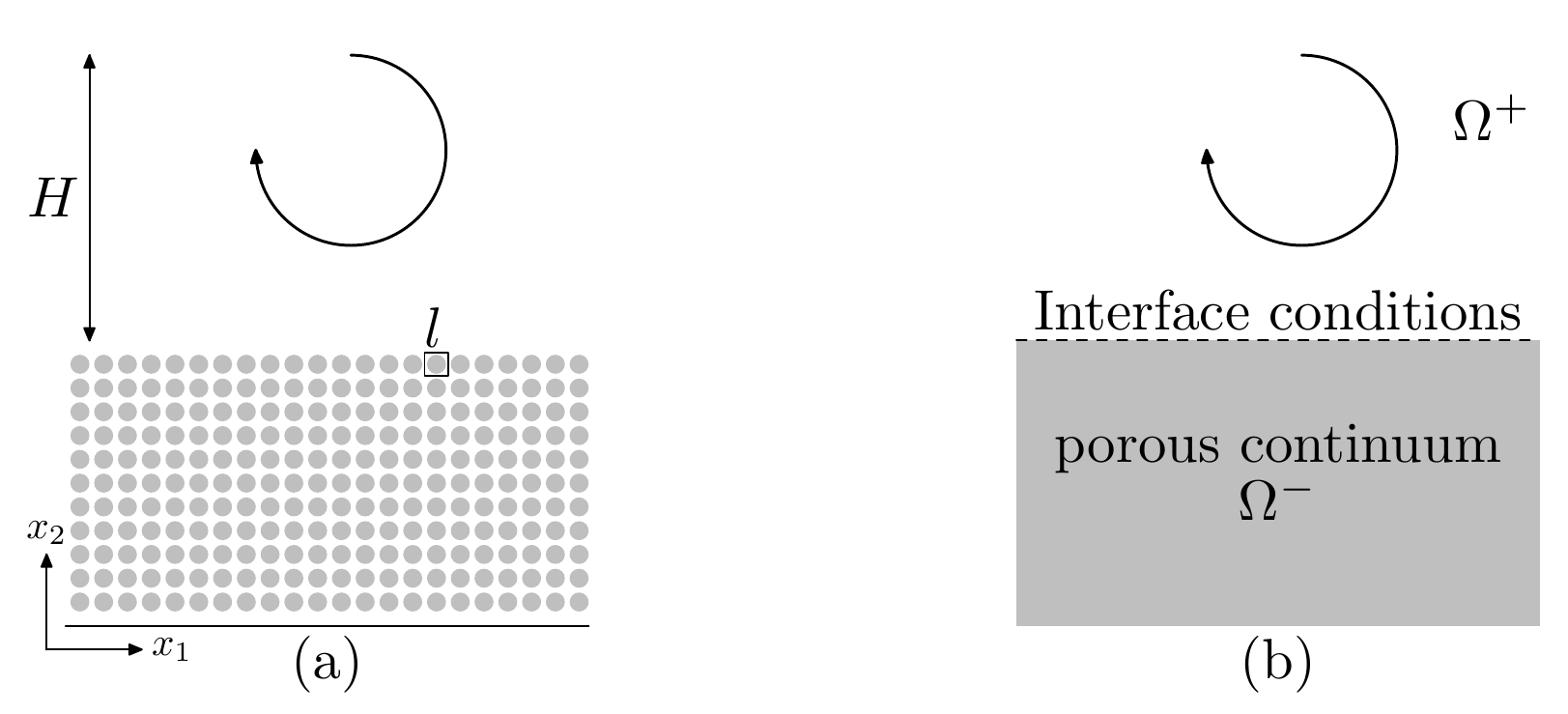}
\caption{Descriptions of flows over a porous domain. (a)~microscopic, and (b)~macroscopic. Interface conditions on velocity and pressure must be specified in the macroscopic description.}
\label{fig:intro-porous}
\end{figure}

 \subsection{Flow over porous walls}
 \label{sec:main-poro}
 Consider an incompressible viscous flow over a porous medium, as shown in figure~\ref{fig:intro-porous}(a). The configuration
 has two distinct length scales: microscale~($l$) denoting the length scale of voids in the porous medium, and macroscale~($H$) characterizing the relevant length scale of the fluid flow. 
 
 %For majority of porous media problems, there exists a strong scale separation i.e., $\epsilon=l/L\ll 1$.% $\epsilon$ is called the scale separation parameter.

 Numerical simulations employing microscopic description~(figure~\ref{fig:intro-porous}a) mandates the resolution of all length scales of the problem. Owing to the inherent multiscale nature of such configurations, microscale simulations are computationally extremely demanding to achieve. This limitation of the microscopic description has motivated the development of macroscopic models~(figure~\ref{fig:intro-porous}b) that govern the behavior of averaged quantities; such models do not require us to resolve all geometrical scales, and are thus feasible for addressing such multiscale problems.

%While the microscopic description of laminar flows over porous media is a multiscale single-physics problem, the macroscopic description is a single-scale multiphysics problem. 
In the macroscopic description~(figure~\ref{fig:intro-porous}b), an interface is introduced above the porous medium, and the domain is split into a free-fluid region~($\Omega^+$) and the porous part~($\Omega^-$). The flow within the porous medium is upscaled to yield the Darcy equation, and the flow in the free-fluid region is governed by the Navier-Stokes equations. The key factor that decides the accuracy of macroscopic models is the appropriate specification of interface conditions that determine all essential features of transport phenomena across the two domains.

In this work, we derive the fluid-porous interface conditions by using multiscale homogenization approach, as detailed in section~\ref{sec:homog}. We invoke the following assumptions.
\begin{enumerate}
\item The porous medium consists of ordered periodic solid inclusions of characteristic length scale $l$, also called as microscopic length scale.
\item The porous medium is homogeneous.
\item The characteristic macroscopic length scale ($H$) is much larger than $l$, i.e., $l \ll H$. This separation of scales is quantified by the parameter, $\epsilon=l/H \ll 1$.
\item Viscous effects dominate inertia near the surface.
\item Characteristic time scale of microscopic flow is much smaller than that of macroscopic flow. This implies that the microscopic flows readjust itself very quickly to any changes in the macroscopic flow. As a consequence, even when the macroscopic flow is unsteady, the behavior of effective microscopic flow need not contain time-dependent terms.
%\item The radius of curvature of the interface is much larger than $l$.
 \end{enumerate}
 
 For flows over porous media, interface conditions on velocity and pressure are needed for complete specification of the problem. Generalized interface conditions that are derived using multiscale homogenization approach are given in equation~\eqref{eqn:epsdim}. For clarity, we present here the interface conditions for 2D flows with horizontal interface.
 The condition for tangential velocity is
 \begin{align}
     u_1&=\underbrace{\mathcal{L}_{11}l\left(\frac{\partial u_1}{\partial x_2}+\frac{\partial u_2}{\partial x_1}\right)}_{\textrm{Slip term}}-\underbrace{\frac{\mathcal{K}_{11}l^2}{\mu}\frac{\partial p^-}{\partial x_1}-\frac{\mathcal{K}_{12}l^2}{\mu}\frac{\partial p^-}{\partial x_2}}_{\textrm{Darcy-like term}} \nonumber \\
     &\ +\underbrace{l^2\left(\mathcal{M}_{111}+\mathcal{K}_{11}\left<B_1^-\right>\right)\left(\frac{\partial ^2u_1}{\partial x_1\partial x_2}+\frac{\partial ^2u_2}{\partial x_1^2}\right)}_{\textrm{term accounting for variation of shear}}. \label{eqn:simpleporoa}
 \end{align}
 The combination of the slip-term and the Darcy-like term in equation~\eqref{eqn:simpleporoa} forms the classical Beavers-Joseph condition. The difference is that $\mathcal{K}_{ij}$ appearing in the above equations denote interface permeability tensor, which is different than the interior permeability used in the Darcy equation. Our model has one additional contribution for $u_1$, which accounts for variation of shear. % and is of order $\mathcal{O}(\epsilon)$. 
 The condition for interface normal velocity is
 \begin{equation}
     u_2=-\underbrace{\frac{\mathcal{K}_{21}l^2}{\mu}\frac{\partial p^-}{\partial x_1}-\frac{\mathcal{K}_{22}l^2}{\mu}\frac{\partial p^-}{\partial x_2}}_{\textrm{Darcy-like term}}+\underbrace{l^2\left(\mathcal{M}_{211}+\mathcal{K}_{21}\left<B_1^-\right>\right)\left(\frac{\partial ^2u_1}{\partial x_1\partial x_2}+\frac{\partial ^2u_2}{\partial x_1^2}\right)}_{\textrm{term accounting for variation of shear}}. \label{eqn:simpleporob}
 \end{equation}
 The transpiration velocity ($u_2$) has also a Darcy-like term, and a high-order term that accounts for the variation of shear along the interface. The latter term is a result of mass conservation, as elaborated in \cite{lacisjfm1}.
 The pressure jump across the interface $\llbracket p \rrbracket=p-p^-$ is
 \begin{align}
     \llbracket p \rrbracket &= \underbrace{\mu\mathcal{B}_1\left(\frac{\partial u_1}{\partial x_2}+\frac{\partial u_2}{\partial x_1}\right)}_{\textrm{PJ-1}}+\underbrace{2\mu\frac{\partial u_2}{\partial x_2}}_{\textrm{PJ-2}}\underbrace{-\mathcal{A}_1l\frac{\partial p^-}{\partial x_1}-\mathcal{A}_2l\frac{\partial p^-}{\partial x_2}}_{\textrm{PJ-3}} \nonumber \\
&\ +\underbrace{\mu l\left(\mathcal{C}_{11}+2\mathcal{L}_{11}+\mathcal{A}_1\left<B_1^-\right>\right)\left(\frac{\partial^2 u_1}{\partial x_1\partial x_2}+\frac{\partial^2 u_2}{\partial x_1^2}\right)}_{\textrm{PJ-4}}.
\label{eqn:simpleporoc}
 \end{align}
The pressure jump derived in the present work has four terms, as marked in equation~\eqref{eqn:simpleporoc}. The PJ-1 term accounts for interface normal force due to the slip velocity \citep{lacisjfm1}; this term is non-zero only for anisotropic porous surfaces and it has been previous derived by \cite{mikelic2012}. The PJ-2 term is a direct consequence of normal stress balance at the interface. The PJ-3 term accounts for the normal force induced at the interface due to wall-normal velocity at the interface \citep{lacisjfm1}. The PJ-4 term is the additional term that arises in our derivation. Here, PJ-3 and PJ-4 terms are the higher order terms, as will be shown in section~\ref{sec:homog}.

In the above equations, $\mathcal{L}_{ij}$, $\mathcal{K}_{ij}$, $\mathcal{M}_{ijk}$, $\mathcal{B}_{i}$, $\mathcal{A}_{i}$, $\mathcal{C}_{ij}$  and $\left<B_1^-\right>$ denote flow-independent constitutive coefficients that are dictated by the microscopic structure of the fluid-porous interface. One of the main feature of this work is that all these coefficients can be computed by solving a number of Stokes problems in a reduced computational domain, as will be explained in section~\ref{sec:coeffcompute}. This renders the entire formulation  free of any empirical parameters.

\subsection{Flow over rough walls}
\begin{figure}
\centering
\includegraphics[trim = 0cm 0cm 0cm 0cm, clip, height=4cm]{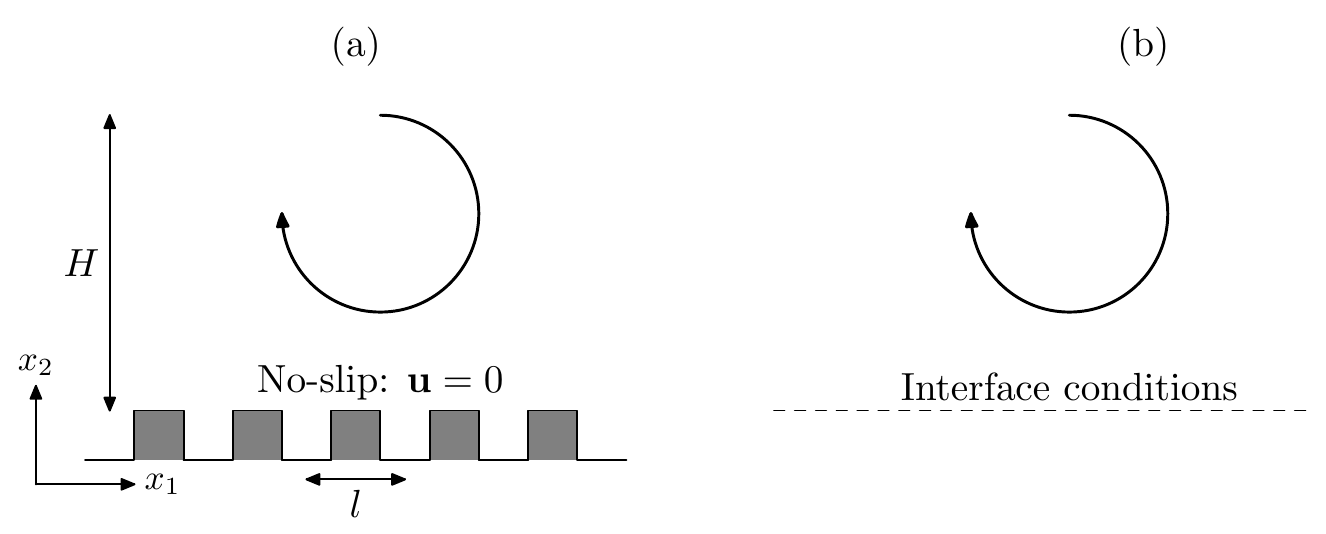}
\caption{Descriptions of flows over a rough wall. (a)~microscopic, and (b)~macroscopic. Interface conditions on velocity  must be specified in the macroscopic description.}
\label{fig:intro-rough}
\end{figure}

  Macroscopic models in this case involve replacing the original rough wall with an equivalent smooth surface that is located slightly above the roughness elements, as shown in figure~\ref{fig:intro-rough}. The domain below the interface is cut off; the effect of geometrical perturbations is converted into perturbations in boundary conditions at the smooth wall, in such a way that these modified boundary conditions represent the averaged effect of roughness on the macroscale flow accurately.
  
 Effective models developed for porous media, in general, contain all essential components for modeling flow over rough walls. The pressure jump condition given in equation~\eqref{eqn:simpleporoc} is redundant as  there is no medium below the interface. Moreover, as will be discussed in section~\ref{sec:coeffcompute}, $\mathcal{K}_{12}=\mathcal{K}_{21}=\mathcal{K}_{22}=\mathcal{M}_{111}=0$. After performing the aforementioned simplifications, the interface conditions for 2D flows over rough walls are as follows:
\begin{align}
u_1&=\mathcal{L}_{11}l\left(\frac{\partial u_1}{\partial x_2}+\frac{\partial u_2}{\partial x_1}\right)-\frac{\mathcal{K}_{11}l^2}{\mu}\frac{\partial p}{\partial x_1} + 2l^2 \mathcal{M}_{121}\frac{\partial ^2 u_2}{\partial x_1\partial x_2}, \\
u_2&=l^2\mathcal{M}_{211}\left(\frac{\partial ^2u_1}{\partial x_1\partial x_2}+\frac{\partial ^2u_2}{\partial x_1^2}\right),
\label{eqn:simplerough}
\end{align}
where $p$ is the free-fluid pressure. We can observe that when compared to interface conditions for porous media (equation~\ref{eqn:simpleporoa}), an additional term for $u_1$ involving $\mathcal{M}_{121}$ appears for rough walls. This arises when rewriting the pore pressure gradient ($\partial p^-/\partial x_1$) in terms of the corresponding gradient of free-fluid pressure. The exact details can be found in Appendix~2. 

The accuracy of these conditions are demonstrated by performing microscopic geometry resolved simulations and comparing the results with our macroscopic model in section~\ref{sec:results}.

%========================================================================================
\section{Multiscale homogenization}
\label{sec:homog}
\begin{figure}
\centering
\includegraphics[trim = 3.25cm 18.5cm 11cm 2.3cm, clip, height=6cm]{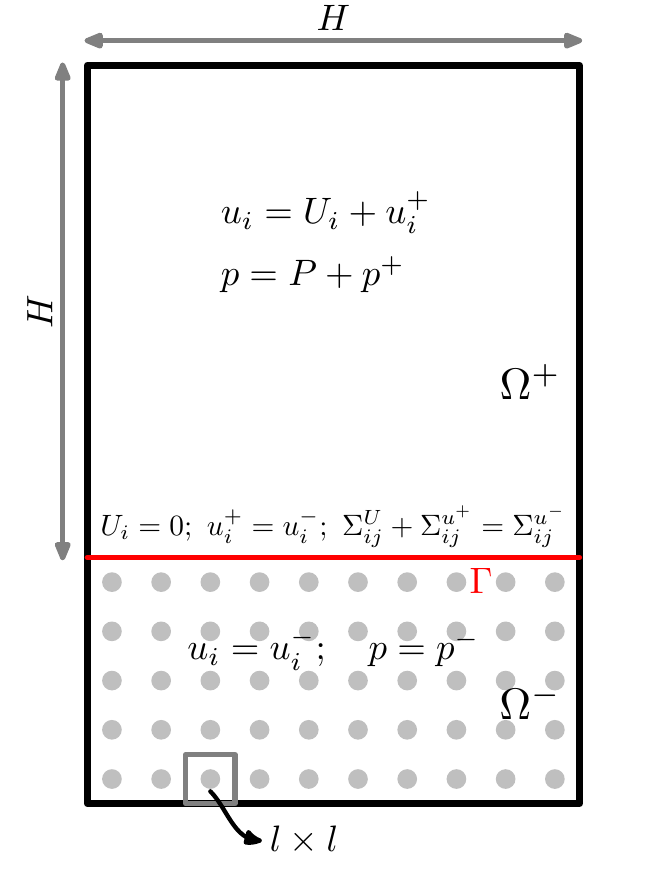}
\caption{Flow over a porous block: Notation and decomposition of flow fields.}
\label{fig:homogeni}
\end{figure}

In this section, we derive coupling conditions at the fluid-porous interface by employing multiscale homogenization technique. We consider isothermal incompressible single-phase flow of a Newtonian viscous fluid over an ordered homogeneous rigid porous medium. It is saturated by the same fluid of constant density $\rho$ and dynamic viscosity $\mu$ (figure~\ref{fig:homogeni}). In the microscopic description, such a fluid flow is described by the Navier-Stokes equations:
\begin{subequations}\label{eqn:apns}
\begin{align}
u_{i,i}&=0,\\
\rho\left(u_{i,t}+u_ju_{i,j}\right)&=-p_{,i}+\mu u_{i,jj},
\end{align}
\end{subequations}
where  $(\bm{\cdot})_{,t}$ and  $(\bm{\cdot})_{,i}$ denote time- and spatial-derivates respectively; $(\bm{\cdot})_{,t}=\frac{\partial (\bm{\cdot})}{\partial
 t}$, $(\bm{\cdot})_{,i}=\frac{\partial (\bm{\cdot})}{\partial x_i}$, and $(\bm{\cdot})_{,ij}=\frac{\partial (\bm{\cdot})}{\partial x_i \partial x_j}$. On the wetted surfaces of the porous medium, the flow obeys no-slip condition: $u_i=0$.

 The physical microscopic configuration contains two distinct length scales: microscopic~($l$) and macroscopic~($H$). As per our assumptions, introduced in section~\ref{sec:main}, scale separation parameter is very small: $\epsilon=l/H \ll 1$. In the macroscopic description, (shown in figure~\ref{fig:homogeni}) a virtual interface~($\Gamma$) is introduced, and the domain is decomposed into free-fluid part~($\Omega^+$) and the porous region($\Omega^-$).

To denote the equations appearing in the same form on both domains, for example
\begin{align*}
a^+_{i,j}&=c^+_{ij}\ \ \ &\textrm{in }\Omega^+,\\
a^-_{i,j}&=c^-_{ij}\ \ \ &\textrm{in }\Omega^-,
\end{align*}
we use the following notation for compactness
\begin{equation*}
a^{\pm}_{i,j}=c^{\pm}_{ij}\ \ \ \textrm{in }\Omega^{\pm}.
\end{equation*}
The homogenization process employed in the present work involves the following six steps.

\subsection*{Step 1: Decomposition of the flow field}
We decompose the flow field in the free fluid region into the sum of flow fields without the presence of porous medium ($U_i,P$) and perturbations introduced due to the porous medium ($u^+_i,p^+$):
\begin{equation}
u_i=U_i+u^+_i;\ \ \ p=P+p^+\ \ \ \textrm{in }\Omega^+.
\label{eqn:decompff}
\end{equation}
Within the porous medium, flow quantities ($u^-_i,p^-$) are only due to the porous block:
\begin{equation}
u_i=u^-_i;\ \ \ p=p^-\ \ \ \textrm{in }\Omega^-.
\label{eqn:decompporo}
\end{equation}
On the fluid-porous interface, the following conditions are met 

\begin{equation}
U_i=0;\ \ \  u^+_i=u^-_i; \ \ \ \Sigma^{U}_{ij}n_j+\Sigma^{u^+}_{ij}n_j=\Sigma^{u^-}_{ij}n_j\ \ \  \textrm{on }\Gamma,
\label{eqn:strdec}
\end{equation}
where the stress tensors are defined as,
\begin{equation}
\begin{aligned}
\Sigma^{U}_{ij}&=-P\delta_{ij}+\mu\left(U_{i,j}+U_{j,i}\right),\\
\Sigma^{u^+}_{ij}&=-p^+\delta_{ij}+\mu\left(u^+_{i,j}+u^+_{j,i}\right),\\
\Sigma^{u^-}_{ij}&=-p^-\delta_{ij}+\mu\left(u^-_{i,j}+u^-_{j,i}\right).
\end{aligned}
\end{equation}
This decomposition will be used in step 3 for obtaining coupled equations governing the perturbation quantities.

\subsection*{Step 2: Order estimates}
It is, both experimentally and numerically, well established that the porous medium introduces a non-zero slip velocity~($U^s$) at the interface $\Gamma$. Since this slip velocity is the manifestation of the fluctuating velocities of porous medium near the interface, it is natural to assume that
\begin{equation}
u^{\pm}\sim U^s.
\end{equation}
Using this estimate, and also assuming that the perturbation shear stress above the interface is much smaller than the free fluid shear stress $u^+ / l \ll U^f / H$, we can write the following estimate based on the stress balance (equation~\ref{eqn:strdec}),
\begin{equation}
\frac{u^-}{l}\sim \frac{U^s}{l}\sim \frac{U^f}{H},
\end{equation}
where $U^f$ is the characteristic fast flow velocity. The above expression yields
\begin{equation}
U^s\sim\epsilon U^f.
\end{equation}
The pressure fluctuations scales naturally with the macroscopic pressure difference ($\Delta P$)
\begin{equation}
p^{\pm}\sim \Delta P.
\end{equation}
Moreover, we assume that the pressure forces balances viscous effects
\begin{equation}
    \frac{\Delta P}{l}\sim \frac{\mu U_s}{l^2},
\end{equation}
which provides the following estimate
\begin{equation}
U_s\sim \frac{l\Delta P}{\mu}.
\label{eqn:delp-u-relation}
\end{equation}
This inherently leads to the assumption of negligible inertia.

\subsection*{Step 3: Non-dimensionalization of the governing equations}
Using the above estimates,  we introduce the following non-dimensionalization of variables in the whole domain, with overbar denoting dimensionless quantities,
\begin{equation}
x_i=\overline{x}_il,\ \ \ u_i=\overline{u}_iU^s,\ \ \ p=\overline{p}\Delta P=\frac{\overline{p}U_s\mu}{l},\ \ \ t=\overline{t}T.
\label{eqn:nondimvariables}
\end{equation}
Here, $T$ denotes the time scale associated with $(U_i,P)$, and we have used equation~\eqref{eqn:delp-u-relation} for pressure. Introducing the above non-dimensionalization and removing overbar for convenience, the Navier-Stokes equations become
\begin{equation}
\begin{aligned}
Re_s\left(St\, u_{i,t}+u_ju_{i,j}\right)&=-p_{,i}+u_{i,jj},\\
u_{i,i}&=0,
\end{aligned}
\label{eqn:non-dim}
\end{equation}
where $Re_s=\rho U^s l/\mu$ is the slip Reynolds number, and $St=l/U^sT$ is the Strouhal number, which relates time scales associated with fast flow and fluctuations. Introducing the decomposition from step 1, we get the governing equations for the fast flow
\begin{equation}
\begin{aligned}
U_{i,i}&=0\ \ \ &\textrm{in }\Omega^+,\\
Re_s \left(St\, U_{i,t}+U_jU_{i,j}\right)&=-P_{,i}+U_{i,jj} &\textrm{in }\Omega^+,\\
U_i&=0 &\textrm{on }\Gamma,
\end{aligned}
\label{eqn:fastflow}
\end{equation}
and the following set of coupled equations for perturbation quantities
\begin{equation}
\begin{aligned}
u^{\pm}_{i,i}&=0&\textrm{in }\Omega^{\pm},\\
Re_s \left(St\, u^+_{i,t}+u^+_ju^+_{i,j}+u^+_jU_{i,j}+U_ju^+_{i,j}\right)&=-p^+_{,i}+u^+_{i,jj} &\textrm{in }\Omega^+,\\
Re_s \left(u^-_{i,t}+u^-_ju^-_{i,j}\right)&=-p^-_{,i}+u^-_{i,jj} &\textrm{in }\Omega^-,\\
u^+&=u^-&\textrm{on }\Gamma,\\
\Sigma^{U}_{ij}n_j+\Sigma^{u^+}_{ij}n_j&=\Sigma^{u^-}_{ij}n_j&\textrm{on }\Gamma,
\end{aligned}
\label{eqn:pertcoupled}
\end{equation}
where the stress tensors are defined as,
\begin{equation}
\begin{aligned}
\Sigma^{U}_{ij}&=-P\delta_{ij}+\left(U_{i,j}+U_{j,i}\right),\\
\Sigma^{u^+}_{ij}&=-p^+\delta_{ij}+\left(u^+_{i,j}+u^+_{j,i}\right),\\
\Sigma^{u^-}_{ij}&=-p^-\delta_{ij}+\left(u^-_{i,j}+u^-_{j,i}\right).
\end{aligned}
\label{eqn:strnondim}
\end{equation}
By solving the coupled system of PDEs given in equation~\eqref{eqn:pertcoupled} in the interface region (a domain sized only a fraction of full domain as discussed later), we can quantify the effect of the porous medium on the overlying free flow. For such a problem, the computational domain will carry zero stress and the interior Darcy solution at the top and bottom boundaries respectively. However the direct solution of perturbation quantities is not practically feasible, therefore in the next steps, we employ formal homogenization to analytically quantify the averaged effect of perturbation introduced by the porous medium.

\subsection*{Step 4: Multiscale expansion of perturbation quantities}
In most practical applications involving porous media, there exists a clear scale separation, $\epsilon = l/H \ll 1$. This enables us to employ the method of multiple scales by introducing fast and slow coordinates $x$ and $X=\epsilon x$, respectively.  The perturbation quantities are assumed to be functions of both coordinates i.e., $u^{\pm}_i=u^{\pm}_i(x,X)$ and $p^{\pm}=p^{\pm}(x,X)$. According to the chain rule, the derivative is written as,
\begin{equation}
\frac{\partial (\cdot)}{\partial x}=\frac{\partial (\cdot)}{\partial x}+\epsilon\frac{\partial (\cdot)}{\partial X}=(\cdot)_{,i}+\epsilon(\cdot)_{,I}.
\end{equation}
Here, the first and second term on the right side denote derivative with respect to microscale and macroscale coordinates. We assume that fluctuation quantities can be written in the form of perturbation expansion as follows
\begin{equation}
\begin{aligned}
u^{\pm}_i(x,X) &=\epsilon ^{0}u^{\pm(0)}_i(x,X)+\epsilon ^{1}u^{\pm(1)}_i(x,X)+\epsilon ^{2}u^{\pm(2)}_i(x,X)+...,\\
p^{\pm}(x,X) &=\epsilon ^{0}p^{\pm(0)}(x,X)+\epsilon ^{1}p^{\pm(1)}(x,X)+\epsilon ^{2}p^{\pm(2)}(x,X)+...
\end{aligned}
\label{eqn:expansion}
\end{equation}
Substituting the above expansion into the governing equations of fluctuation quantities (equation~\ref{eqn:pertcoupled}), and collecting the terms corresponding to different orders of $\epsilon$, we obtain the following hierarchy of equations. We make use of assumptions (4) and (5) presented in section~\ref{sec:main-poro}, which leads to $Re_s \le \epsilon^2$ and $St\le 1$, respectively.

\noindent \textit{Continuity equation:}
\begin{align}
\label{eqn:c0}
\mathcal{O}\left(\epsilon ^{0}\right):\ \ \ & u_{i,i}^{\pm (0)}=0,\\
\label{eqn:c1}
\mathcal{O}\left(\epsilon ^{1}\right):\ \ \ & u_{i,i}^{\pm (1)}+u_{i,I}^{\pm (0)}=0.
\end{align}

\noindent \textit{Momentum equations:}
\begin{align}
\label{eqn:m0}
\mathcal{O}\left(\epsilon ^{0}\right):\ \ \ & -p^{\pm(0)}_{,i}+u^{\pm (0)}_{i,jj}=0,\\
\label{eqn:m1}
\mathcal{O}\left(\epsilon ^{1}\right):\ \ \ & -p^{\pm(1)}_{,i}+u^{\pm (1)}_{i,jj}=-\left[-p^{\pm(0)}_{,I}+2u^{\pm (0)}_{i,Jj}\right].
\end{align}
\noindent \textit{Interface conditions:}
\begin{align}
\label{eqn:vbc0}
\mathcal{O}\left(\epsilon ^{0}\right):\ \ \ & u^{+(0)}_i=u^{-(0)}_i,\\
\label{eqn:vbc1}
\mathcal{O}\left(\epsilon ^{1}\right):\ \ \ & u^{+(1)}_i=u^{-(1)}_i,\\
\label{eqn:strbc0}
\mathcal{O}\left(\epsilon ^{0}\right):\ \ \ & \Sigma^{U}_{ij}n_j+\Sigma^{u^{+(0)}}_{ij}n_j=\Sigma^{u^{-(0)}}_{ij}n_j,\\
\label{eqn:strbc1}
\mathcal{O}\left(\epsilon ^{1}\right):\ \ \ & \left\llbracket -p^{(1)}\delta_{ij}n_j+\left(u_{i,j}^{(1)}+u_{j,i}^{(1)}\right)n_j+\left(u_{i,J}^{(0)}+u_{j,I}^{(0)}\right)n_j\right\rrbracket=0,
\end{align}
where the stress tensors $\Sigma_{ij}$ are defined in equation~\eqref{eqn:strnondim}, and the jump identity across the interface is defined as, $\left\llbracket\left(\bm{\cdot}\right)\right\rrbracket=\left(\bm{\cdot}\right)^+-\left(\bm{\cdot}\right)^-$.

\subsection*{Step 5: $\mathcal{O}(\epsilon ^0)$-formulation}
The leading order  governing equations for perturbation quantities are given by equations~\eqref{eqn:c0} and \eqref{eqn:m0} together with coupling conditions \eqref{eqn:vbc0} and \eqref{eqn:strbc0}. They constitute a Stokes problem in terms of $\left(u^{\pm(0)},p^{\pm(0)}\right)$ as shown below:

\begin{subequations}\label{ap:0order}
\begin{align}
u_{i,i}^{\pm (0)}&=0\ \ \ & \textrm{in } \Omega^{\pm},\\
-p^{{\pm} (0)}_{,i}+u^{{\pm} (0)}_{i,jj}&=0 & \textrm{in } \Omega^{\pm},\\
u^{+(0)}_i&=u^{-(0)}_i& \textrm{on } \Gamma, \\
\Sigma^{U}_{ij}n_j+\Sigma^{u^{+(0)}}_{ij}n_j&=\Sigma^{u^{-(0)}}_{ij}n_j& \textrm{on } \Gamma.
\label{eqn:0orderstr}
\end{align}
\end{subequations}
We can infer from equation~\eqref{eqn:0orderstr} that the flow is driven by the known traction induced at the interface due to  fast flow ($\Sigma^{U}_{ij}n_j$). The fast flow is decoupled from the perturbation quantities, and the traction can be obtained by solving equation~\eqref{eqn:fastflow}. Using this fact, and owing to the linearity of the problem, the solution to this system of equations can be constructed as follows
\begin{equation}
\begin{aligned}
u^{\pm (0)}_i&=L^{\pm}_{ij}(x)S_{j}(X),\\
p^{+(0)}&=B^+_{i}(x)S_{i}(X),\\
p^{-(0)}&=B^-_{i}(x)S_{i}(X)+P^-(X),
\end{aligned}
\label{eqn:0ansatz}
\end{equation}
where $S_{i}=\left(U_{i,j}\vert_{\Gamma}+U_{j,i}\vert_{\Gamma}\right)n_j$ is the traction due to viscous stresses of fast flow, in which $n_j$ denotes unit normal vector to the interface. The integration constant $P^-(X)$ is dependent only on macroscale coordinates, such that at the interface it reaches the value of fast flow pressure: $P^-(X)\vert_{\Gamma}=P\vert_{\Gamma}$. Such constants in first two of the above equations are set to zero respectively by (i)~no-slip condition on the wetted surfaces of porous solids, and (ii)~zero-stress condition at the top of the domain in microscale problems, as will be discussed with section~\ref{sec:coeffcompute}. In principle, one could retain all constants and all forcing terms in the solution; this would, however, lead to trivial microscale problems confirming that these terms indeed are zero.

Substituting the velocity Ansatz given by equation~\eqref{eqn:0ansatz} into the multiscale expansion given in equation~\eqref{eqn:expansion}, the perturbation velocity in the free-fluid domain can be written as
\begin{equation}
u_i^+=L_{ij}^+S_j+\mathcal{O}(\epsilon),
\label{eqn:ui+temp}
\end{equation}
and the perturbation pressure in both domains as
\begin{equation}
\begin{aligned}
p^+&=B^+_{j}S_j+\mathcal{O}(\epsilon),\\
p^-&=B^-_{j}S_j+P^-+\mathcal{O}(\epsilon).
\end{aligned}
\label{eqn:p+-temp}
\end{equation}
Inserting equation~\eqref{eqn:ui+temp} into the decomposition given in equations~\eqref{eqn:decompff}, we get
\begin{equation}
u_i=U_i+L_{ij}^+S_j+\mathcal{O}(\epsilon).
\label{eqn:ui+temp1}
\end{equation}
Note that $u_i^-$ is not considered in the above steps because we are interested in the effect of the porous medium on the free-fluid region, which can be obtained by considering $u_i^+$ alone. A complete specification of coupling conditions at the fluid-porous interface requires conditions on the free-fluid velocities, and a condition relating the stresses in both domains.

Taking average of equations~\eqref{eqn:ui+temp1} and \eqref{eqn:p+-temp}, and using the fact that $U_i\vert_{\Gamma}=0$ and $P^-\vert_{\Gamma}=P\vert_{\Gamma}$, we get
\begin{align}
\label{eqn:zerobcinteru}\left<u_i\right>^s&=\left<L^+_{ij}\right>(U_{j,k}\vert_{\Gamma}+U_{k,j}\vert_{\Gamma})n_k+\mathcal{O}(\epsilon), \\
\left\llbracket p \right\rrbracket &=\left[\left< B_{j}^+\right>-\left< B_{j}^- \right>\right](U_{j,k}\vert_{\Gamma}+U_{k,j}\vert_{\Gamma})n_k+\mathcal{O}(\epsilon),
\label{eqn:zerobcinterp}
\end{align}
where superscript $s$ in velocity indicates surface averaging as explained below, and the pressure jump is given by $\left\llbracket p \right\rrbracket=P+\left< p^+\right>-\left< p^-\right>$. From the decomposition introduced in step 1, it is clear that $P+p^+$ and $p^-$ represent pressure in free-fluid and porous domain respectively. The averaging operator used in the above equations involves a surface integral on interface and a volume integral over the respective domain for quantities associated with velocities and  pressure respectively,
\begin{align}
\left<a\right>^s(\textbf{X})&=\frac{1}{l^2}\int_{-l/2}^{l/2}\int_{-l/2}^{l/2}{a^+(\textbf{x}-\textbf{X})d\Gamma}\ \ \ &\textrm{for velocity,}\\
\left<b^{\pm}\right>(\textbf{X})&=\frac{1}{\Omega_{\textrm{avg}}}\int_{-l/2}^{l/2}\int_{-l/2}^{l/2}\int_{-l/2}^{l/2}{b^{\pm}(\textbf{x}-\textbf{X})d\Omega^{\pm}}\ \ \ &\textrm{for pressure}.
\end{align}
Detailed discussion about the averaging is provided in section~\ref{sec:coeffcomputeL}. For brevity, the superscript in  averaged velocity is neglected hereafter.

In order to make equations~\eqref{eqn:zerobcinteru} and \eqref{eqn:zerobcinterp} into useful conditions at the fluid porous interface, the fast scale quantities $(U_i,P)$ should be written in terms of the actual flow field quantities $(u_i,p)$. This can be obtained by differentiating the following
\begin{equation}
    \left<u_i\right>=U_i+\left<u_i^{+(0)}\right>+\mathcal{O}(\epsilon),
\end{equation}
which yields
\begin{equation}
\left<u_i\right>_{,j}=U_{i,j}+\left<u_i^{+(0)}\right>_{,j}+\mathcal{O}(\epsilon).
\end{equation}
Substituting the Ansatz for $u^{+(0)}$ in the above expression, we get
\begin{equation}
\left<u_i\right>_{,j}=U_{i,j}+\left<L^+_{ik}\right>_{,j}S_k+\mathcal{O}(\epsilon),
\end{equation}
where the macroscale derivative is absorbed into $\mathcal{O}(\epsilon)$ term. Since $\left<L^+_{ik}\right>$ is constant (section~\ref{sec:coeffcompute}), its spatial derivative is zero. Hence we can write
\begin{equation}
\left<u_i\right>_{,j}=U_{i,j}+\mathcal{O}(\epsilon).
\label{eqn:gradlead}
\end{equation}
Substituting the above expression into equations \eqref{eqn:zerobcinteru} and \eqref{eqn:zerobcinterp}, and neglecting the averaging operator for clarity, we get the final form of the leading order interface conditions:
\begin{equation}
\begin{aligned}
u_i&=\mathcal{L}_{ij}(u_{j,k}\vert_{\Gamma}+u_{k,j}\vert_{\Gamma})n_k+\mathcal{O}(\epsilon),\\
\left\llbracket p \right\rrbracket &=\mathcal{B}_{j}(u_{j,k}\vert_{\Gamma}+u_{k,j}\vert_{\Gamma})n_k+\mathcal{O}(\epsilon),
\label{eqn:zerobc}
\end{aligned}
\end{equation}
where $\mathcal{L}_{ij}=\left<L^+_{ij}\right>$ and $\mathcal{B}_{i}=\left< B_{i}^+\right>-\left< B_{i}^- \right>$. The second order tensor $\mathcal{L}^{\pm}_{ij}$ and vector $\mathcal{B}^{\pm}_{i}$ are constitutive coefficients of the fluid-porous interface that are dependent only on the geometry of the porous material near the interface. Computation of these coefficients are discussed in detail in section~\ref{sec:coeffcompute}.

The aforementioned interface condition for velocity and a part of pressure jump condition have already been reported in \cite{mikelic2000,mikelic2009,mikelic2012,carraro2018}. As will be explained in section~\ref{sec:coeffcompute}, $\mathcal{L}_{2j}=0$, and as a result $u_2=0$. As a consequence, the flow within the porous medium and  in the overlying region are completely decoupled from each other; the effect of porous medium is only to introduce a slip velocity at the interface. This suggests that  existing models do not capture full physical features of the transport phenomena at the fluid-porous interface.

Based on mass conservation arguments, it has been shown in \cite{lacisjfm1} that $u_2=0$ only when the tangential velocity is constant along the entire length of the interface. When $\partial u_1/\partial x_1\neq 0$, then mass conservation requires that $u_2 \neq 0$. In order to rigorously introduce this physics in our model, we will solve next order perturbation problem in the following.

\subsection*{Step 6: $\mathcal{O}(\epsilon ^1)$ formulation}
Equations~\eqref{eqn:c1} and \eqref{eqn:m1} together with interface conditions \eqref{eqn:vbc1} and \eqref{eqn:strbc1} form the perturbation problem for next order:
\begin{equation*}
\begin{aligned}
u^{{\pm} (1)}_{i,i}&=-u^{{\pm} (0)}_{i,I}\ \ \ & \textrm{in } \Omega^{\pm},\\
-p_{,i}^{{\pm} (1)}+u^{{\pm}(1)} _{i,jj}&=-\left[-p^{{\pm}(0)}_{,I}+2u^{{\pm}(0)}_{i,jJ}\right] & \textrm{in } \Omega^{\pm},\\
u^{+(1)}_i&=u^{-(1)}_i& \textrm{on } \Gamma, \\
\Sigma_{ij}^{u^{+(1)}}n_j+\left(u^{+(0)}_{i,J}+u^{+(0)}_{j,I}\right)n_j&=\Sigma_{ij}^{u^{-(1)}}n_j+\left(u^{-(0)}_{i,J}+u^{-(0)}_{j,I}\right)n_j& \textrm{on } \Gamma.
\end{aligned}
\end{equation*}
This is a Stokes problem for the unknowns $\left(u^{\pm(1)},p^{\pm(1)}\right)$, forced by the solution of $\mathcal{O}(\epsilon ^0)-$problem. 

Substituting the Ansatz for $\mathcal{O}(\epsilon ^0)$ problem (equation~\ref{eqn:0ansatz}) in to the above equation and using the fact that (i)~$L_{ij}$ and $B_i$ are independent of macroscale coordinate $X$, and (ii)~$S_j$ is independent of microscale coordinate $x$, we obtain the following system of equations for $\mathcal{O}(\epsilon ^1)$ problem:
\begin{subequations}\label{eqn:1eqn}
\begin{align}
u^{{\pm}(1)}_{i,i}&=-L_{ik}^{\pm}D_{ki}\ \ \ & \textrm{in } \Omega^{\pm},\\
-p_{,i}^{+(1)}+u^{+(1)} _{i,jj}&=-\Sigma^{L'^{+}}_{ij}D_{kj} & \textrm{in } \Omega^+,\\
-p_{,i}^{-(1)}+u^{-(1)} _{i,jj}&=P^-_{,I}-\Sigma^{L'^{-}}_{ij}D_{kj} & \textrm{in } \Omega^-,\\
u^{+(1)}_i&=u^{-(1)}_i& \textrm{on } \Gamma, \\
\Sigma_{ij}^{u^{+(1)}}n_j&=\Sigma_{ij}^{u^{-(1)}}n_j& \textrm{on } \Gamma,
\end{align}
\end{subequations}
where $D_{ij}=S_{i,J}$ and $\Sigma^{L'^{\pm}}_{ij}=-B^{\pm}_{k}\delta_{ij}+2L^{\pm}_{ik,j}$. Due to linearity, the solution of the above problem can be written as
\begin{subequations}\label{eqn:1ansatz}
\begin{align}
u_i^{\pm(1)}&=-K_{ij}^{\pm}P^-_{,J}+M_{ijk}^{\pm}D_{jk},\\
p^{\pm(1)}&=-A_{i}^{\pm}P^-_{,I}+C_{ij}^{\pm}D_{ij},
\end{align}
\end{subequations}
where $K_{ij}$, $L_{ij}$, $A_i$ and $C_{ij}$ are constitutive tensors that depend on the structure of the porous medium at the interface.

In order arrive at the $\mathcal{O}(\epsilon)$ interface conditions, we use the decomposition and multiscale expansion introduced in equations~\eqref{eqn:decompff} and \eqref{eqn:expansion}, respectively,
\begin{align}
\label{eqn:uiepstemp}
u_i&=U_i+u_i^{+(0)}+\epsilon u_i^{+(1)}+\mathcal{O}(\epsilon ^2)\ \ \ &\textrm{on}\ \Omega^+, \\
\label{eqn:ppepstemp}
p^+&=P+p^{+(0)}+\epsilon p^{+(1)}+\mathcal{O}(\epsilon ^2)\ \ \ &\textrm{on}\ \Omega^+, \\
p^-&=p^{-(0)}+\epsilon p^{-(1)}+\mathcal{O}(\epsilon ^2)\ \ \ &\textrm{on}\ \Omega^-.
\label{eqn:pmepstemp}
\end{align}
Averaging equation~\eqref{eqn:uiepstemp} at the interface, substituting the Ansatz of the problem, and using the fact that $U_i\vert_{\Gamma}=0$, we get
\begin{equation}
\begin{aligned}
\left<u_i\right>&=\mathcal{L}_{ij} S_j \vert_{\Gamma}+\epsilon \mathcal{M}_{ijk}D_{jk}\vert_{\Gamma}-\epsilon \mathcal{K}_{ij}P^-_{,J}\vert_{\Gamma}+\mathcal{O}(\epsilon ^2), \\
\left\llbracket p \right\rrbracket &=\mathcal{B}_iS_i+\epsilon\mathcal{C}_{ij}D_{ij}-\epsilon\mathcal{A}_iP^-_{,I}+\mathcal{O}(\epsilon ^2).
\end{aligned}
\label{eqn:highbctemp}
\end{equation}
The above pressure jump condition is obtained by taking the average of difference between equations~\eqref{eqn:ppepstemp} and \eqref{eqn:pmepstemp}. The notations introduced in these equations are $\mathcal{L}_{ij}=\left<L^+_{ij}\right>$, $\mathcal{M}_{ijk}=\left<M^+_{ijk} \right>$, $\mathcal{K}_{ij}=\left<K^+_{ij}\right>$,  $\mathcal{B}_i=\left< B_{i}^+\right>-\left< B_{i}^- \right>$, $\mathcal{C}_{ij}=\left< C_{ij}^+\right>-\left< C_{ij}^- \right>$, and $\mathcal{A}_i=\left< A_{i}^+\right>-\left< A_{i}^- \right>$. They are constitutive coefficients that depend only on the geometry of fluid-porous interface.
In equation~\eqref{eqn:highbctemp}, all terms of the RHS contain fast flow quantities $(U_i,P)$. In order to transfer the equation into interface conditions, we need to express these terms ($S_j$, $D_{ij}$, and $P^-_{,J}$) in terms of their counterparts defined with actual fields $(u_i,p)$. These expressions are obtained here as follows.
\subsubsection*{Relation between $U_{i,j}$ and $u_{i,j}$}
Averaging equation~\eqref{eqn:uiepstemp}, and taking derivative, we get
\begin{equation}
\left<u_i\right>_{,j}=U_{i,j}+\left<u^{+(0)}_i \right>_{,j}+\epsilon\left<u^{+(0)}_i\right>_{,J}+\epsilon\left<u_i^{+(1)}\right>_{,j}+\mathcal{O}(\epsilon ^2).
\label{eqn:uUtemp}
\end{equation}
From our earlier discussions in the leading order problem, the second term on the right hand side is zero, and the equation becomes
\begin{equation}
\left<u_i\right>_{,j}=U_{i,j}+\epsilon\left<u_i^{+(0)}\right>_{,J}+\epsilon\left<u_i^{+(1)}\right>_{,j}+\mathcal{O}(\epsilon ^2).
\label{eqn:rel1temp}
\end{equation}
Substituting the Ansatz for $u_i^{+(0)}$ and  $u_i^{+(1)}$ in the above equation,
\begin{equation}
\left<u_i\right>_{,j}=U_{i,j}+\epsilon\left(\mathcal{L}_{ik}S_k\right)_{,J}+\epsilon\left<M^+_{ilm}\right>_{,j}D_{lm}+\mathcal{O}(\epsilon ^2),
\end{equation}
and using the fact that $\mathcal{L}_{ij}$ is macroscale independent, we can write
\begin{equation}
\left<u_i\right>_{,j}=U_{i,j}+\epsilon\mathcal{L}_{ik}D_{kj}+\epsilon\left<M^+_{ilm}\right>_{,j}D_{lm}+\mathcal{O}(\epsilon ^2).
\label{eqn:gradhigh}
\end{equation}
Substituting the above expression in the definition of $S_j=\left(U_{j,k}+U_{k,j}\right)n_k$, we get
\begin{equation}
\begin{split}
S_j=&s_j-\epsilon\mathcal{L}_{km}D_{mj}n_k-\epsilon\mathcal{L}_{jm}D_{mk}n_k-\epsilon\left<M^+_{klm}\right>_{,j}D_{lm}n_k\\
&-\epsilon\left<M^+_{jlm}\right>_{,k}D_{lm}n_k
+\mathcal{O}(\epsilon ^2),
\end{split}
\label{eqn:expr1}
\end{equation}
where $s_j=(u_{j,k}+u_{k,j})n_k$ denotes traction due to viscous stresses of the physical fluid variables. In order to gain additional confidence in the derived relationship between gradients of fastflow and the full velocity field at leading order (equation~\ref{eqn:gradlead}) and at the higher order order (equation~\ref{eqn:gradhigh}), we provide validation of these expressions in Appendix~1.
\subsubsection*{Relation between $U_{i,jk}$ and $u_{i,jk}$}
Taking derivative of equation~\eqref{eqn:rel1temp}, we obtain
\begin{equation}
U_{i,kJ}=\frac{1}{\epsilon}u_{i,kj}+\mathcal{O}(\epsilon ^2)
\end{equation}
This leads to the following relation
\begin{equation}
D_{ij}=\frac{1}{\epsilon}d_{ij}.
\label{eqn:expr2}
\end{equation}
where $d_{ij}=s_{i,j}$. The above expression will be used to express velocity diffusion tensor $D_{ij}$ in terms of physical fluid variables.
\subsubsection*{Relation between $P^-_{,I}$ and $p^-_{,i}$}
Consider the multiscale expansion for pressure perturbation inside the porous medium,
\begin{equation}
p^-=p^{-(0)}+\epsilon p^{-(1)}+\mathcal{O}(\epsilon ^2).
\end{equation}
Averaging the above equation at the interface, and taking its derivative we get
\begin{equation}
\left<p^-\right>_{,i}=\epsilon\left<p^{-(0)}\right>_{,I}+\mathcal{O}(\epsilon ^2).
\end{equation}
Substituting the Ansatz for $p^{-(0)}$ from equation~\eqref{eqn:0ansatz}, we get the required relation,
\begin{equation}
\epsilon \,P^-_{,I} = \left<p^-\right>_{,i}-\epsilon\,\mathcal{B}_k^-D_{ki}+\mathcal{O}(\epsilon ^2).
\label{eqn:expr3}
\end{equation}

\subsubsection*{Final interface conditions}
By substituting equations~\eqref{eqn:expr1}, \eqref{eqn:expr2}, and \eqref{eqn:expr3} into equation~\eqref{eqn:highbctemp}, and neglecting the averaging operator for clarity, we get the higher order interface conditions:
\begin{equation}
\begin{aligned}
u_i&=\mathcal{L}_{ij}\left(u_{j,k}+u_{k,j}\right)n_k+\hat{\mathcal{M}}_{ijk}\left(u_{j,lk}+u_{l,jk}\right)n_l-\mathcal{K}_{ij}p^-_{,j}+\mathcal{O}(\epsilon^2),\\
\llbracket p \rrbracket &=\mathcal{B}_i\left(u_{i,j}+u_{j,i}\right)n_j+\hat{\mathcal{C}}_{jk}\left(u_{j,lk}+u_{l,jk}\right)n_l-\mathcal{A}_ip^-_{,i}+\mathcal{O}(\epsilon^2),
\end{aligned}
\label{eqn:epsnondim}
\end{equation}
where,
\begin{align*}
\hat{\mathcal{M}}_{ijk}&=\mathcal{M}_{ijk}
-\mathcal{L}_{ik}\mathcal{L}_{mj}n_m-\mathcal{L}_{im}\mathcal{L}_{mj}n_k
-\mathcal{L}_{il}\left<M^+_{ljk}\right>_{,m}n_m
-\mathcal{L}_{il}\left<M^+_{mjk}\right>_{,l}n_m
+\mathcal{K}_{ik}\left<B_j^-\right>,\\
\hat{\mathcal{C}}_{jk}&=\mathcal{C}_{jk}-\mathcal{B}_k\mathcal{L}_{mj}n_m-\mathcal{B}_m\mathcal{L}_{mj}n_k
-\mathcal{B}_l\left<M^+_{mjk}\right>_{,l}n_m
-\mathcal{B}_l\left<M^+_{ljk}\right>_{,m}n_m
+\mathcal{A}_k\left<B^-_j\right>.
\end{align*}

In the dimensional setting, the interface conditions presented in equation~\eqref{eqn:epsnondim} become:
\begin{subequations}
\begin{empheq}[box=\fbox]{align}
u_i&=\mathcal{L}_{ij}l\left(u_{j,k}+u_{k,j}\right)n_k+\hat{\mathcal{M}}_{ijk}l^2\left(u_{j,lk}+u_{l,jk}\right)n_l-\frac{\mathcal{K}_{ij}l^2}{\mu}p^-_{,j},\\
\llbracket p \rrbracket &=\mu\mathcal{B}_i\left(u_{i,j}+u_{j,i}\right)n_j+\mu l\hat{\mathcal{C}}_{jk}\left(u_{j,lk}+u_{l,jk}\right)n_l-l\mathcal{A}_ip^-_{,i}.
\end{empheq}
\label{eqn:epsdim}
\end{subequations}
In section 2, a simplified form of the above boundary condition is presented for the case of two-dimensional flows over rough and porous wall. The computation of all the constitutive coefficients are elaborated in the following section.
 %========================================================================================
\section{Computation of constitutive coefficients}
\label{sec:coeffcompute}
\begin{figure}[!h]
\centering
\subfloat[]{\includegraphics[trim = 0cm 0cm 0cm 0cm, clip, height=3cm]{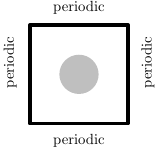}} \hspace{3cm}
\subfloat[]{\includegraphics[trim = 0cm 0cm 0cm 0cm, clip, height=5cm]{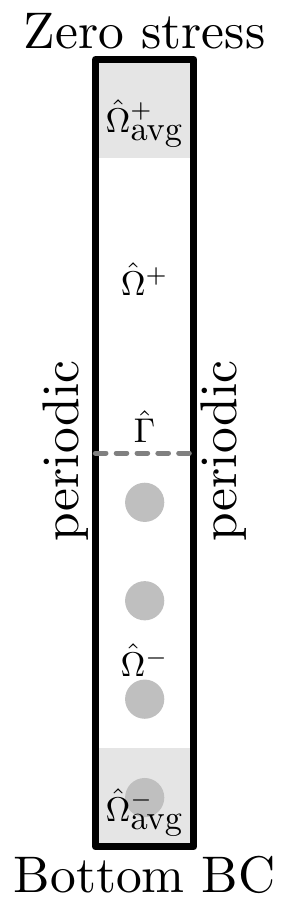}}
\caption{Domain used to compute constitutive coefficients: (a)~Interior cell, (b)~Interface cell. $\hat{\Omega}^+$ and $\hat{\Omega}^-$ denote free-fluid and porous domain of the interface cell, respectively. $\hat{\Gamma}$ is the interface. $\hat{\Omega}^+_{\textrm{avg}}\subset \hat{\Omega}^+$ and $\hat{\Omega}^-_{\textrm{avg}}\subset \hat{\Omega}^-$ are the domain used for averaging constitutive coefficients for pressure.}
\label{fig:interfcell}
\end{figure}

The interface conditions presented in the previous section contain a number of constitutive coefficients, which carry all information about the microscale geometry of complex surfaces. This section details the computation of these coefficients for a porous medium example. The method of computing interior permeability tensor ($\mathcal{\textbf{K}}^i$) (as appearing in the Darcy equation given in equation~\ref{eqn:darcy}) by solving a Stokes problem on a periodic unit cell  (figure~\ref{fig:interfcell}a) is well established~\citep{meibook}. Hence we focus only on the procedure to compute interface coefficients in equation~\eqref{eqn:epsnondim}.

In order to compute the interface coefficients, we construct an interface cell (figure~\ref{fig:interfcell}b), which contains one slice of periodic repetitive elements in horizontal direction. In the vertical direction, the interface cell consists of five repetitive solid inclusions of porous media below the interface; the domain extent above the interface is $5l$. The general methodology of computing coefficients, as will be shown below, requires to solve Stokes systems in the interface cell with appropriate coupling conditions at the interface. The flow is driven either by forcing on the interface, or by volume forcing within the porous domain. As far as the boundary conditions on the interface cell are concerned, we enforce zero traction at the top boundary, and periodicity on the vertical sides of the interface cell. The general guideline for boundary condition on the bottom surface is as follows: no-slip condition is applied if the problem is interface-forced; solution obtained at the top surface of the interior problem (figure~\ref{fig:interfcell}a) is enforced, if the flow is driven by volume forcing. These boundary conditions are equivalent to those to be applied for equation~\eqref{eqn:pertcoupled}, which governs the behavior of perturbation quantities.

To compute all the constitutive coefficients appearing in equation~\eqref{eqn:epsnondim}, we need to solve three sets of Stokes problems, as explained below. The configurations of first and third problem are discussed in~\cite{lacis2016}, however in that study the focus was only on computation of coefficients arising in the velocity boundary conditions. %Here, we provide complete discussion on computation of coefficients appearing in interface velocities and pressure jump.

\begin{figure}
\centering
\includegraphics[trim = 0cm 5.5cm 0cm 0cm, clip, height=18cm]{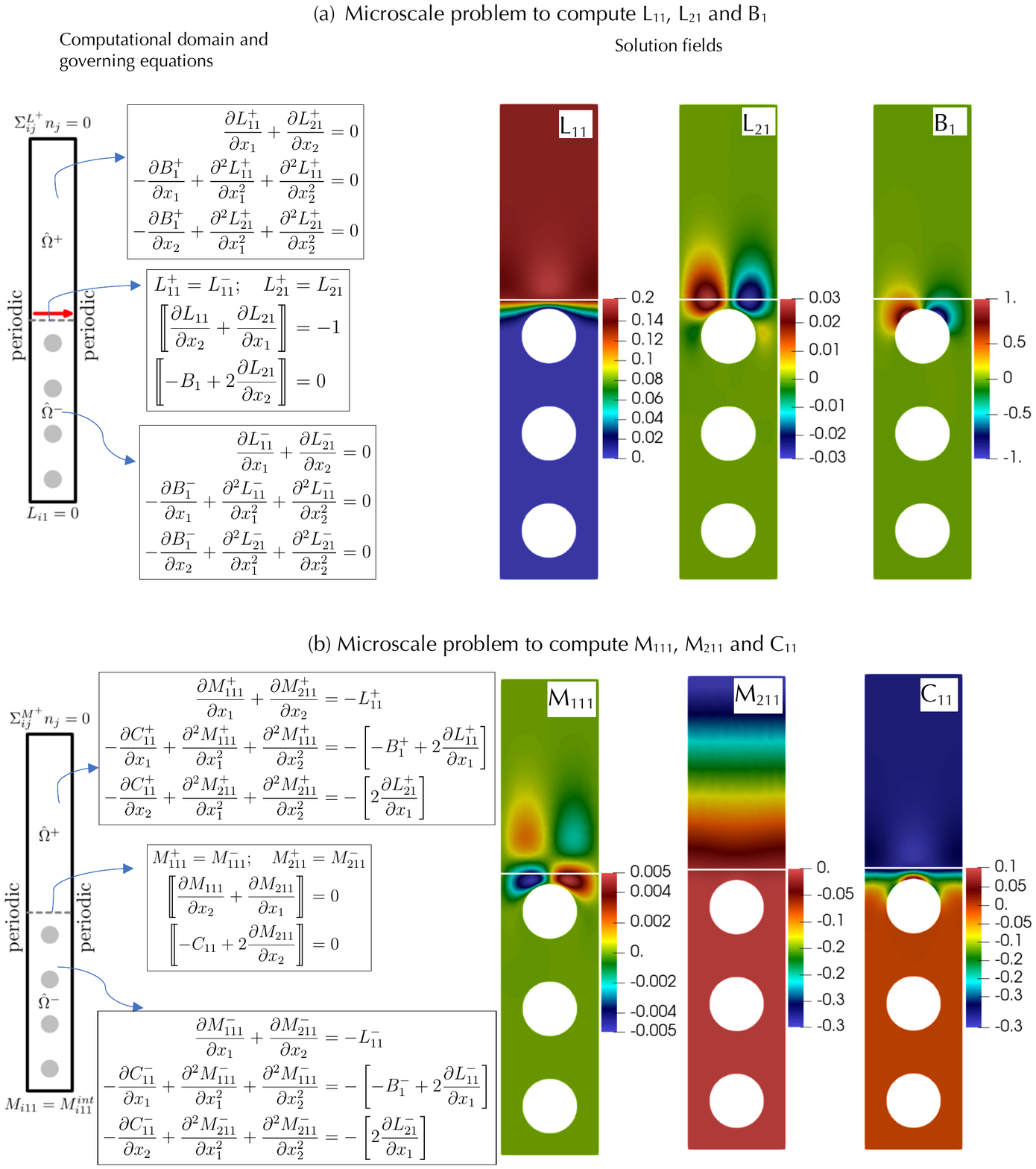}
\caption{Computation of coefficients $\mathcal{L}_{ij}$, $\mathcal{B}_{i}$, $\mathcal{M}_{ijk}$, and $\mathcal{C}_{ij}$. (a)~Solution with unit interface forcing enables computation of  $\mathcal{L}_{ij}$ and $\mathcal{B}_{i}$ (b)~Fields of $L_{ij}$ and $B_i$ act as source terms in the computation of  $\mathcal{M}_{ijk}$, and $\mathcal{C}_{ij}$. Zoomed view of solution fields with 3 inclusions are shown.}
\label{fig:lproblem}
\end{figure}

\subsection{Computation of $\mathcal{L}_{ij}$ and $\mathcal{B}_i$}
\label{sec:coeffcomputeL}
Substituting the Ansatz of $\mathcal{O}(\epsilon ^0)$, given in equation~\eqref{eqn:0ansatz}, into the respective governing equations~\eqref{ap:0order} and requiring the resulting equations to hold for any arbitrary surface shear forcing, we obtain the following Stokes problem for $L_{ij}$ and $B_i$:
\begin{subequations}
\begin{align}
L_{ik,i}^{\pm}&=0\ \ \ & \textrm{in } \hat{\Omega}^{\pm},\\
-B_{k,i}^{\pm}+L^{\pm} _{ik,jj}&=0 & \textrm{in } \hat{\Omega}^{\pm},\\
\left\llbracket L_{ik}\right\rrbracket &=0& \textrm{on } \hat{\Gamma},\\
\left\llbracket -B_{k}\delta_{ij}n_j+\left(L_{ik,j}+L_{jk,i}\right)n_j \right\rrbracket &=-\delta_{ik}& \textrm{on } \hat{\Gamma}.
\label{eqn:lstrbc}
\end{align}
\label{eqn:lprob}
\end{subequations}
Equation~\eqref{eqn:lstrbc} states that the flow is driven by a unit forcing applied at the interface. Owing to the fact that there is only Darcy velocity in the interior, on the bottom side of the interface cell no-slip conditions are enforced. On the top of the interface cell, the zero stress condition for perturbation velocity leads to zero stress condition for $L$ and $B$ variables. It can be directly inferred from the above equation system that $L_{ik}$ physically denotes the velocity component in $i-$direction due to interface forcing in direction $k$. Moreover, $B_k$ can be interpreted as the pressure jump produced in the microscale problem due to unit interface forcing in direction $k$.

The microscale problem (equation~\ref{eqn:lprob}) with unit forcing applied parallel to the interface is illustrated in figure~\ref{fig:lproblem}(a). The solution of this problem provides the fields of $L_{i1}$ (velocity solution) and $B_1$ (pressure solution) on the entire domain. From these fields, the constitutive coefficients can be computed by employing appropriate averaging as explained below.
\begin{subequations}
\begin{align}
\mathcal{L}_{ij}&=\frac{1}{l}\int_{\hat{\Gamma}}{L^+_{ij} dx_1},\\
\mathcal{B}_i&=\left<B^+_i\right>-\left<B^-_i\right>=\frac{1}{\hat{\Omega}^+_\textrm{avg}}\int_{\hat{\Omega}^+_\textrm{avg}}{B^+_i d\hat{\Omega}}-\frac{1}{\hat{\Omega}^-_\textrm{avg}}\int_{\hat{\Omega}^-_\textrm{avg}}{B^-_i d\hat{\Omega}},
%\mathcal{B}_i&=\left<B^+_i\right>-\left<B^-_i\right>=\frac{1}{l^2}\int_{\hat{\Omega}^+_\textrm{avg}}{B^+_i d\hat{\Omega}}-\frac{1}{l^2}\int_{\hat{\Omega}^-_\textrm{avg}}{B^-_i d\hat{\Omega}},
\end{align}
\label{eqn:lbavg}
\end{subequations}
where $\hat{\Gamma}$, $\Omega^+_\textrm{avg}$ and $\Omega^-_\textrm{avg}$ is defined in figure~\ref{fig:interfcell}(b). The above averaging for pressure jump is justified due to the following reason. The interface introduced in the macroscopic representation (figure~\ref{fig:intro-porous}b) is a lumped representation of the interface cell used in this section. Hence, the relevant quantities averaged over $\hat{\Omega}^+_\textrm{avg}$ and $\hat{\Omega}^-_\textrm{avg}$ (in the microscopic interface cell) represent interface constitutive coefficients  on the free-fluid and porous domain respectively. The interface cell and the associated averaging can also be interpreted from the point of view of matched asymptotic solutions: if the cell is viewed as the boundary layer (inner solution) between two macroscopic outer solutions (free fluid and porous domain), then the matching should be done at $+\infty$ for the free fluid and at $-\infty$ for the porous domain. These are analogous, respectively, to averaging in $\Omega^+_\textrm{avg}$ and $\Omega^-_\textrm{avg}$ for an interface cell of a finite size.

When the forcing is in interface normal direction, since the bottom boundary carries no-slip condition,  the only possibility to maintain the periodicity along vertical sides is that $L_{i2}=0$ everywhere, and hence $\mathcal{L}_{i2}=0$~\citep{lacis2016}. Since there is no flow, and a zero stress condition is applied at the top boundary $\left<B_2^+\right>=0$. Moreover, stress balance on the interface cell lead to $\left<B_2^-\right>=-1$, which gives us $\mathcal{B}_2=1$. These values are independent of the type of solid inclusions. In summary, for a 2D interface, the only microscale problem required to compute all effective non-zero $\mathcal{L}_{ij}$ and $\mathcal{B}_{i}$  coefficients is the one presented in figure~\ref{fig:lproblem}(a). Moreover, this is the only microscale problem that need to be solved if we employ $\mathcal{O}(\epsilon ^0)$ interface conditions presented in equation~\eqref{eqn:zerobc}.

To compute additional constitutive coefficients appearing in the higher order interface conditions presented in equation~\eqref{eqn:epsnondim}, two additional sets of microscale problems need to be solved. We can observe from equation~\eqref{eqn:1ansatz} that $\mathcal{O}(\epsilon)$-problem is driven by two sources: velocity diffusion tensor $D_{ij}$ and the macroscopic pore pressure gradient ($P^-_{,J}$). Due to linearity of the equations the effects of these sources can be computed separately, and can be superimposed to find their combined effect. This fact will be used to arrive at the two corresponding microscale problems, as described below.

\subsection{Computation of $\mathcal{M}_{ijk}$ and $\mathcal{C}_{ij}$}
Microscale equations governing the coefficients $\mathcal{M}$ and $\mathcal{C}$ can be found by neglecting $P^-_{,J}$ from Ansatz, given by equation~\eqref{eqn:1ansatz}, and substituting $u_i^{\pm(1)}=M_{ijk}^{\pm}D_{jk}$
and $p^{\pm(1)}=C_{ij}^{\pm}D_{ij}$ into the $\mathcal{O}(\epsilon ^1)$ governing equations. This yields the following system of equations:
\begin{subequations}\label{eqn:MCproblem}
\begin{align}
M^{\pm}_{ilm,i}&=-L^{\pm}_{ml} &\textrm{in } \hat{\Omega}^{\pm},\\
-C^{{\pm}}_{lm,i}+M^{\pm}_{ilm,jj}&=-\left[-B^{\pm}_l\delta_{im}+2L^{\pm}_{il,m}\right]\ \ \ & \textrm{in } \hat{\Omega}^{\pm},\\
\left\llbracket M_{ijk}\right\rrbracket &=0& \textrm{on } \hat{\Gamma}, \\
\left\llbracket \Sigma_{ij}^{M}n_j \right\rrbracket &=0& \textrm{on } \hat{\Gamma},
%\Sigma_{ij}^{M^+}n_j&=\Sigma_{ij}^{M^-}n_j& \textrm{on } \hat{\Gamma}
\end{align}
\end{subequations}
where $\Sigma_{ij}^{M^{\pm}}=-C^{\pm}_{lm}\delta_{ij}+\left(M^{\pm}_{ilm,j}+M^{\pm}_{jlm,i}\right)$. The solution of the $\mathcal{O}(\epsilon^0)$ microscale problem acts as source/sink in the continuity equation as well as the body force term in the momentum equation. For this microscale problem, the zero stress condition warrants additional comments. We recall the zero stress $-p^+\delta_{ij}n_j+\left(u^+_{i,j}+u^+_{j,i}\right)n_j=0$ in terms of multiscale perturbation expansion, and collect the corresponding $\mathcal{O}(\epsilon)$ terms to get the following expression
\begin{equation}
-p^{+(1)}\delta_{ij}n_j+\left(u^{+(1)}_{i,j}+u^{+(1)}_{j,i}\right)n_j=-\left(u^{+(0)}_{i,J}+u^{+(0)}_{j,I}\right)n_j.
\end{equation}
Substituting the ansatz for $(u^{+(0)}_i,p^{+(0)})$ and $(u^{+(1)}_i,p^{+(1)})$ into the above expression, we will get the following boundary condition at the top of the cell for this microscale problem
\begin{equation}
-C_{lm}^+\delta_{ij}n_j+\left(M^+_{ilm,j}+M^+_{jlm,i}\right)n_j=-\left(L^+_{il}\delta_{mj}+L^+_{jl}\delta_{mi}\right)n_j
\label{eqn:mtopbc}
\end{equation}
The source term appearing in the right hand side of the above equation is forced by the solution of leading order microscale problem. Such source terms are absent in other two interface cell problems.

Physically ${M}_{ijk}$ can be interpreted as  the velocity components obtained due to $L_{kj}$ acting as the source/sink terms, and $C_{jk}$ quantifies the corresponding pressure jump. As presented in section~\ref{sec:main}, for a 2D problem,  non-zero terms appearing in the interface conditions contain only three coefficients: $\mathcal{M}_{111}$, $\mathcal{M}_{211}$ and $\mathcal{C}_{11}$, whose computation is schematically illustrated in figure~\ref{fig:lproblem}(b). Since these problems are driven by volume forcing, at the bottom boundary, the corresponding solution obtained from an interior cell is enforced. After solving this microscale problem, $\mathcal{M}_{ijk}$ and $\mathcal{C}_{jk}$  are computed by appropriate averaging discussed for the previous microscale problem.

\begin{figure}
\centering
\subfloat[]{\includegraphics[trim = 0cm 0cm 0cm 0cm, clip, height=4cm]{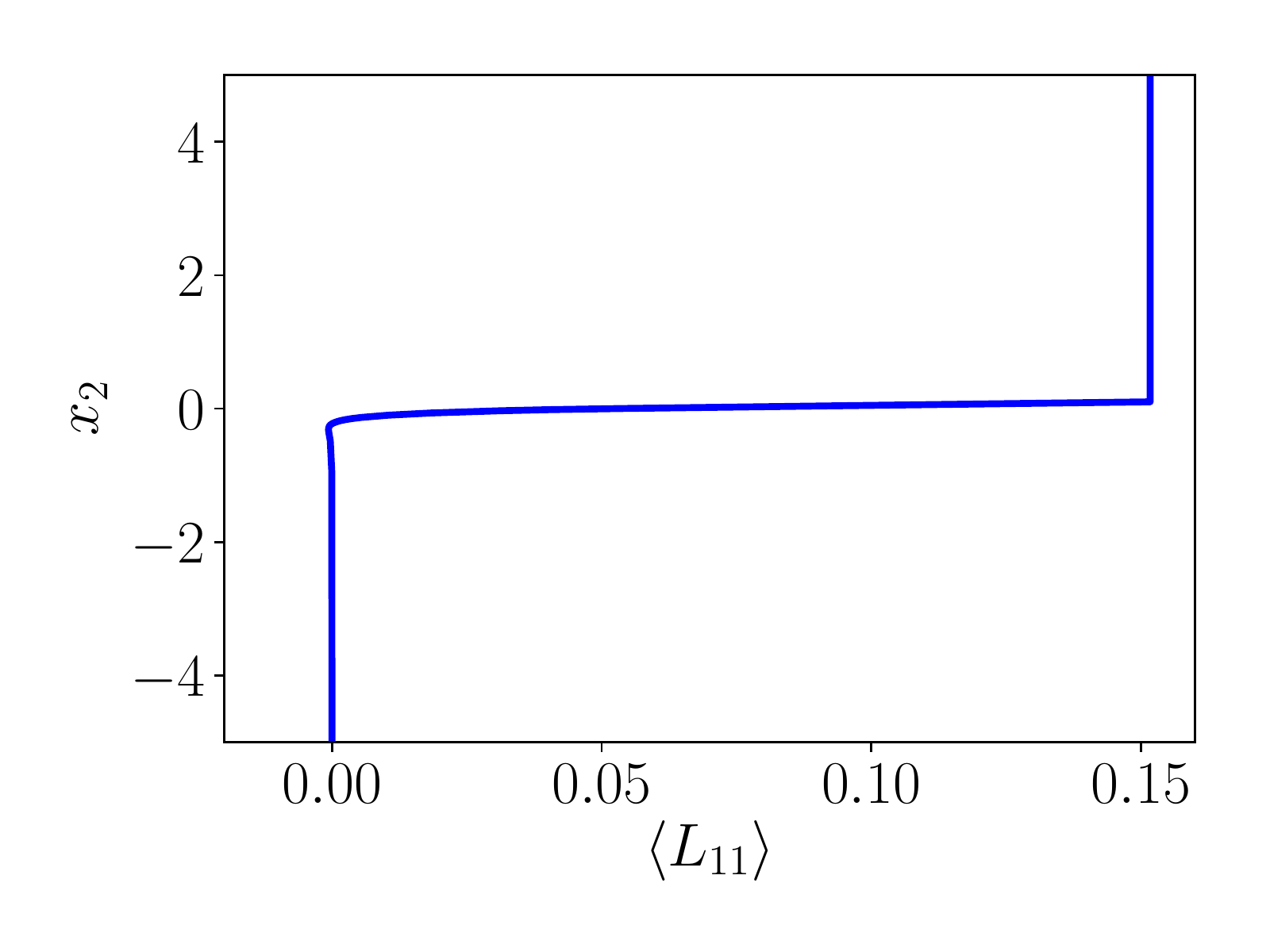}}
\subfloat[]{\includegraphics[trim = 0cm 0cm 0cm 0cm, clip, height=4cm]{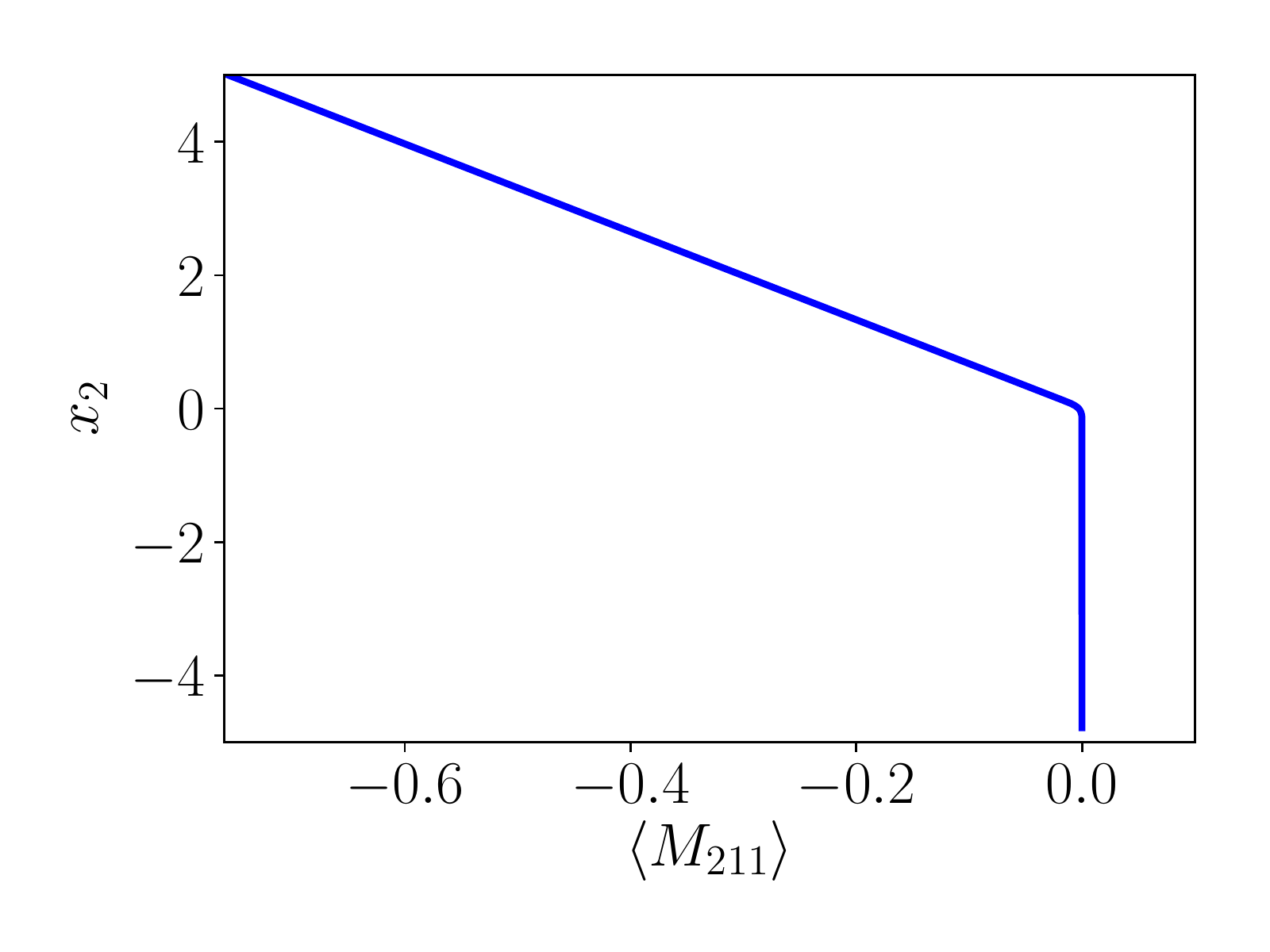}}
\caption{Microscale variation of averaged coefficients (a)~$\langle L_{11}\rangle$ stays constant in the free-fluid region, (b)~$\langle M_{211}\rangle$ displays linear variation in the free-fluid region.}
\label{fig:mcoefline}
\end{figure}

In section~\ref{sec:homog}, we have used the identities $\left<L^+_{ij}\right>_{,m}=0$ and $\left<M^+_{ijk}\right>_{,m}\neq 0$. We demonstrate this by plotting variation of $\langle L_{11}\rangle$ and $\langle M_{211}\rangle$ in figure~\ref{fig:mcoefline}.

\begin{figure}
\centering
\includegraphics[trim = 0cm 6cm 0cm 0cm, clip, height=18cm]{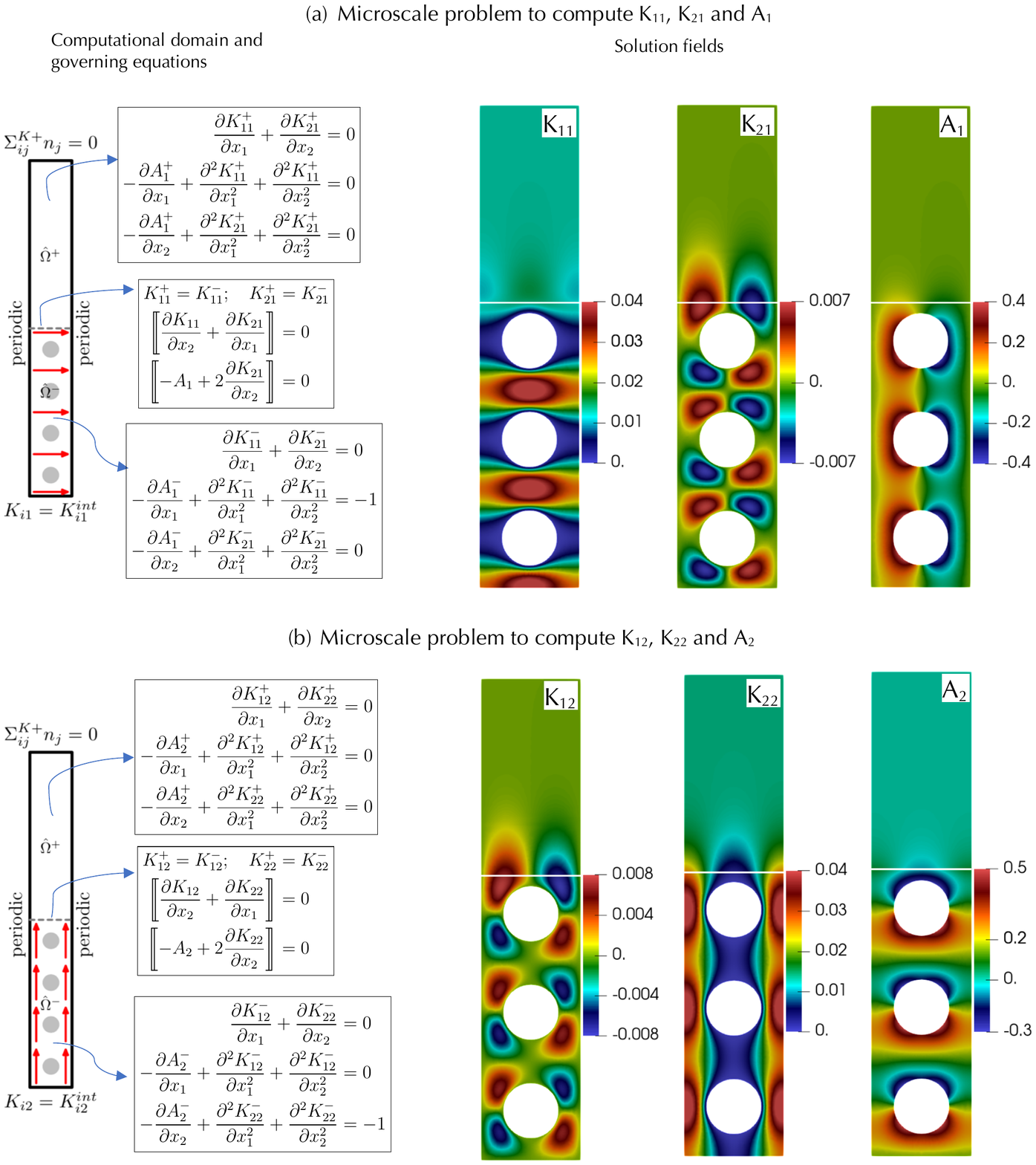}
\caption{Computation of coefficients $\mathcal{K}_{ij}$ and $\mathcal{A}_i$. Zoomed view of solution fields with 3 inclusions are shown.}
\label{fig:kproblem}
\end{figure}

\subsection{Computation of $\mathcal{K}_{ij}$ and $\mathcal{A}_{i}$}
 These microscale problems are obtained by neglecting $D_{ij}$, and substituting  $u_i^{\pm(1)}=-K_{ij}^{\pm}P^-_{J}$ and $p^{\pm(1)}=-A_{i}^{\pm}P^-_{,I}$ in equation~\eqref{eqn:1eqn},
\begin{subequations}\label{eqn:KAproblem}
\begin{align}
K^{\pm}_{ik,i}&=0 &\textrm{in } \hat{\Omega}^{\pm},\\
-A^{+}_{k,i}+K^+_{ik,jj}&=0\ \ \ & \textrm{in } \hat{\Omega}^+,\\
-A^{-}_{k,i}+K^-_{ik,jj}&=-\delta_{ik}\ \ \ & \textrm{in } \hat{\Omega}^-,\\
K^{+}_{ik}&=K^{-}_{ik}& \textrm{on } \hat{\Gamma},\\
\Sigma_{ij}^{K^+}n_j&=\Sigma_{ij}^{K^-}n_j& \textrm{on }. \hat{\Gamma}
\end{align}
\end{subequations}
Here $\Sigma_{ij}^{K^{\pm}}=-A^{\pm}_{k}\delta_{ij}+\left(K^{\pm}_{ik,j}+K^{\pm}_{jk,i}\right)$. The tensor $K_{ij}$ represents the velocity field in direction $i$ due to volume forcing within the porous medium in the direction $j$. Details of the microscale problems necessary to compute all terms of $\mathcal{K}_{ij}$ and $\mathcal{A}_{i}$ are depicted in figure~\ref{fig:kproblem}.
 After the solving these microscale problems, the necessary constitutive coefficients can be obtained by performing appropriate averaging:
 \begin{subequations}
\begin{align}
\mathcal{K}_{ij}&=\frac{1}{l}\int_{\hat{\Gamma}}{K^+_{ij} dx_1},\\
\mathcal{A}_i&=\left<A^+_i\right>-\left<A^-_i\right>=\frac{1}{\hat{\Omega}^+_\textrm{avg}}\int_{\hat{\Omega}^+_\textrm{avg}}{A^+_i d\hat{\Omega}}-\frac{1}{\hat{\Omega}^-_\textrm{avg}}\int_{\hat{\Omega}^-_\textrm{avg}}{A^-_i d\hat{\Omega}}.
\end{align}
\label{eqn:kaavg}
\end{subequations}
As explained in this section, by solving these three sets of microscale problems, we can compute all the coefficients appearing in our formulations apriori. %Since the domain to solve these equations is only a tiny fraction of the total computational domain, the solution process is very efficient, and hence it does not add noticeable computational overhead to the solution of the equations describing the macroscopic description. (moved to conclusion)

\subsection{Salient points}
A few notable points in the computation of the interface coefficients are summarized here. These points are used in arriving at the 2D interface conditions presented in section~\ref{sec:main} from the general conditions given in equation~\eqref{eqn:epsnondim}.
\begin{itemize}
\item When the problem is driven by forcing at the interface, at the bottom wall of the interface cell, no-slip conditions are specified. When volume forcing is applied at the porous part of the interface cell, then the interior solution is prescribed.
\item $\mathcal{L}_{i2}=0$ because due to the combination of no-slip condition at the bottom and periodicity of vertical boundaries, mass conservation requires zero velocity everywhere. This is elaborated in \cite{lacis2016}.
\item Irrespective of the geometry of the considered porous medium, $\left<B_2^-\right>=-1$ because the forcing is applied in the vertical direction at the interface, and since flow does not occur, the pressure difference balances the applied interface forcing. This leads to $\mathcal{B}_2=1$, because $\left<B_i^+\right>=0$.
\item The source term appearing in the boundary condition at the top surface of interface cell, given in equation~\eqref{eqn:mtopbc}, will contribute only to the problem that corresponds to the computation of $(\mathcal{M}_{i12}, \mathcal{C}_{12})$. The respective stress condition is given below
\begin{equation*}
\frac{\partial M^+_{112}}{\partial x_2}+\frac{\partial M^+_{212}}{\partial x_1}=-L^+_{11}
\end{equation*}
\item $\mathcal{M}_{12i}=\mathcal{K}_{1i}$ and $\mathcal{M}_{22i}=\mathcal{K}_{2i}$ (corresponding pressure terms $\mathcal{C}_{21}=\mathcal{A}_1$ and $\mathcal{C}_{22}=\mathcal{A}_2$). Although these problems are governed by different equations, effectively these are driven by corresponding unit volume forcing in the porous domain. Together with these conditions, the fact that $\left<B_2^-\right>=-1$ is used to set the following terms to zero:
\begin{align*}
\mathcal{M}_{12i}+\mathcal{K}_{1i}\left< B_2^-\right>&=0,  \\
\mathcal{M}_{22i}+\mathcal{K}_{2i}\left< B_2^-\right>&=0 , \\
\mathcal{C}_{2i}+\mathcal{A}_i\left< B_2^-\right>&=0.
\end{align*}
\item Based on several tests we performed over various geometries, we observe that numerically the following expressions always give zero result:
\begin{align*}
\mathcal{M}_{112}+\mathcal{K}_{12}\left< B_1^-\right>&=0,\\
\mathcal{M}_{212}+\mathcal{K}_{22}\left< B_1^-\right>&=0,\\
\mathcal{C}_{12}+\mathcal{A}_2\left< B_1^-\right>&=0.
\end{align*}
\item The interface permeability tensor $\mathcal{K}_{ij}$ is in general non-symmetric.
\item The following coefficients are non-zero only for anisotropic porous media: $\mathcal{K}_{12}$, $\mathcal{K}_{21}$, $\mathcal{B}_{1}$, $\left<B_1^-\right>$, $\mathcal{A}_1$, and  $\mathcal{M}_{111}$.
\item In order to compute constitutive coefficients for rough surfaces, the bottom boundary of the interface cell will be the physical geometry of rough surface. No-slip condition will be applied irrespective of the type of forcing applied. As a result of the no-slip condition, the mass conservation yields $\mathcal{K}_{i2}=0$. Moreover, for 2D rough surfaces, also due to mass conservation $\mathcal{K}_{21}=0$ and $\mathcal{M}_{111}=0$.
\end{itemize}
 %========================================================================================
\section{Results and discussion}
\label{sec:results}
In this section, we demonstrate the effectiveness of our macroscopic model for flow over rough and porous surfaces. For all test cases, we perform geometry-resolved microscopic simulations, denoted as DNS, which are used as reference values to compute the errors associated with macroscopic models. In addition to the leading order $\mathcal{O}(\epsilon^0)-$, and higher order $\mathcal{O}(\epsilon^1)-$models, we also consider modified Beavers-Joseph model (denoted as BJ model hereafter) in our numerical studies.% as BJ model is widely used in the literature.

%We consider four macroscopic models of porous media: (i)~$\mathcal{O}(\epsilon^0)$ model given in equation~\eqref{eqn:zerobc}, \textcolor{red}{(ii)~modified Beavers-Joseph model (denoted as BJ model hereafter),} (iii)~Transpiration resistance model presented in~\cite{lacisjfm1}, and (iv)~$\mathcal{O}(\epsilon^1)$ model given in equation~\eqref{eqn:epsnondim}. \textcolor{red}{In case of rough walls,  BJ model is equivalent to$\mathcal{O}(\epsilon^0)$ model, and hence three models are considered.}

Interface conditions proposed by \cite{beavers1967} are used extensively in modeling flows over porous media. Here, we consider modified Beavers-Joseph model  which is described by BJ slip condition, mass conservation, and stress balance at the interface. Mathematically, these are given as
\begin{align}
u_1&=\mathcal{L}_{11}l\left(\frac{\partial u_1}{\partial x_2}+\frac{\partial u_2}{\partial x_1}\right)-\frac{K^i_{11}l^2}{\mu}\frac{\partial p^-}{\partial x_1}-\frac{K^i_{12}l^2}{\mu}\frac{\partial p^-}{\partial x_2}\\
\label{eqn:bju2}
u_2&=-\frac{K^i_{21}l^2}{\mu}\frac{\partial p^-}{\partial x_1}-\frac{K^i_{22}l^2}{\mu}\frac{\partial p^-}{\partial x_2}\\
\label{eqn:bjpc}
\llbracket p \rrbracket &= 2\mu\frac{\partial u_2}{\partial x_2}
\end{align}
where $K^i_{nm}$ denote interior permeability tensor, which can be computed by solving a unitcell problem as described, for example, in \cite{meibook}. The constant $\alpha$ appearing in the classical \cite{beavers1967} condition can be related to our slip coefficient, $\mathcal{L}_{11}l=\sqrt{K^i}/\alpha$. This equivalence can also be found in other works~\citep{mikelic2000,lacis2016}. \cite{beavers1967} expressed $\alpha$ as an empirical parameter, and its relation with $\mathcal{L}_{11}$ is essential to employ their boundary conditions to model porous media of given microscale geometry. Since $\alpha$ is related to tangential velocity at the interface, we use $K^i=K^i_{11}+K^i_{12}$ for the estimate of permeability, and hence $\alpha=\sqrt{K^i_{11}+K^i_{12}}/\mathcal{L}_{11}l$. $\alpha$ values for the considered test cases are presented in tables~\ref{tab:coeff-porous} and \ref{tab:por-iso-loc}.

We call the above formulation as `modified', because \cite{beavers1967} conducted experiments on unidirectional channel flow with a porous bed. In their setup, $u_2$ is zero everywhere in the domain; moreover, the pressure condition at the fluid-porous interface was not addressed. In order to enable an appropriate comparison with the proposed method for 2D flows, we combine  Beavers-Joseph slip condition (given in equation~\ref{eqn:bj}) together with continuity of transpiration velocity (equation~\ref{eqn:bju2}) and balance of normal stresses (equation~\ref{eqn:bjpc}) at the interface. 
%While constitutive coefficients were treated as empirical parameters in \cite{beavers1967}, in our comparison we compute all coefficients appearing in the above equations by solving interface cell problems presented in section~\ref{sec:coeffcompute}. %The constant $\alpha$ as described in \cite{beavers1967} is computed by related it to our slip length parameter, $\alpha=\sqrt{K^i}/\mathcal{L}_{11}l$.

In all of the above macroscopic models, the averaged influence of microscopic flow fields is represented by the interface conditions. The microscopic geometrical details of the surfaces are lumped into the constitutive coefficients appearing in the interface conditions. Solution process of the macroscopic flow field involves two steps: computation of coefficients, and the solution of the coupled two-domain problem. All the simulations presented in this paper are performed using \texttt{Freefem++}~\citep{freefem}, an open-source finite element  package.

 In order to compare the results of DNS with macroscopic simulations, we eliminated the microscopic flow field details by performing ensembled averaging of the flow field. Samples for the ensembled averaging are generated by moving the solid inclusions in the porous medium (or roughness elements). For each successive sampling, the solid inclusions are moved to a small horizontal distance, in such a way that after the required number of sampling, the inclusions are traversed by one microscopic length $l$.
 We emphasize here that although DNS can be performed for simple configurations considered here, for real-life porous media applications, DNSs are prohibitively expensive. 

\subsection{Stokes flow over porous media}
In order to quantify the accuracy of our macroscopic description, we consider a steady 2D incompressible Stokes flow within a modified lid driven cavity, in which the bottom half of the cavity is filled with an ordered rigid porous bed. The geometry is similar to that considered in~\cite{sipp2007}, but added with a porous bed (figure~\ref{fig:poro-dns}). It consists of a channel ABGI to which a cavity is attached. The fluid flows through the channel inlet AB with a uniform velocity, and a laminar boundary layer develops from point C. The fluid passing through this  channel drives the fluid flow within the cavity, and induces a large vortex similar to the classical lid-driven cavity problem. The parameters used in the simulations are $H=10$~units, $U_0=100$~units, and $\mu=1$~units.

\subsubsection{Order of convergence}
\label{sec:results-poro-conv}
\begin{figure}
\centering
\includegraphics[trim = 0cm 0cm 0cm 0cm, clip, height=7cm]{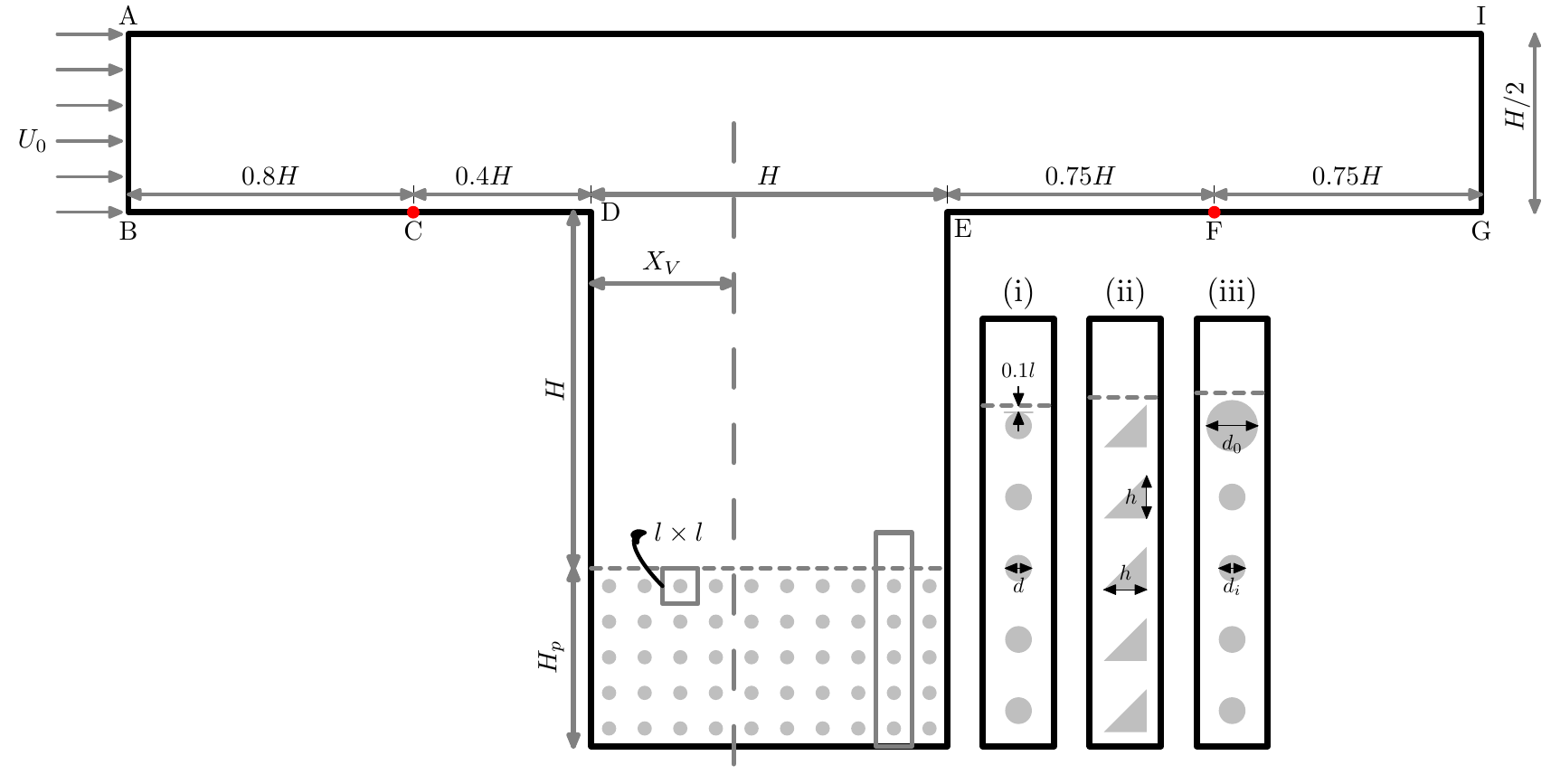}
\caption{Geometry used to study flow over porous media. Cavity flow is driven by the flow through the channel. $H_p$ is set in such a way that the porous medium contains 5 solid inclusions. The interface is located at 0.1$l$ from the top surface of solid inclusions for results presented in section~\ref{sec:accuracy}. Boundary conditions in AB is $u_1=100$ and $u_2=0$. CD, EF, and cavity surfaces are no-slip walls. On BC, FG and AI, $u_2=0$ and zero shear stress conditions are enforced. (i)~isotropic porous medium with $d=0.5642l$, (ii)~anisotropic porous medium with $h=0.6l$, and (iii)~isotropic porous medium whose inclusions at the interface are different from that of interior, $d_o=0.5046$ and $d_i=0.2523$.}
\label{fig:poro-dns}
\end{figure}

In order to show convergence of various models, we simulate the flow within the geometry given in figure~\ref{fig:poro-dns} with the following scale separation parameters, $\epsilon$=0.1, 0.05, 1/30, and 0.025.  $H_p$ is set to be 5 which is an integer multiple of microscale $l$, and the porous medium is filled with repetitive unit cells containing circular solid inclusions as shown in figure~\ref{fig:poro-dns}. The solid inclusions are located at the centre of the unit cells. The number of unit cells  varies with $\epsilon$. Macroscale $H=10$ in all cases, and the microscale $l$ is adjusted to achieve different $\epsilon$ values.

In order to compute the order of convergence, we define the following norm
\begin{equation}
||g||_2=\left[\int_{\overline{\Gamma}}{\left(g_{mac}-g_{DNS}\right)^2 dx}\right]^{1/2}
\end{equation}
where $g_{mac}$ and $g_{DNS}$ denote quantities obtained from macroscopic model and DNS, respectively. $\overline{\Gamma}$ is fraction of the total fluid-porous interface. We invoked the assumption of $l-$periodicity in the microscale solution fileds in the  vicinity of the interface, as given in section~\ref{sec:coeffcompute}. However, due to the no-slip condition at the cavity wall, this assumption breaks down at the intersection between the fluid-porous interface and the cavity walls. Hence, in order to eliminate the error associated with the no-slip condition, we compute error over $\overline{\Gamma}$ in which some portion at either end of the interface were cut off. While length of the interface is $H$, length of $\overline{\Gamma}$ is $0.8H$. A similar argument is presented in \cite{carraro2013} for their non-periodic test case.

\begin{figure}
\centering
\subfloat[Tangential velocity]{\includegraphics[trim = 0cm 0cm 0cm 0cm, clip, height=5cm]{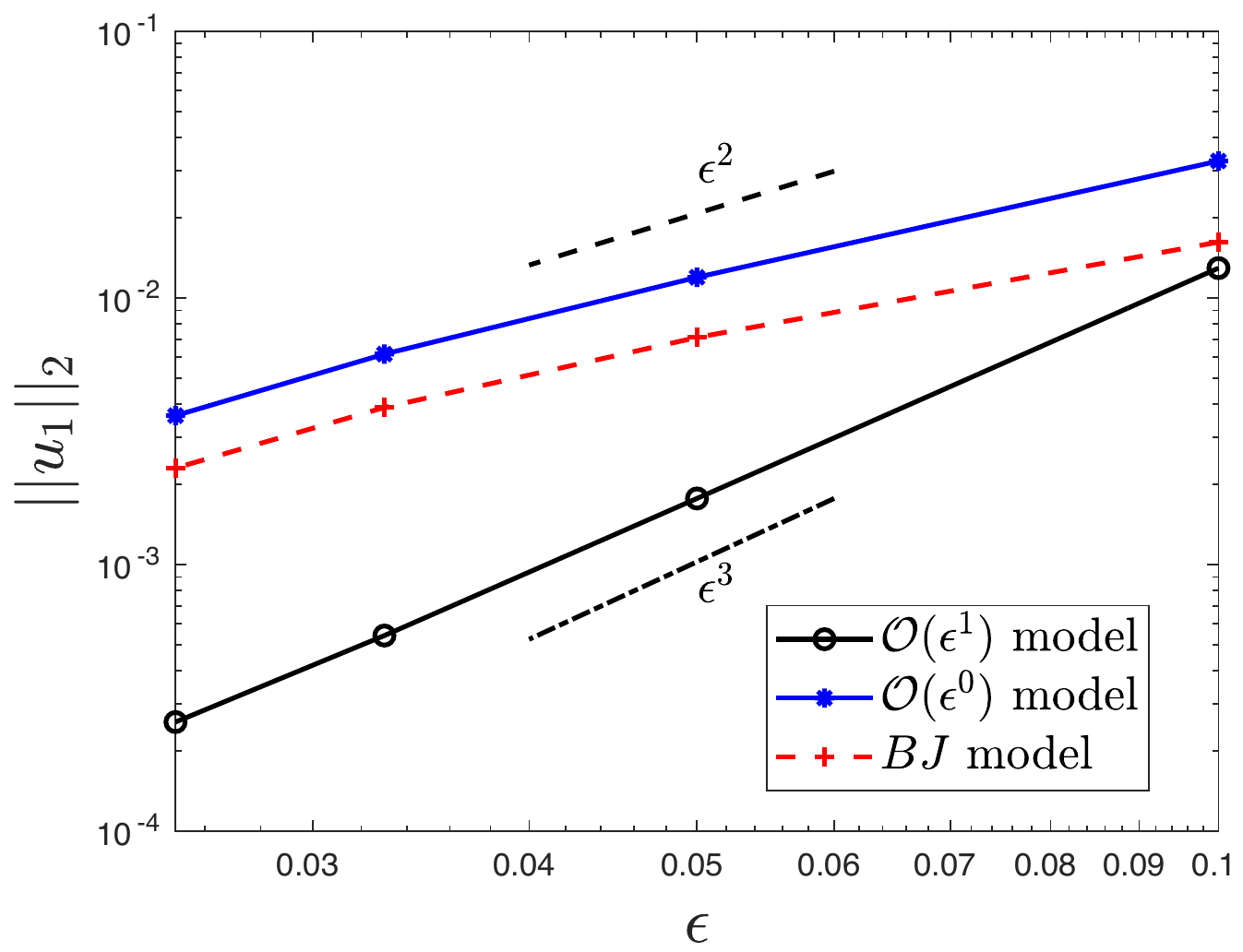}}
\subfloat[Transpiration velocity]{\includegraphics[trim = 0cm 0cm 0cm 0cm, clip, height=5cm]{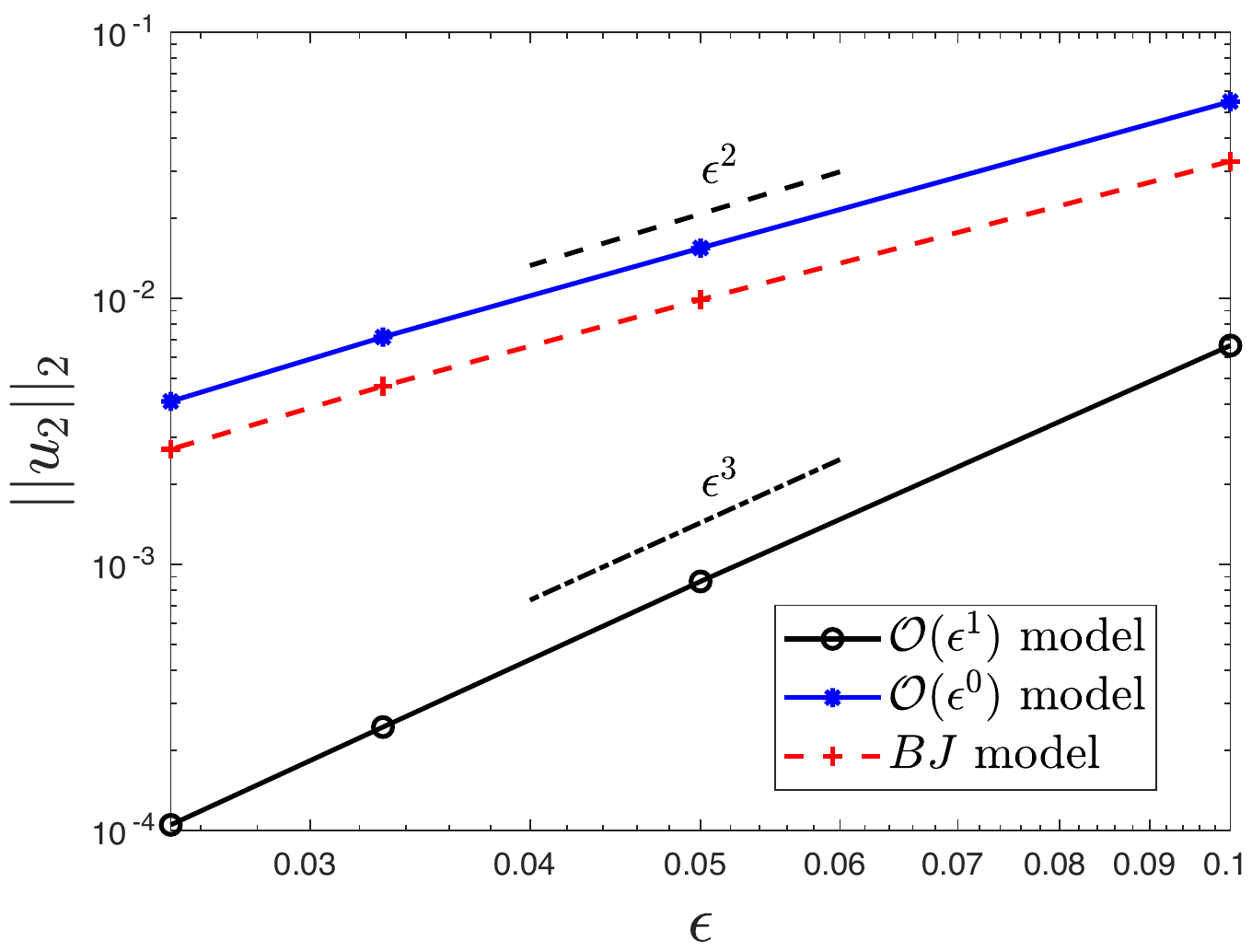}}
\caption{Convergence plots for porous medium with isotropic solid inclusions.}
\label{fig:poro-conv}
\end{figure}

Convergence of tangential and transpiration velocities for all three models are presented in figure~\ref{fig:poro-conv}. It is evident from the figure that for both velocity components the leading order $\mathcal{O}(\epsilon^0)$ model and the higher order $\mathcal{O}(\epsilon^1)$ model shows second and third order convergence, respectively, with respect to $\epsilon$. As far as absolute errors are concerned, the higher order model proposed in this paper produces consistently smaller error than the other two models. The error incurred by the Beavers-Joseph model is smaller than the leading order, but larger than the higher order homogenized model.

\subsubsection{Accuracy}
\label{sec:accuracy}
The order of convergence of macroscopic schemes were discussed in the previous section, by comparing the error norm at the interface. In this section, we will study the accuracy of macroscopic models in detail. In all simulations discussed in this section, $\epsilon=0.1$. and $H_p$ is chosen such that there are five rows of solid inclusions in the porous medium, and the interface is located at a vertical distance of $0.1l$ from the top surface of porous solid inclusions (figure~\ref{fig:poro-dns}).

The tensorial form of our interface conditions makes it possible to use our macroscopic model for flows over isotropic as well as anisotropic surfaces. In order to demonstrate this point, as shown in figure~\ref{fig:poro-dns}, we consider three different classes of microscale geometries: (i)~isotropic, consisting of circular solid inclusions, (ii)~anisotropic, consisting of triangular inclusions, and (iii)~layered porous medium, whose interface geometry is different than that of the interior.
 Auxiliary microscopic problems, described in section~\ref{sec:coeffcompute}, are solved to compute the constitutive coefficients for all these geometries. The coefficients are listed in table~\ref{tab:coeff-porous}.

\begin{table}[h!]
  \begin{center}
    \caption{Constutive coefficients for the geometries considered in flows over porous media, represented in figure~\ref{fig:poro-dns}. Except $\mathcal{L}_{11}$, all other terms in $\mathcal{L}_{ij}$ are zero. For completeness, interior permeability tensor $K^i$ and $\alpha$ for BJ condition are also presented.}
    \label{tab:coeff-porous}
    \begin{tabular}{ l c c  c  c  c c  c  c c}
    \noalign{\smallskip}\hline
      & $K^i$ & $\alpha$ & $\mathcal{L}_{11}$ & $\mathcal{B}$ & $\mathcal{K}$ & $\mathcal{A}$ & $\mathcal{M}_{111}$ & $\mathcal{M}_{211}$ & $\mathcal{C}_{11}$\\ \hline
      \addlinespace[1.5ex]
      isotropic & $\begin{bmatrix}  0.01378 & 0 \\ 0 &  0.01378 \end{bmatrix}$  & 0.7743   & 0.1516  & $\begin{pmatrix}  0 \\ 1 \end{pmatrix}$ & $\begin{bmatrix}  0.01295 & 0 \\ 0 &  0.01378 \end{bmatrix}$ & $\begin{pmatrix}  0 \\ -0.14389 \end{pmatrix}$ & 0 & -0.01295 & -0.2912 \\
      \addlinespace[3ex]
      anisotropic & $\begin{bmatrix}  0.01238 & 0.00245 \\ 0.00245 &  0.01233 \end{bmatrix}$  & 0.6739 & 0.1807 & $\begin{pmatrix}  0.3831 \\ 1 \end{pmatrix}$ & $\begin{bmatrix}  0.01943 &  0.00472 \\ 0.00245 &  0.01233 \end{bmatrix}$ & $\begin{pmatrix}  0.03637 \\ 0.08754 \end{pmatrix}$ & 0.000354 & -0.01755 & -0.4185\\
      \addlinespace[3ex]
      layered     & $\begin{bmatrix}   0.06431 & 0 \\ 0 &   0.06431 \end{bmatrix}$ & 1.6489 & 0.1538 & $\begin{pmatrix}  0 \\ 1 \end{pmatrix}$ & $\begin{bmatrix}  0.01329 &  0 \\ 0 &  0.06431 \end{bmatrix}$ & $\begin{pmatrix}  0 \\ -2.569 \end{pmatrix}$ & 0 & -0.01329 & -0.30142 \\ 
    \noalign{\smallskip}\hline
    \end{tabular}
  \end{center}
\end{table}

\begin{figure}
\centering
\subfloat[Tangential velocity]{\includegraphics[trim = 0cm 0cm 0cm 0cm, clip, height=5cm]{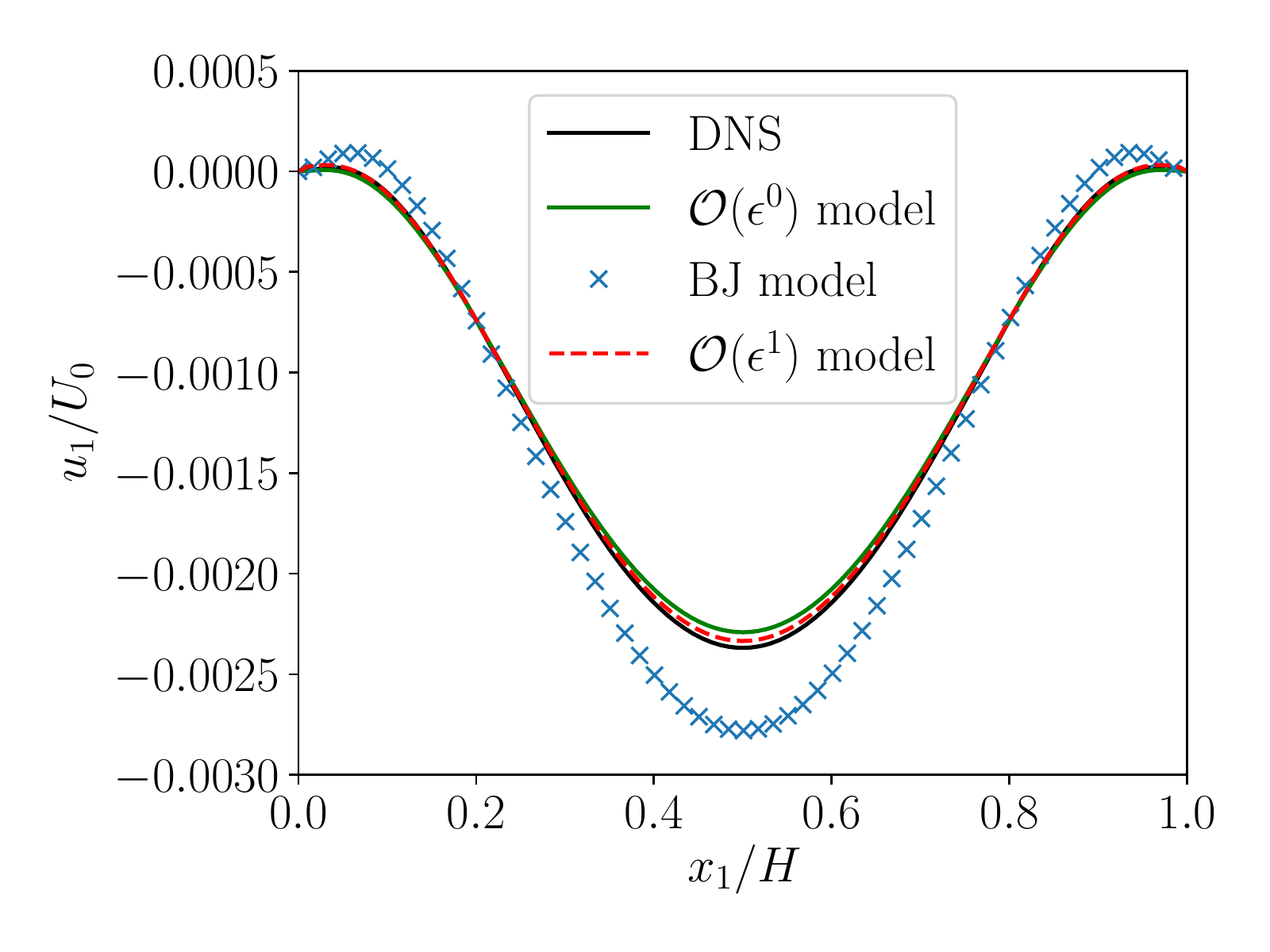}}
\subfloat[Transpiration velocity]{\includegraphics[trim = 0cm 0cm 0cm 0cm, clip, height=5cm]{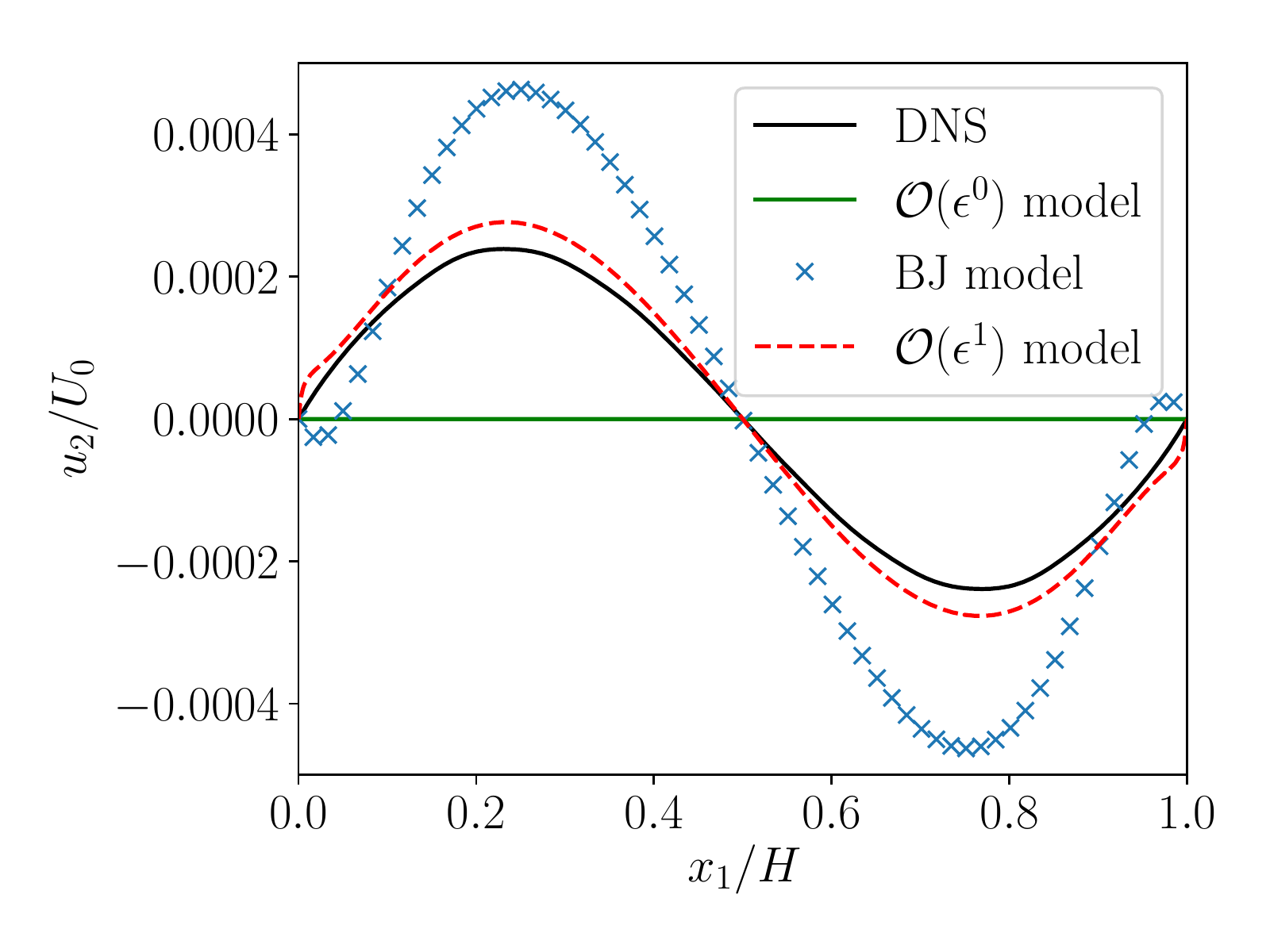}}
\caption{Velocities along the interface for cavity with channel: layered porous inclusions.}
\label{fig:poro-res}
\end{figure}

The vortex generated within the cavity induces a negative shear along the entire length of the interface, and as a consequence it introduces a negative slip velocity at the interface. This is illustrated in figure~\ref{fig:poro-res}(a), which shows tangential velocity along the interface for the layered porous case. It can be seen that both $\mathcal{O}(\epsilon^0)$ and $\mathcal{O}(\epsilon^1)$  models approximate the tangential velocity accurately, while $\mathcal{O}(\epsilon^1)$ model performing slightly better than the leading order model. The error introduced by the BJ model is larger in this case because the BJ model involves a filtration velocity term that is proportional to interior permeability. The interface permeability, which quantifies the interface region more accurately, is approximately twice that of the interior permeability (Table~\ref{tab:coeff-porous}). Hence adding the contribution of filtration velocity leads to larger error than $\mathcal{O}(\epsilon^0)$-model. However, this observation is specific to this layered medium example. For isotropic, and anisotropic porous media, BJ model is more accurate than $\mathcal{O}(\epsilon^0)$-model as can be seen in table~\ref{tab:poro-res}. The error introduced by a macroscopic model in computing a quantity $q$ is expressed as,
\begin{equation*}
e_q=\left|\frac{q_{mac}-q_{DNS}}{q_{DNS}}\right|
\end{equation*}
where $q_{mac}$ and $q_{DNS}$ denote quantities obtained from macroscopic model and DNS, respectively.

For the considered configuration, the slip velocity is spatially varying along the interface, and hence non-zero transpiration velocity arises owing to mass conservation~\citep{lacisjfm1}. By comparing figures~\ref{fig:poro-res}(a) and (b), three regions in the transpiration velocity curve can be seen (i)~$u_2 >0$ when $\frac{du_1}{dx_1}<0$, (ii)~$u_2 <0$ when $\frac{du_1}{dx_1}>0$, and (iii)~$u_2 =0$ when $\frac{du_1}{dx_1}=0$. These observations are  consistent with the  physical arguments presented in~\cite{lacisjfm1}. As far as the performance of macroscale models are concerned, the $\mathcal{O}(\epsilon^0)$ model predicts zero transpiration, owing to the fact that $\mathcal{L}_{2j}=0$. Moreover, since the term accounting for spatial variation of shear in the transpiration velocity is absent, BJ model induces a very large error. Hence, in certain cases, these models will not be able to capture transport phenomena across fluid-porous interfaces. It has been shown in \cite{lacisjfm1} that owing to this limitation, the $\mathcal{O}(\epsilon^0)$ model produces unphysical results in a turbulent flow simulation. As can be seen from the figure, $\mathcal{O}(\epsilon^1)$ produces accurate predictions of transpiration along the entire length of the interface. The accuracy of transpiration resistance model \citep{lacisjfm1} with respect to the proposed higher order model is presented in Appendix~3.

In order to quantify the errors introduced in all macroscopic models, we compare minimum tangential velocity ($u_1^{min}$) and maximum transpiration velocity ($u_2^{max}$) of the interface quantities. This comparison is presented for all three porous cases in table~\ref{tab:poro-res}. It can be seen that the tangential velocity is accurately predicted by all three macroscopic models. While transpiration velocity is well-approximated only by $\mathcal{O}(\epsilon^1)$ model.

\begin{table}[h!]
 \begin{center}
    \caption{Stokes flow over cavity with porous bed. Here BJ denotes Beavers-Joseph model. Since $u_2=0$ for $\mathcal{O}(\epsilon^0)$ model, it produces 100\% error in approximating $u_2^{max}$.}
    \label{tab:poro-res}
    \begin{tabular}{ l c c c c c c c}
    \noalign{\smallskip}\hline
      & & & \multicolumn{3}{c}{Error in $u_1^{min}$ (\%)} &   \multicolumn{2}{c}{Error in $u_2^{max}$  (\%)} \\ \cline{4-8}
        & $u_1^{min}/U_0$ DNS & $u_2^{max}/U_0$ DNS & $\mathcal{O}(\epsilon^0)$ & BJ & $\mathcal{O}(\epsilon^1)$  & BJ & $\mathcal{O}(\epsilon^1)$  \\ \hline
      Isotropic & -0.0023502&  $ 1.4104 \times 10^{-4}  $ & 3.71 & 0.89 & 0.25 & 26.14 &  16.28 \\
      Anisotropic & -0.00277153&  $ 1.710 \times 10^{-4}  $ & 4.85 & 1.57 & 0.86 & 45.98 & 17.39  \\
      Layered & -0.00236893&  $ 2.38923 \times 10^{-4}$ & 3.28 & 17.37 & 1.44 & 93.76 &  15.83 \\ 
      \noalign{\smallskip}\hline
    \end{tabular}
  \end{center}
\end{table}

BJ conditions are the widely used interface conditions to model coupled fluid-porous problems. From figure~\ref{fig:poro-res}, we can see that while these conditions provide a good approximation to the slip velocity, BJ model does not accurately approximate the transpiration component. This can be inferred from Table~\ref{tab:poro-res}, which shows the relative error between the present model, and BJ model is the largest for the layered porous medium. This is expected because for the layered porous medium, BJ model produces a larger error in approximating pressure jump across the interface. Moreover, the seepage velocity within the porous medium is also approximated with a larger error by the BJ model. A similar conclusion is presented in the recent study by \cite{eggenweiler2020}. These two points are discussed in the following. %does not account for the contribution for spatial variation of shear to the transpiration velocity and pressure jump. These contributions are quantified by constitutive coefficients $\mathcal{M}_{211}$ and $\mathcal{C}_{11}$ that are larger for anisotropic case when compared to other two geometries.

\begin{figure}[!h]
\centering
\includegraphics[height=3.5cm]{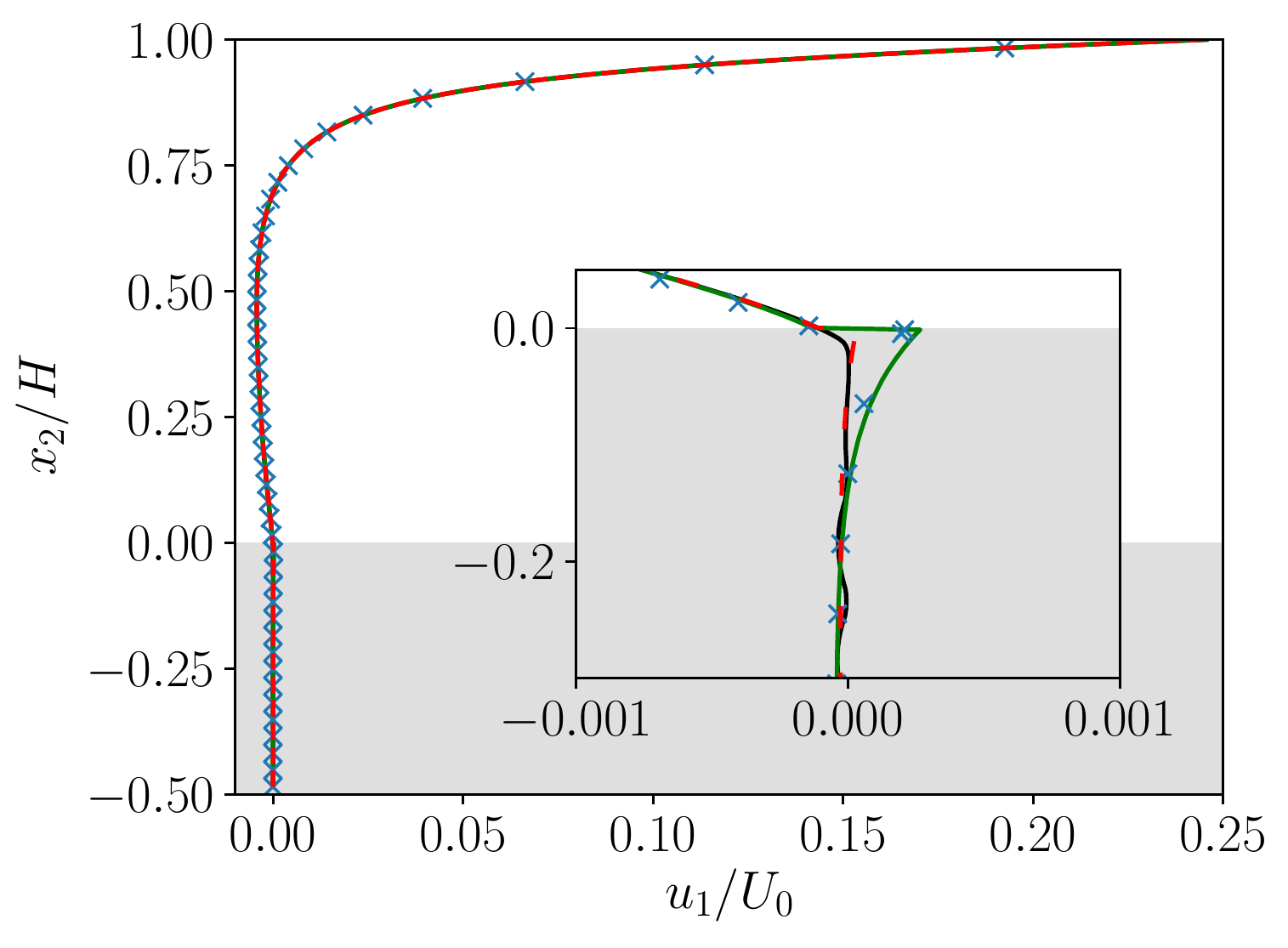}
\includegraphics[height=3.5cm]{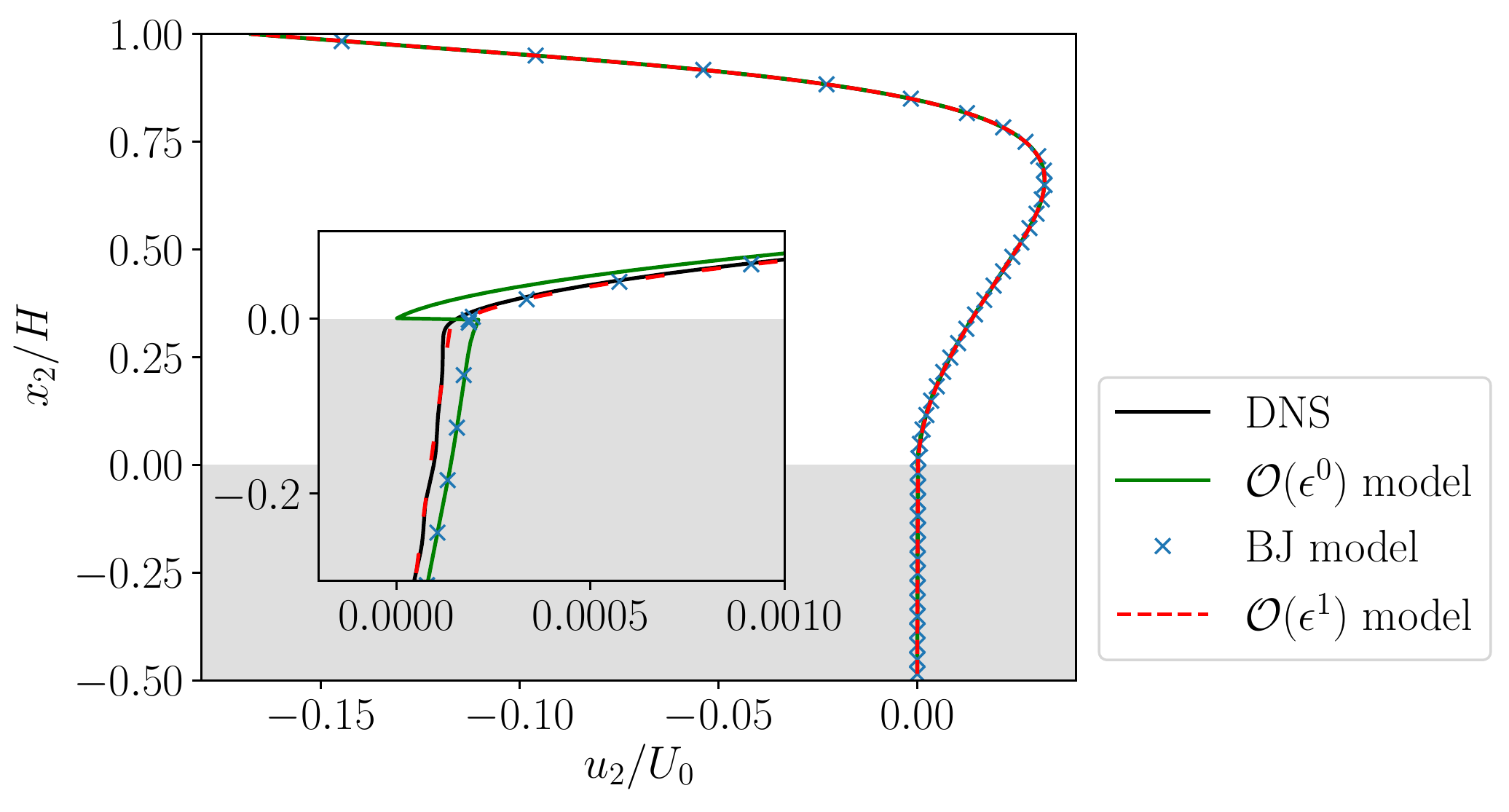}\\
(a) Velocity variation along $X_V=0.1 H$\\[0.2cm]
\includegraphics[height=3.5cm]{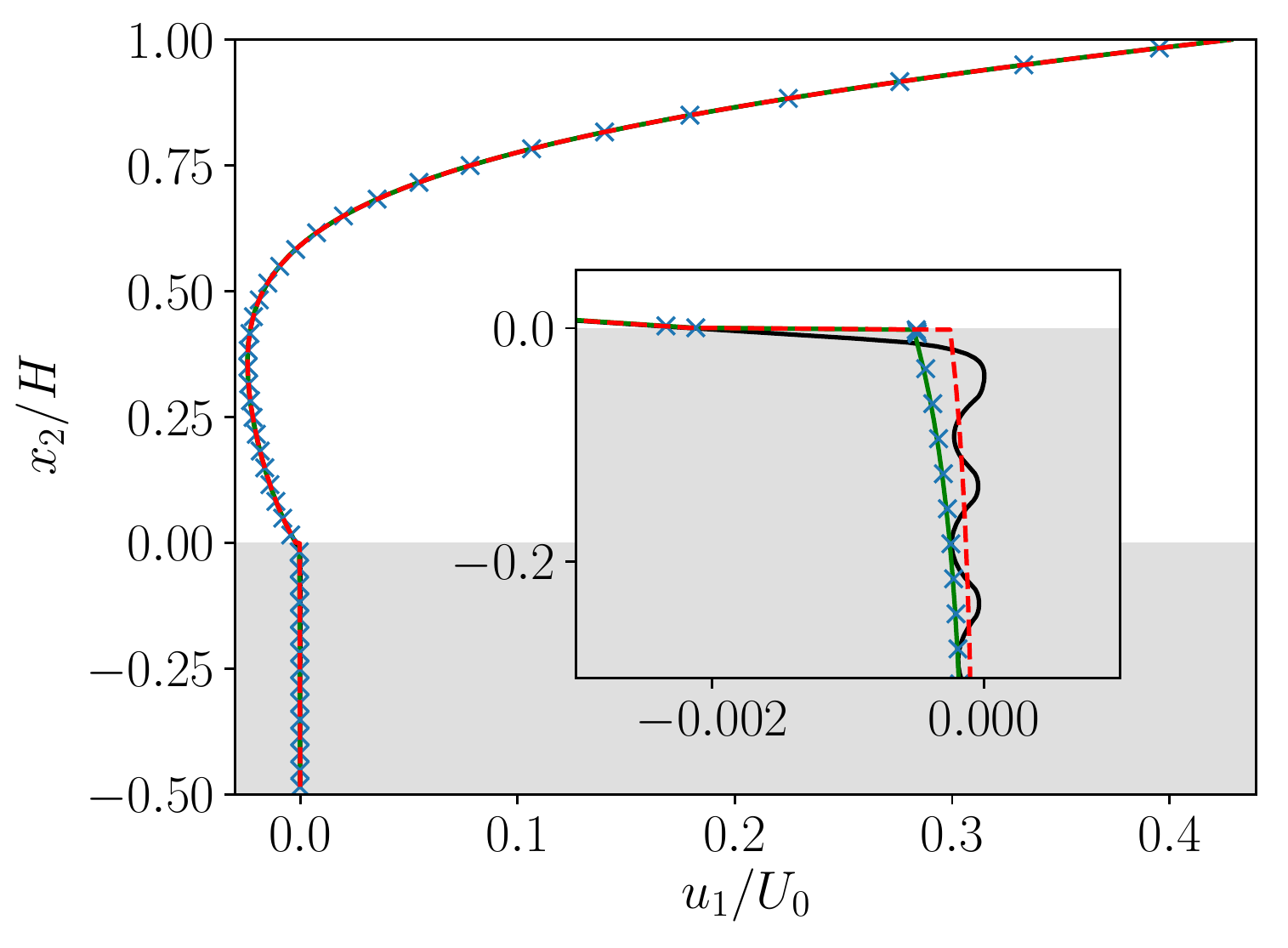}
\includegraphics[height=3.5cm]{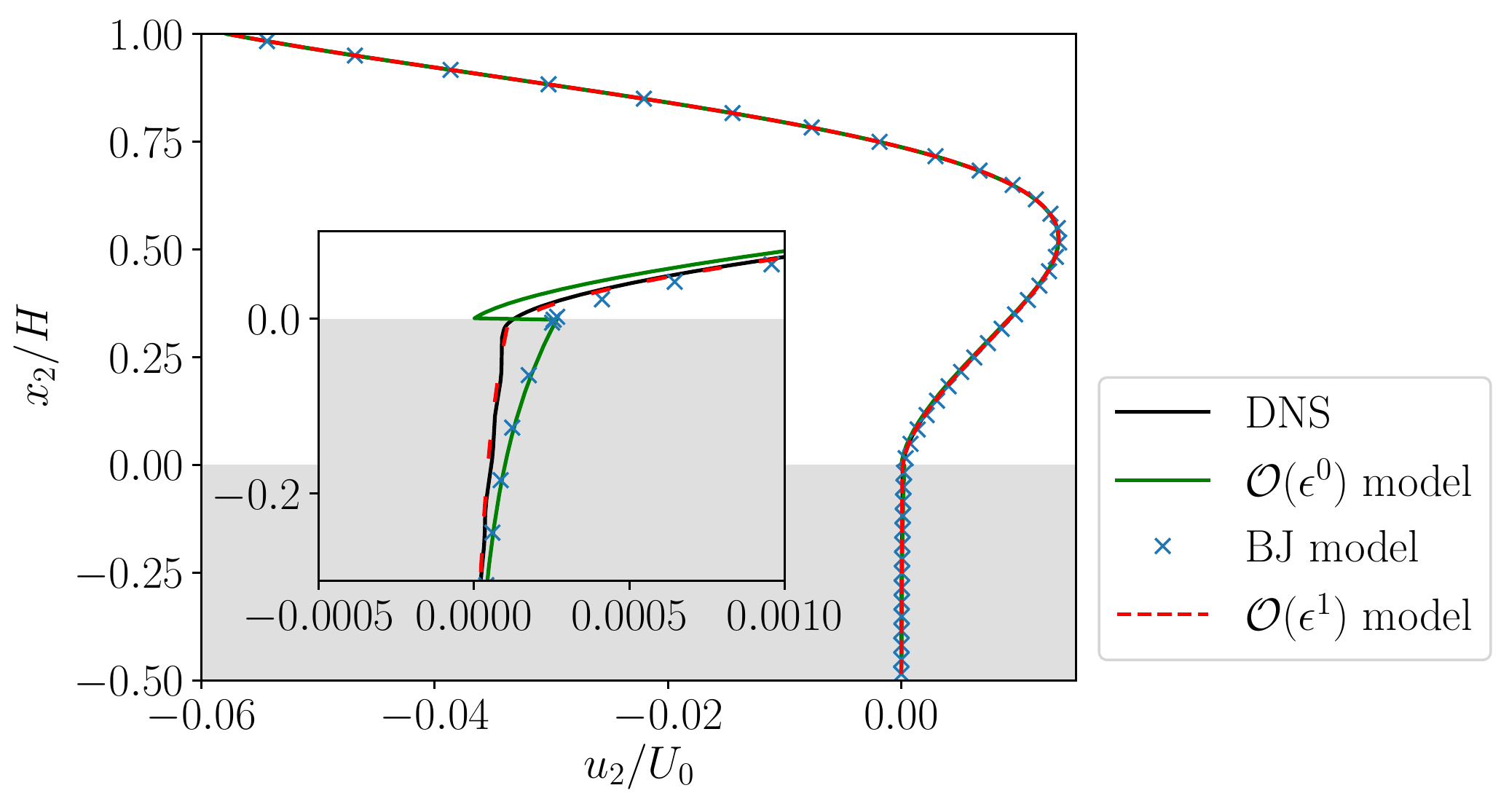}\\
(b) Velocity variation along $X_V=0.4 H$\\[0.2cm]
\includegraphics[height=3.5cm]{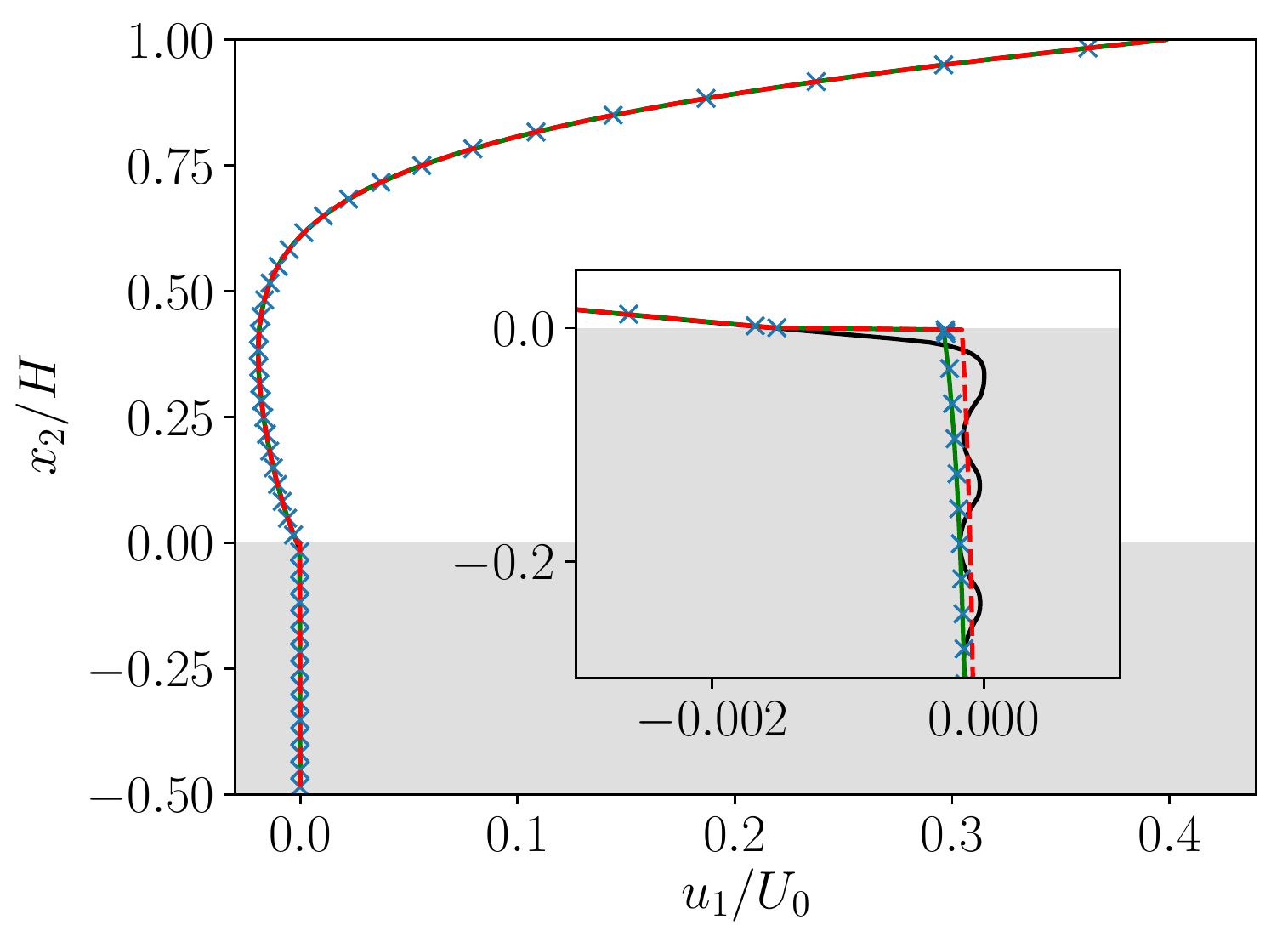}
\includegraphics[height=3.5cm]{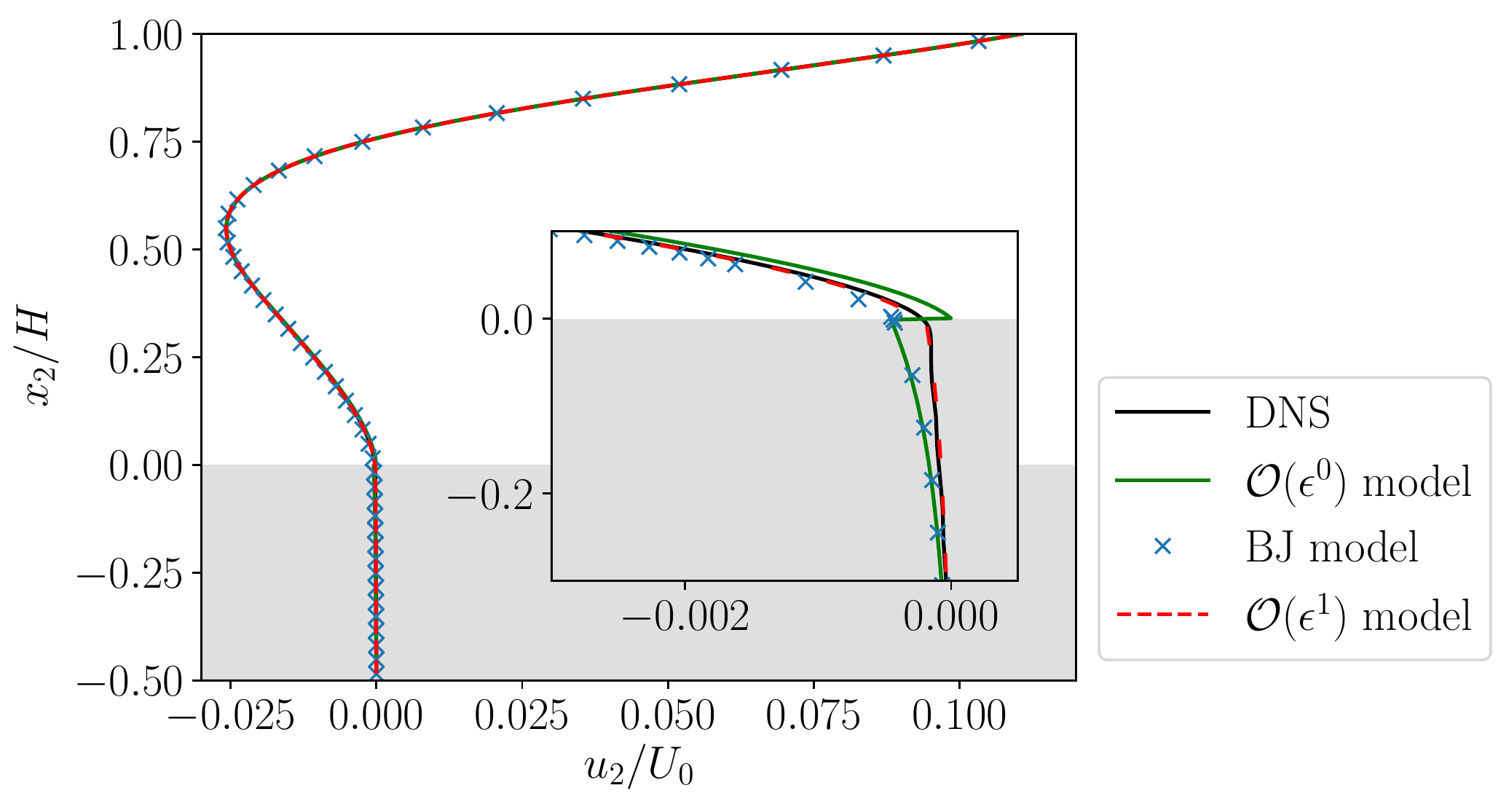}\\
(c) Velocity variation along $X_V=0.7 H$\\[0.2cm]
\caption{Variation of velocity along three different locations of vertical dashed lined in figure~\ref{fig:poro-dns} for layered porous medium. Plots in the left and right columns depict tangential and transpiration velocities respectively.}
\label{fig:poro-bl}
\end{figure}

To investigate how closely the macroscopic models approximate boundary layer over porous walls, and to study the behavior of different models within the porous domain in the vicinity of the interface, we plot velocity variation along the vertical line as shown in figure~\ref{fig:poro-dns}. We considered the vertical line at three different locations $X_V$=0.1H, 0.4H, and 0.7H; the corresponding plots are shown in figure~\ref{fig:poro-bl}. The following conclusions can be drawn irrespective of the value of $X_V$:
\begin{itemize}
\item All macroscopic models provide accurate predication of slip velocity at the interface.
\item While $\mathcal{O}(\epsilon ^1)$ models predicts variation of seepage velocity through the porous medium accurately, BJ and $\mathcal{O}(\epsilon ^0)$ models produce curves that depart from DNS values. This effect can be clearly seen from figure~\ref{fig:poro-bl}(a). As has been pointed out by \cite{nakshatrala2019}, majority of the macroscopic descriptions focus only on the free-fluid part, and do not consider the variation of velocities within the porous medium. We emphasize here that the proposed homogenized model not only provide accurate prediction of velocities at the interface, it also enables correct prediction of velocity components within the porous medium as well.
\item Due to the fact that $u_2=0$ in $\mathcal{O}(\epsilon ^0)$ model,  transpiration velocity is discontinuous; it jumps from seepage velocity in the porous domain to zero velocity at the interface.
\item $\mathcal{O}(\epsilon ^1)$ model provides accurate approximation of $u_2$. This highlights the importance of the contribution of variation of shear along the interface to the transpiration velocity.
\end{itemize}

\begin{figure}
\centering
\includegraphics[trim = 0cm 0cm 0cm 0cm, clip, height=6cm]{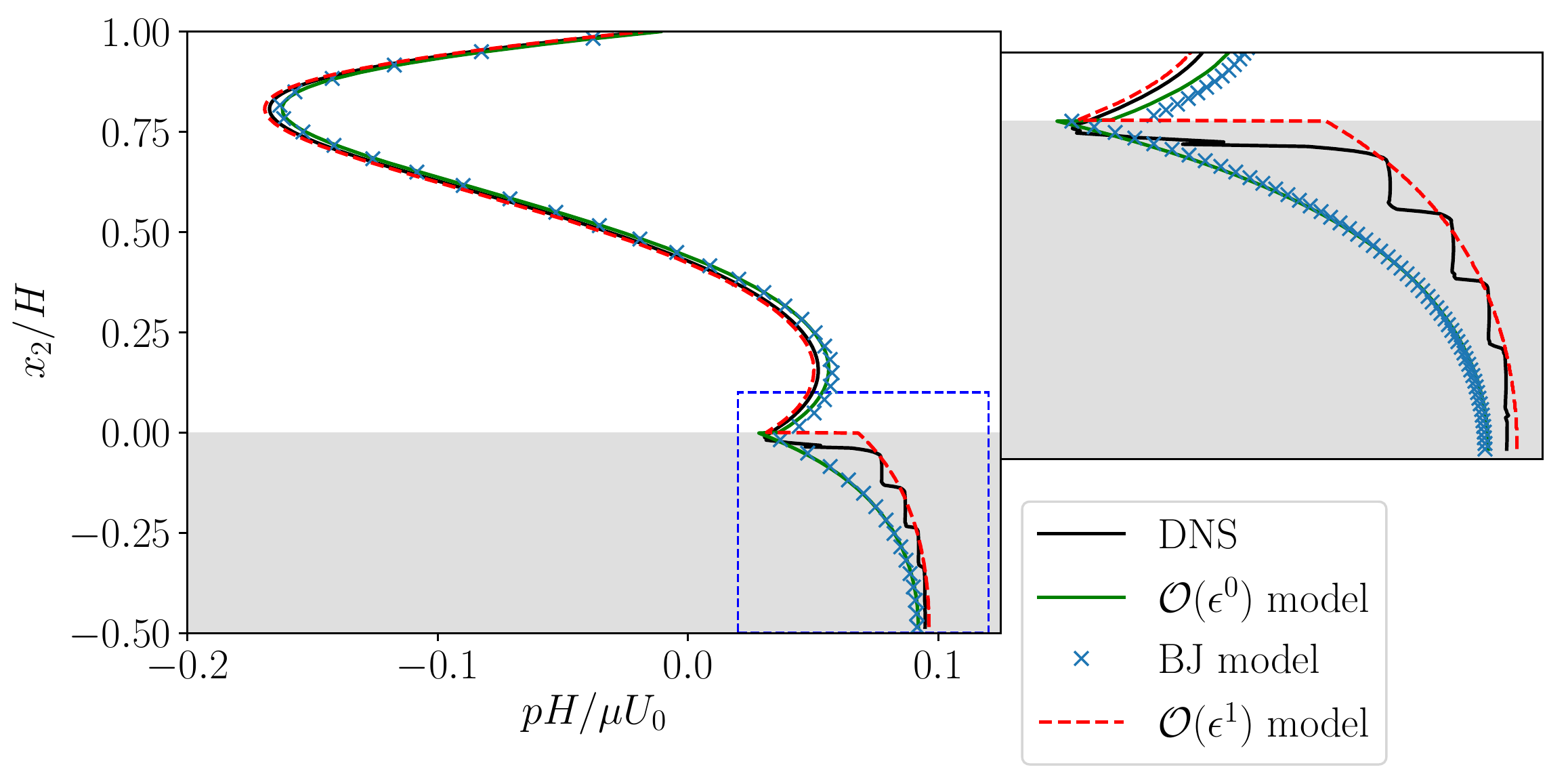}
\caption{Pressure variation along the vertical dashed line for layered porous medium, with $X_V=0.4H$, as shown in figure~\ref{fig:poro-dns}.}
\label{fig:poro-pres}
\end{figure}

It has been shown that the $\mathcal{O}(\epsilon^1)$ model can predict both velocity components accurately. For capturing transport phenomena across fluid-porous interfaces, the model needs to predict the pressure variation across the interface, in addition to velocity components. In order to show this, in figure~\ref{fig:poro-pres}, we plot pressure variation along the dashed vertical line shown in  figure~\ref{fig:poro-dns} with $X_V=0.4H$. It is  evident that pressure jump exists across the fluid-porous interfaces. While the $\mathcal{O}(\epsilon^0)$ and BJ models poorly approximate pressure jump, $\mathcal{O}(\epsilon^1)$ model provides very good prediction of pressure variation. Same conclusions can be drawn by looking at pressure plots at other values of $X_V$.

The microscale geometry considered in case~(iii) is  not fully homogeneous. The assumption of homogeneity, as described in section~\ref{sec:main} is not strictly valid, and this assumption has been used in the derivation of macroscopic interface conditions. Despite this, the results presented above clearly demonstrates that the present method provides accurate results. However, it is not our intention to claim that our method is applicable to non-homogeneous media. The microscale geometry considered in case (iii) is special that the non-homogeneity is restricted only to the interface region, and does not change along the interface. This effect has been captured by our interface cell problems as described in section~\ref{sec:coeffcompute}.

%\begin{table}[h!]
% \begin{center}
%    \caption{Stokes flow over cavity with porous bed. \textcolor{red}{Since $u_2=0$ for $\mathcal{O}(\epsilon^0)$ model, it produces 100\% error in approximating $u_2^{max}$.}}
%    \label{tab:poro-res}
%    \begin{tabular}{ l c c c c c c c}
%    \noalign{\smallskip}\hline
%      & & & \multicolumn{3}{c}{Error in $u_1^{min}$ (\%)} &   \multicolumn{2}{c}{Error in $u_2^{max}$  (\%)} \\ \cline{4-8}
%        & $u_1^{min}/U_0$ DNS & $u_2^{max}/U_0$ DNS & $\mathcal{O}(\epsilon^0)$ model & TR model & $\mathcal{O}(\epsilon^1)$ model  & TR model & $\mathcal{O}(\epsilon^1)$ model  \\ \hline
%      Isotropic & $-2.3502 \times 10^{-3}$&  $ 1.4104 \times 10^{-4}  $ & 3.71 & 3.85 & 0.12 & 13.90 & 8.30 \\
%      Anisotropic & $-2.77153 \times 10^{-3}$&  $ 1.710 \times 10^{-4}  $ & 4.85 & 5.00 & 0.21 & 3.80 & 3.54  \\
%      Layered & $-2.36893 \times 10^{-3}$&  $ 2.38923 \times 10^{-4}$ & 3.28 & 3.59 & 1.61 & 15.33 & 6.47 \\ 
%      \noalign{\smallskip}\hline
%    \end{tabular}
%  \end{center}
%\end{table}

Results presented in the above discussions correspond to the interface located at $\frac{x_2^{\Gamma}}{l}=0.1$. However, the better accuracy of the present model when compared to other existing models is not limited to one particular choice of interface location alone. In order to show this, we moved the interface location up to $\frac{x_2^{\Gamma}}{l}=0.5$ in uniform steps. The error introduced in computing  $u_1^{min}$ and $u_2^{max}$ for the isotropic porous medium configuration is shown in Table~\ref{tab:por-iso-loc}. It is directly evident from the tabulated values that the present model produced consistently more accurate results than other considered models. We have observed the same behavior for layered and anisotropic porous geometries considered in figure~\ref{fig:poro-dns}.  As explained earlier, none of the constitutive parameters considered in the paper are empirical. They are computed by solving  interface cell problems, as explained in section~\ref{sec:coeffcompute}. For each interface location, corresponding interface cell problems need to be solved, and respective coefficients can be obtained by averaging the solution field obtained from interface cells. Furthermore, it is remarkable that one can choose the interface location arbitrarily up to distances $\mathcal{O}(l)$, without introducing significant error. Previously a similar observation was made numerically by \cite{lacis2016} and \cite{lacisjfm1}, and theoretically by \cite{mikelic2012}. Nevertheless, to reduce volume over which the simplified Darcy modelling is imposed, as a 'rule-of-thumb' we suggest placing the interface as close to the solid structure as possible.
Additional details on the variation of constitutive coefficients with interface location are provided in section~\ref{sec:results-rough}

%\begin{table}[h!]
%  \begin{center}
%    \caption{\textcolor{red}{Stokes flow over cavity with layered inclusions.}}
%    \label{tab:por-lay-loc}
%    \begin{tabular}{l c c c c c c c}
%    \noalign{\smallskip}\hline
%      &\multicolumn{4}{c}{Error in $u_1^{min}$ (\%)} &   \multicolumn{3}{c}{Error in $u_2^{max}$  (\%)} \\ \cline{2-5} \cline{6-8}
%      $\frac{x_2^{\Gamma}}{l}$& $\mathcal{O}(\epsilon^0)$ model & BJ model & TR model & $\mathcal{O}(\epsilon^1)$ model  & BJ & TR model & $\mathcal{O}(\epsilon^1)$ model  \\ \hline
%%      0.0 & $ -1.36305\times 10^{-3}$&  $ 3.753\times 10^{-5} $ & 7.61$\pm$+0.1  & 7.63$\pm$+0.09 & 4.35$\pm$+0.1 & 21.13$\pm$0.39 & 17.04$\pm$0.42 \\
%      0.1 & 3.28 &  17.37 & 3.59 & 1.61 & 93.76 & 15.33 & 7.19 \\
%      0.2 & 4.28 &  7.84 & 4.69 & 1.83 & 41.20 & 10.76 & 2.36 \\
%      0.3 & 5.03 &  3.44 & 5.57 & 1.86 & 3.00 & 6.10 & 0.95 \\
%      0.4 & 5.48 &  0.98 & 6.18 & 1.67 & 22.34 & 2.37 & 3.01 \\
%      0.5 & 5.66 &  0.47 & 6.55 & 1.32 & 39.22 & 0.40 & 4.22 \\
%      \noalign{\smallskip}\hline
%    \end{tabular}
%  \end{center}
%\end{table}

\begin{table}[h!]
  \begin{center}
    \caption{Interface location study for Stokes flow over cavity with isotropic inclusions. Variation of $\alpha$ associated with Beavers-Joseph model is also reported.}
    \label{tab:por-iso-loc}
    \begin{tabular}{l c c c c c c}
    \noalign{\smallskip}\hline
      & &\multicolumn{3}{c}{Error in $u_1^{min}$ (\%)} &   \multicolumn{2}{c}{Error in $u_2^{max}$  (\%)} \\ \cline{3-5} \cline{6-7}
      $\frac{x_2^{\Gamma}}{l}$& $\alpha$ & $\mathcal{O}(\epsilon^0)$ model & BJ model & $\mathcal{O}(\epsilon^1)$ model  & BJ &  $\mathcal{O}(\epsilon^1)$ model  \\ \hline
%      0.0 & $ -1.36305\times 10^{-3}$&  $ 3.753\times 10^{-5} $ & 7.61$\pm$+0.1  & 7.63$\pm$+0.09 & 4.35$\pm$+0.1 & 21.13$\pm$0.39 & 17.04$\pm$0.42 \\
      0.1 & 0.7743 & 3.71 &  0.89 &  0.25  & 26.14 &  16.28  \\
      0.2 & 0.4665 & 4.70 &  2.00 &   1.49  & 55.41 &  13.17  \\
      0.3 & 0.3338 & 5.44 &  3.56 &  3.19  & 70.96 &   12.23  \\
      0.4 & 0.2599 & 5.88 &  4.45 &  5.34  & 79.52 &   13.22  \\
      0.5 & 0.2128 & 6.07 &  4.92 &  7.97  & 84.61 &  15.75 \\
      \noalign{\smallskip}\hline
    \end{tabular}
  \end{center}
\end{table}

\subsection{Cavity with rough bottom}
\label{sec:results-rough}
In this section, we show that the interface conditions introduced in section~\ref{sec:main} can accurately capture the effect of roughness elements on the overlying free-fluid region.

\begin{figure}
\centering
\includegraphics[trim = 0cm 0cm 0cm 0cm, clip, height=7cm]{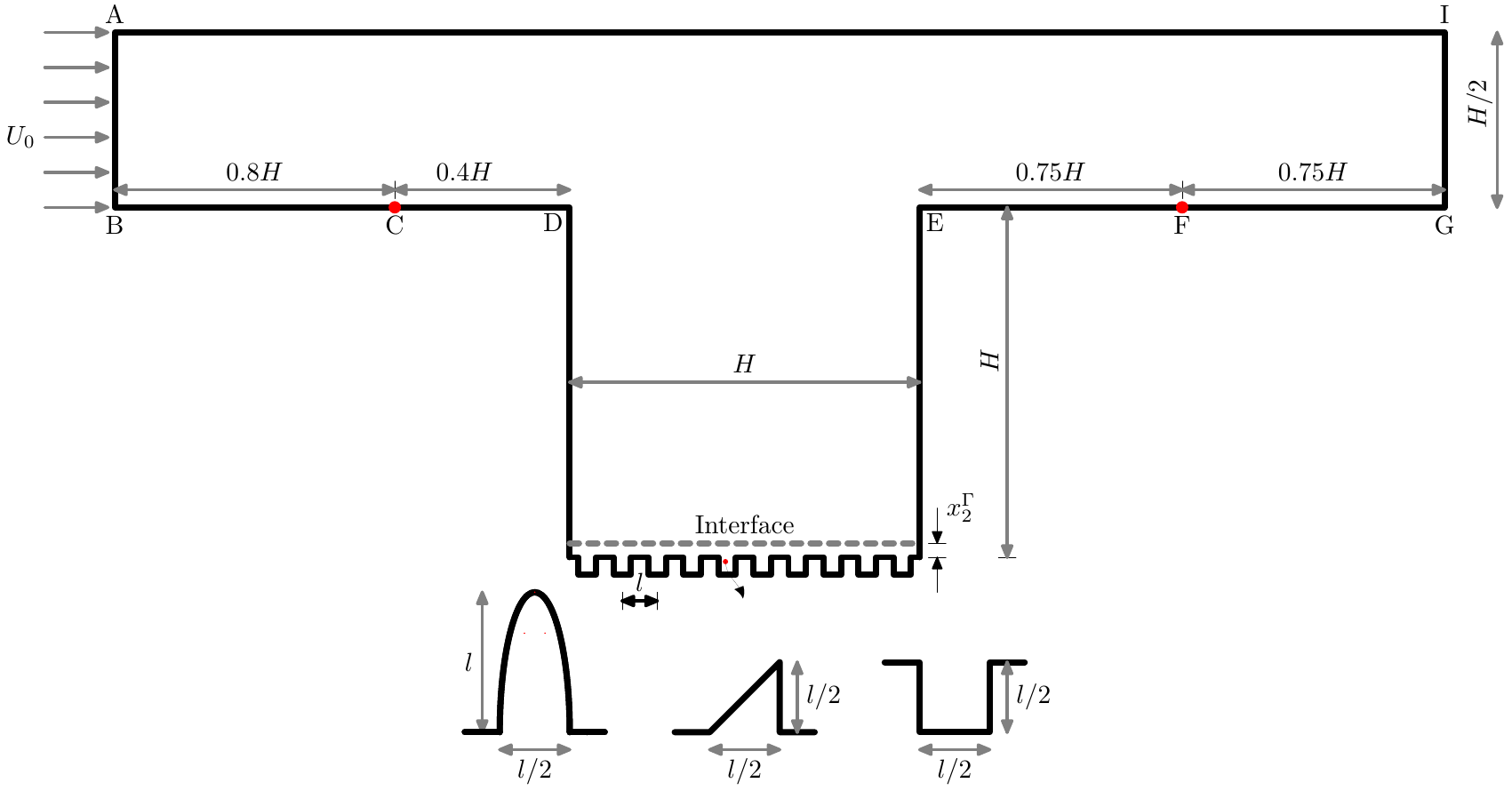}
\caption{Cavity with channel: rough bottom. Three different roughness geometries are also shown.}
\label{fig:cav-roughgeom}
\end{figure}

The configuration considered is same as that in the previous section:   Stokes flow through a channel, which drives the flow inside the cavity. However, instead of the porous block at the bottom, a solid wall with roughness elements is placed (figure~\ref{fig:cav-roughgeom}). The parameters used in the simulations are $H=1$, $U_0=1$, $\mu=0.1$, and $\epsilon(=l/H)=0.1$. We consider three different roughness geometries (semi-elliptical, triangular, and square elements)  as shown in the figure~\ref{fig:cav-roughgeom}. Since BJ model is not relevant for rough surfaces, we replace it with transpiration resistance model \citep{lacisjfm1}, denoted as TR model, for comparison.

\subsubsection*{Semi-elliptical roughness}
\begin{figure}
\centering
\subfloat[Tangential velocity]{\includegraphics[trim = 0cm 0cm 0cm 0cm, clip, height=4cm]{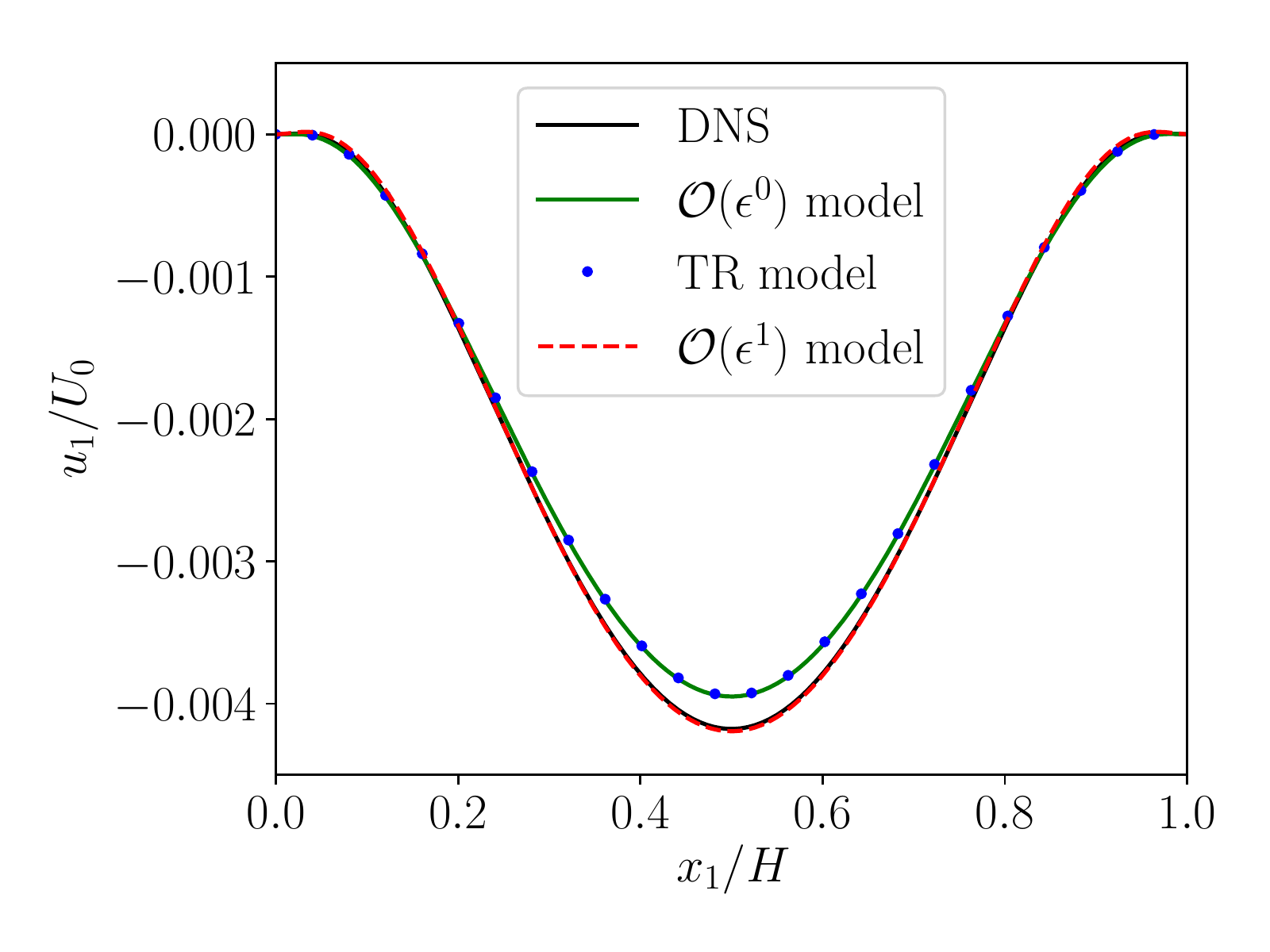}}
\subfloat[Transpiration velocity]{\includegraphics[trim = 0cm 0cm 0cm 0cm, clip, height=4cm]{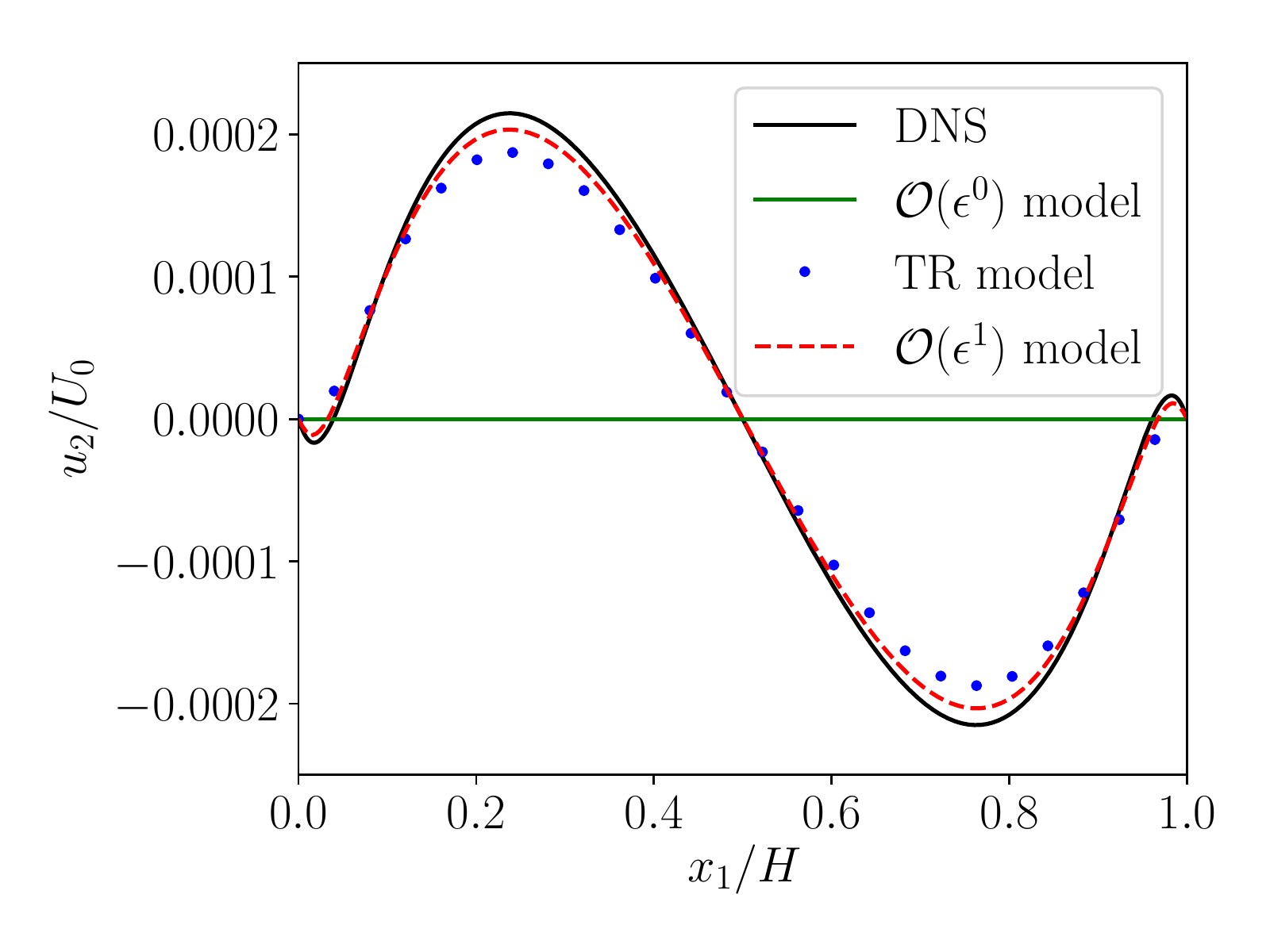}}
\caption{Cavity with channel: semi-elliptical rough bottom.}
\label{fig:cav-rough-res}
\end{figure}

The nature of flow field within the cavity is same as that described in the previous section. The flow through the channel drives the fluid flow inside the cavity within which a large vortex is generated. This vorticity induces a negative shear on the interface. As in the previous example, consistent with our boundary conditions, it induces a negative tangential velocity at the interface, as shown in figure~\ref{fig:cav-rough-res}(a). The shape of transpiration velocity curve~(figure~\ref{fig:cav-rough-res}b) is same as that of previous section, for the same physical reason. We can see that the present model, and the TR model very accurately capture the distribution of velocity components along the entire length of the interface.

In table~\ref{tab:rou-ellip}, we compare minimum tangential velocity ($u_1^{min}$) and maximum transpiration velocity ($u_2^{max}$) produced by different models against the DNS values, for various interface locations. The presented values indicate that the error incurred in the transpiration velocity is usually larger than that of the tangential velocity. Also it is evident that the higher order interface conditions produce results with superior accuracy when compared to the other two models for all interface locations.

\begin{table}[h!]
  \begin{center}
    \caption{Stokes flow over cavity with rough walls with semi-elliptic inclusions. Since $u_2=0$ for $\mathcal{O}(\epsilon^0)$ model, it produces 100\% error in approximating $u_2^{max}$.}
    \label{tab:rou-ellip}
    \begin{tabular}{l c c c c c c c}
    \noalign{\smallskip}\hline
      & & & \multicolumn{3}{c}{Error in $u_1^{min}$ (\%)} &   \multicolumn{2}{c}{Error in $u_2^{max}$  (\%)} \\ \cline{4-8}
      $\frac{x_2^{\Gamma}}{l}$  & $u_1^{min}$ DNS & $u_2^{max}$ DNS & $\mathcal{O}(\epsilon^0)$ model & TR model & $\mathcal{O}(\epsilon^1)$ model  & TR model & $\mathcal{O}(\epsilon^1)$ model  \\ \hline
%      0.0 & $-1.17328 \times 10^{-3}$&  $ 3.102 \times 10^{-5}  $ & 5.14 & 5.16 & 1.80 & 20.21 & 15.99 \\
      0.1 & $-2.71475 \times 10^{-3}$&  $ 9.780 \times 10^{-5}  $ & 4.74 & 4.82 & 0.17 & 13.47 & 6.75 \\
      0.2 & $-4.17754 \times 10^{-3}$&  $ 2.1483 \times 10^{-4}$ & 5.45 & 5.62 & 1.53 & 12.80 & 3.09 \\
      0.3 & $-5.56507 \times 10^{-3}$&  $ 3.7813 \times 10^{-4}$ & 6.03 & 6.33 & 3.01 & 13.17 & 0.60 \\
      0.4 & $-6.88238\times 10^{-3}$&   $ 5.8442 \times 10^{-4}$ & 6.37 & 6.84 & 4.65 & 13.65 & 1.61 \\
      0.5 & $-8.13409\times 10^{-3}$&   $ 8.3075 \times 10^{-4}$ & 6.47 & 7.12 & 6.46 & 14.00 & 3.74 \\
      \noalign{\smallskip}\hline
    \end{tabular}
  \end{center}
\end{table}

The influence of the interface location is taken into account in the macroscopic model by the corresponding change in the values of constitutive coefficients. This is quantified in table~\ref{tab:coefellirough}, which shows that as the interface is moved away from the top surface of solid inclusions, the value of constitutive coefficients increases. This observed behavior is due to the fact that as the interface moves up, it experiences reduced resistance offered by the rough wall to the overlying fluid flow.

\begin{table}[h!]
  \begin{center}
    \caption{Constitutive coefficients for semi-elliptic roughness elements for various interface locations. As discussed in section~\ref{sec:coeffcompute}, $\mathcal{M}_{121}=\mathcal{K}_{11}$.}
    \label{tab:coefellirough}
    \begin{tabular}{l c c c }
    \noalign{\smallskip}\hline
      $\frac{x_2^{\Gamma}}{l}$  & $\mathcal{L}_{11}$ & $\mathcal{K}_{11}$ & $\mathcal{M}_{211}$ \\ \hline
%      0.0 & 0.069765  & 0.0045239 & -0.0045253 \\
      0.1 & 0.169692  & 0.0164874 & -0.0164876 \\
      0.2 & 0.269692  & 0.0384566 & -0.0384569 \\
      0.3 & 0.369692  & 0.0704258 & -0.0704261 \\
      0.4 & 0.469692  & 0.1123950 & -0.1123950 \\
      0.5 & 0.569692  & 0.1643640 & -0.1643650  \\
      \noalign{\smallskip}\hline
    \end{tabular}
  \end{center}
\end{table}

\subsubsection*{Influence of roughness geometry}
In order to show that the effective models produce consistent and predictive results, we present the results for the cavity problem with square and triangular inclusions. The quantitative comparison of $u_1^{min}$ and $u_2^{max}$, presented earlier for semi-elliptical inclusions, are repeated here for square and triangular elements, respectively in tables~\ref{tab:rou-sqr} and \ref{tab:rou-tri}. It can be seen from the tables that similar to the semi-elliptic geometries, the $\mathcal{O}(\epsilon^1)$ model produces more accurate results when compared to the other two models. This confirms that the higher order interface conditions produces more accurate results in a consistent manner, by introducing more physical features of the problem into the formulations.

As discussed earlier, the constitutive coefficients contain all details about the geometrical details of rough/porous interfaces. The difference in the geometrical details of roughness accordingly modified the values of the coefficients (not shown here). 

\begin{table}[h!]
  \begin{center}
    \caption{Stokes flow over cavity with rough walls with square inclusions}
    \label{tab:rou-sqr}
    \begin{tabular}{l c c c c c c c}
    \noalign{\smallskip}\hline
      & & & \multicolumn{3}{c}{Error in $u_1^{min}$ (\%)} &   \multicolumn{2}{c}{Error in $u_2^{max}$  (\%)} \\ \cline{4-8}
      $\frac{x_2^{\Gamma}}{l}$ & $u_1^{min}$ DNS & $u_2^{max}$ DNS & $\mathcal{O}(\epsilon^0)$ model & TR model & $\mathcal{O}(\epsilon^1)$ model  & TR model & $\mathcal{O}(\epsilon^1)$ model  \\ \hline
%      0.0 & $-3.0948 \times 10^{-4}$  &  $3.324 \times 10^{-6}$   & 15.17$\pm$0.14 & 15.17$\pm$0.14  & 14.04\pm$0.15  & 43.46$\pm$0.26 & 42.34\pm$0.26 \\
      0.1 & $-1.92885 \times 10^{-3}$&  $4.227 \times 10^{-5}$   & 2.97 & 3.01 &   0.24     & 7.93  &  3.34 \\
      0.2 & $-3.46361 \times 10^{-3}$&  $1.3432 \times 10^{-4}$ & 4.33 & 4.44 &    1.23    & 9.08  &  1.12 \\
      0.3 & $-4.91824 \times 10^{-3}$&  $2.7559 \times 10^{-4}$ & 5.32 & 5.55 &    2.46    & 10.65 & 0.39  \\
      0.4 & $-6.29809 \times 10^{-3}$&  $4.6247 \times 10^{-4}$ & 5.96 & 6.33 &   3.93     & 11.87 & 2.03 \\
      0.5 & $-7.60808 \times 10^{-3}$&  $6.9176 \times 10^{-4}$ & 6.28 & 6.82 &   5.60     & 12.74 & 3.80 \\
      \noalign{\smallskip}\hline
    \end{tabular}
  \end{center}
\end{table}

\begin{table}[h!]
  \begin{center}
    \caption{Stokes flow over cavity with rough walls with triangular inclusions.}
    \label{tab:rou-tri}
    \begin{tabular}{l c c c c c c c}
    \noalign{\smallskip}\hline
      & & & \multicolumn{3}{c}{Error in $u_1^{min}$ (\%)} &   \multicolumn{2}{c}{Error in $u_2^{max}$  (\%)} \\ \cline{4-8}
      $\frac{x_2^{\Gamma}}{l}$& $u_1^{min}$ DNS & $u_2^{max}$ DNS & $\mathcal{O}(\epsilon^0)$ model & TR model & $\mathcal{O}(\epsilon^1)$ model  & TR model & $\mathcal{O}(\epsilon^1)$ model  \\ \hline
%      0.0 & $ -1.36305\times 10^{-3}$&  $ 3.753\times 10^{-5} $ & 7.61$\pm$+0.1  & 7.63$\pm$+0.09 & 4.35$\pm$+0.1 & 21.13$\pm$0.39 & 17.04$\pm$0.42 \\
      0.1 & $ -2.88761\times 10^{-3}$&  $ 1.1178\times 10^{-4} $ & 5.57 & 5.65 & 0.42 & 15.42 & 8.47 \\
      0.2 & $ -4.33469\times 10^{-3}$&  $ 2.3427\times 10^{-4} $ & 5.97 & 6.15 & 1.24 & 14.14 & 4.22 \\
      0.3 & $ -5.70736\times 10^{-3}$&  $ 4.0237\times 10^{-4} $ & 6.38 & 6.69 & 2.87 & 14.19 & 1.37 \\
      0.4 & $ -7.01077\times 10^{-3}$&  $ 6.1287\times 10^{-4} $ & 6.61 & 7.09 & 4.61 & 14.37 & 1.07 \\
      0.5 & $ -8.24953\times 10^{-3}$&  $ 8.6292\times 10^{-4} $ & 6.62 & 7.29 & 6.49 & 14.55 & 3.36 \\
      \noalign{\smallskip}\hline
    \end{tabular}
  \end{center}
\end{table}

%The ability of the proposed interface conditions to consistently represent the effect of interface locations and geometrical details of the roughness confirms the robustness of the proposed conditions.

\subsection{Backward facing step with rough bottom}
\begin{figure}
\centering
\includegraphics[trim = 0cm 0cm 0cm 0cm, clip, height=3cm]{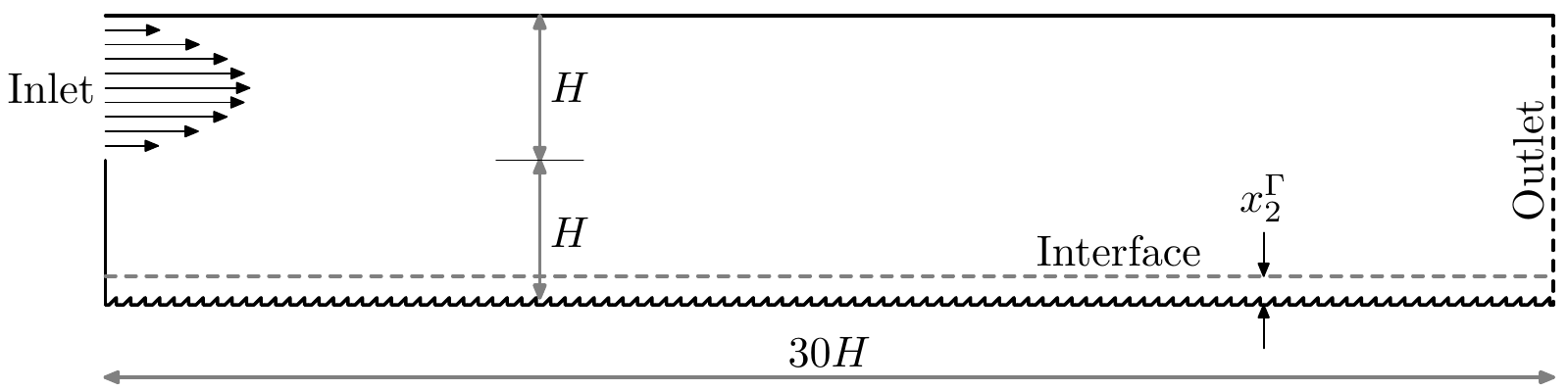}
\caption{Flow over a backward facing step with rough walls}
\label{fig:bwd-geom}
\end{figure}

The previous two examples demonstrated the improved accuracy of $\mathcal{O}(\epsilon^1)$ interface conditions for Stokes flow over rough and porous surfaces. As a further validation, in this section, we report results of a finite Reynolds number incompressible flow over a backward facing step with rough bottom. The configuration and boundary conditions are depicted in figure~\ref{fig:bwd-geom}. The height of the step is $H$, and the length is taken to be 30 times that of height. A parabolic velocity profile is specified at the inlet with averaged velocity given as $U_{\textrm{avg}}$, and the boundary conditions at outlet are $u_2=0$ and zero shear. On all other surfaces, no slip condition is enforced. The geometry of triangular-shaped roughness elements and the interface definition are exactly same as that of the previous example. Scale separation parameter $\epsilon=0.1$ and the Reynolds number, $\textrm{Re}=U_{\textrm{avg}}H/\nu=100$.

\begin{figure}
\centering
\includegraphics[trim = 1.75cm 12cm 1.75cm 2.5cm, clip, height=3cm]{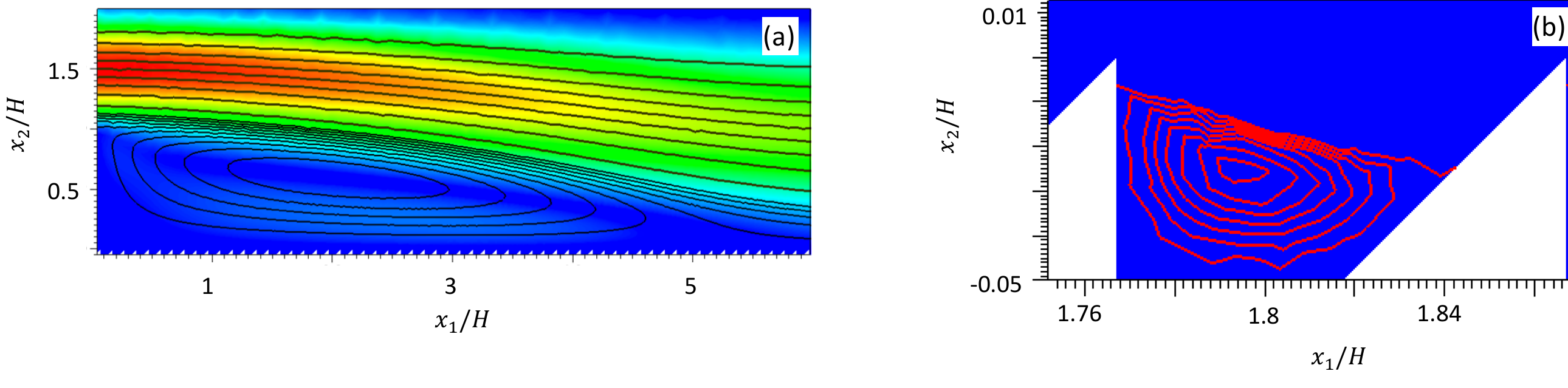}
\caption{Streamlines for flow over a backward facing step. (a)~showing separated vortex, (b)~recirculation region within the roughness elements.}
\label{fig:bwd-stream}
\end{figure}

It is well-known that the step induces flow separation, and as a consequence forms a large vortex, as shown in figure~\ref{fig:bwd-stream}(a). A small distance downstream of the vortex reattachment point, the flow behaves as if it is a fully developed flow through a channel. The vortex behind the step interacts with the triangular-shaped roughness elements at the bottom surface. Due to  inertial effects, as shown in figure~\ref{fig:bwd-stream}(b), a local recirculation is formed within each roughness elements. The velocity magnitude at the crest plane of roughness elements is $U_s\sim 10^{-2}$, and the slip Reynolds number $Re_s \sim 1$.

\begin{figure}
\centering
\subfloat[Tangential velocity]{\includegraphics[trim = 0cm 8cm 0cm 8cm, clip, height=4cm]{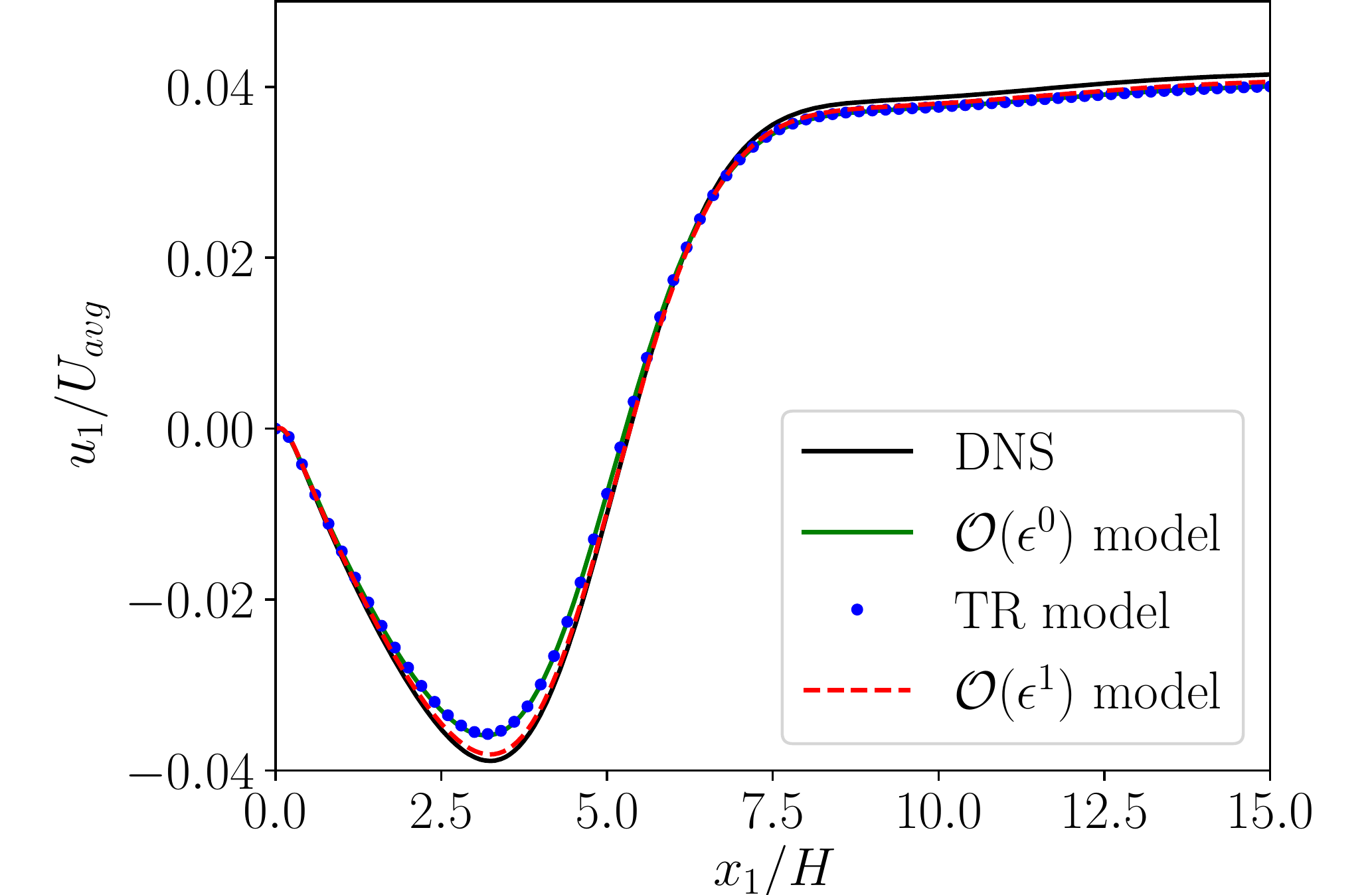}}
\subfloat[Transpiration velocity]{\includegraphics[trim = 0cm 8cm 0cm 8cm, clip, height=4cm]{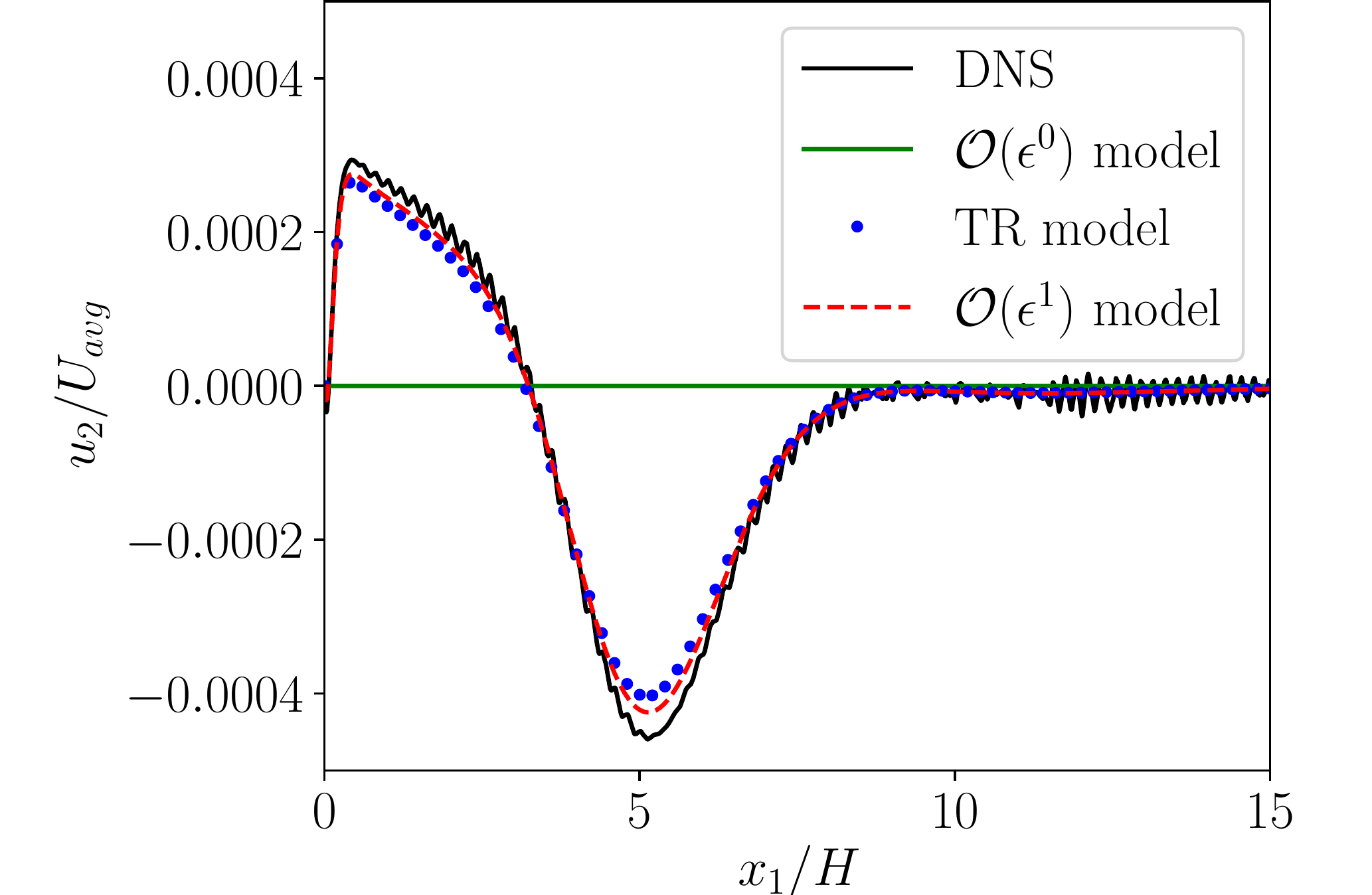}}
\caption{Flow over a backward facing step. Comparison of tangential and transpiration velocity of different models when the interface is located at $x_2^\Gamma=0.2l$}
\label{fig:bwd-results}
\end{figure}

In order to quantify the accuracy of our interface conditions, as in the previous examples, we report tangential and transpiration velocities along the interface located at $x_2^{\Gamma}=0.2l$. It can be seen from figure~\ref{fig:bwd-results}(a) that the $\mathcal{O}(\epsilon^0)$ and TR model produces results of same level of accuracy, while the $\mathcal{O}(\epsilon^1)$-model's accuracy is better. The accuracy of all three effective models for tangential velocity with respect to interface height is presented in table~\ref{tab:bwd}. It can be seen that even for this problem, $\mathcal{O}(\epsilon^1)$-model is more accurate than other two alternatives. Moreover, as can be seen from figure~\ref{fig:bwd-results}(b), the transpiration velocity is also very accurately predicted by TR and $\mathcal{O}(\epsilon^1)$-model with the later producing slightly better results.

These results are surprising because in theory our $\mathcal{O}(\epsilon^1)$-model is valid only when $Re_s < \epsilon^2$; for the present example, $Re_s\sim 1$, in addition to having a small recirculation region within the roughness elements. However, additional numerical examples must be carried out before concluding that our model can produce accurate results in the presence of separation within the roughness elements.

\begin{table}[h!]
  %\begin{center}
    \caption{Backward facing step with triangular inclusions.}
    \label{tab:bwd}
    \begin{tabular}{ l c  c  c  c }
    \noalign{\smallskip}\hline
      & & \multicolumn{3}{c}{Error in $u_1^{min}$ (\%)} \\ \cline{3-5}
      $\frac{x_2^\Gamma}{l}$& $u_1^{min}$ DNS & $\mathcal{O}(\epsilon^0)$ model & TR model & $\mathcal{O}(\epsilon^1)$ model \\ \hline
%      0.0 & $ -1.24602\times 10^{-2}$& 10.75 & 10.81 & 7.94 \\
      0.1 & $ -2.60089\times 10^{-2}$& 7.34 & 7.54 & 3.16 \\
      0.2 & $ -3.88864\times 10^{-2}$& 7.69 & 8.11 & 1.93 \\
      0.3 & $ -5.11105\times 10^{-2}$& 8.57 & 9.27 & 1.33 \\
      0.4 & $ -6.26855\times 10^{-2}$& 9.52 & 10.56 & 0.95  \\
      0.5 & $ -7.36112\times 10^{-2}$& 10.40 & 11.84 & 0.62 \\ 
      \noalign{\smallskip}\hline
    \end{tabular}
%  \end{center}
\end{table}

In order to validate the higher order interface conditions proposed in this paper, we have used 2D flows with flat interfaces that are very commonly used in the literature. The proposed formulations are tensorial in nature, and hence they are straightforwardly applied to 3D flows with complex interfaces. A reduced form of the proposed formulations, called as transpiration resistance model, has been used in our recent paper~\citep{lacisjfm1} to simulate turbulent flow over rough surfaces in 3D. A similar homogenized model~\citep{giuseppe2019} has been used to study flows over curved rough surfaces and it can be inferred from this paper that studies on curved interfaces in 3D require dedicated efforts to design test cases and to quantify the accuracy of the model. Hence, analyses of our model with 3D curved interfaces will be pursued in future.
 %========================================================================================
 \section{Conclusions}
 \label{sec:conc}
In this paper, we presented the derivation of a macroscopic model for flows over porous surfaces. The proposed model enables the simulation of  three-dimensional flows over porous and rough surfaces. The tensorial nature of the proposed interface conditions makes it possible to deal with isotropic, anisotropic, and layered porous structures in 3D. For clarity of exposition, we provided simplified formulations to deal with 2D flows over flat  rough/porous interfaces that are aligned along one of the coordinate axes. By solving a set of auxiliary Stokes problems, we can compute all the constitutive coefficients appearing in the macroscopic interface conditions, making the formulations free of any empirical parameters. Thus, it involves neither  trial-and-error procedure nor empirical curve fitting. Moreover, since the domain to solve the auxiliary equations is only a tiny fraction of the total computational domain, the solution process is very efficient, and hence it does not add noticeable computational overhead to the solution of the equations describing the macroscopic description.  Accuracy of the proposed formulation is studied by comparing the results from our macroscopic model to those obtained from geometry resolved simulations of microscopic texture. These comparisons show that the proposed macroscopic model is  accurate in modeling flows over rough/porous surfaces in the Stokes regime. An additional example shows the predictive capability of the derived interface conditions even when including inertial effects.
These aspects are crucial and make the present formulation a viable tool to map the microstructural details of complex porous/rough surfaces to the associated transport phenomena, without inputs from expensive experiments.
  %========================================================================================
 \section*{Appendix 1: Relation between gradients of fast flow velocity and full velocity}
One of the essential steps in arriving at the final interface conditions is to relate the gradients of fast flow velocities~($U_{i,j}$) to that of full velocities~($\left<u_i\right>_{,j}$). Corresponding leading order $\mathcal{O}(\epsilon)$ and higher order $\mathcal{O}(\epsilon^1)$ relations are described in equations~\eqref{eqn:gradlead} and \eqref{eqn:gradhigh} respectively. In this appendix, we provide validation of these expressions. We compute $\left<u_i\right>_{,j}$ by ensemble average of DNS data, and other terms in these expressions are computed from effective simulations with no-slip conditions at the interface~(equation~\ref{eqn:fastflow}).
 
\begin{figure}
\centering
\subfloat[]{\includegraphics[trim = 0cm 0cm 0cm 0cm, clip, height=5cm]{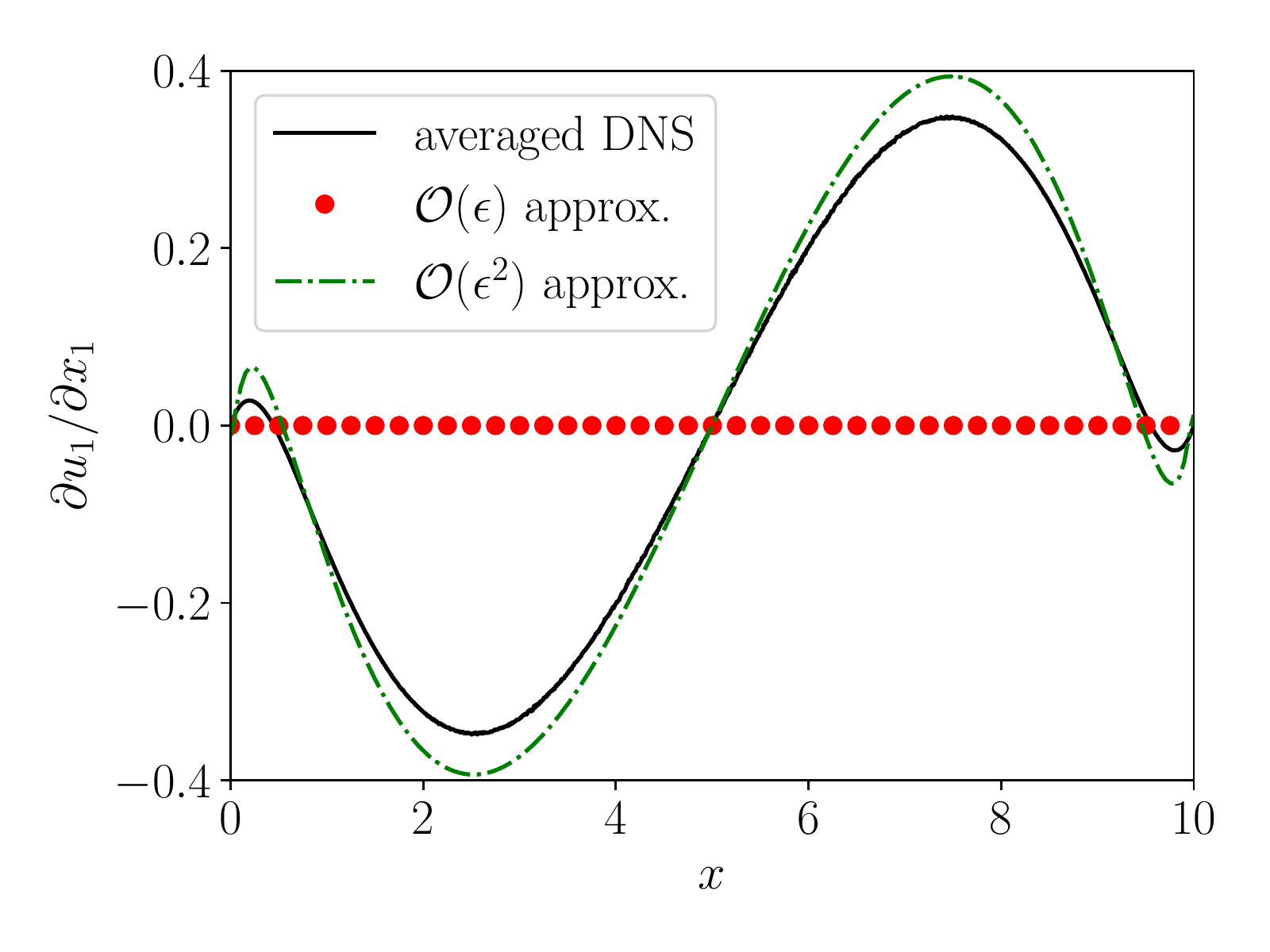}}
\subfloat[]{\includegraphics[trim = 0cm 0cm 0cm 0cm, clip, height=5cm]{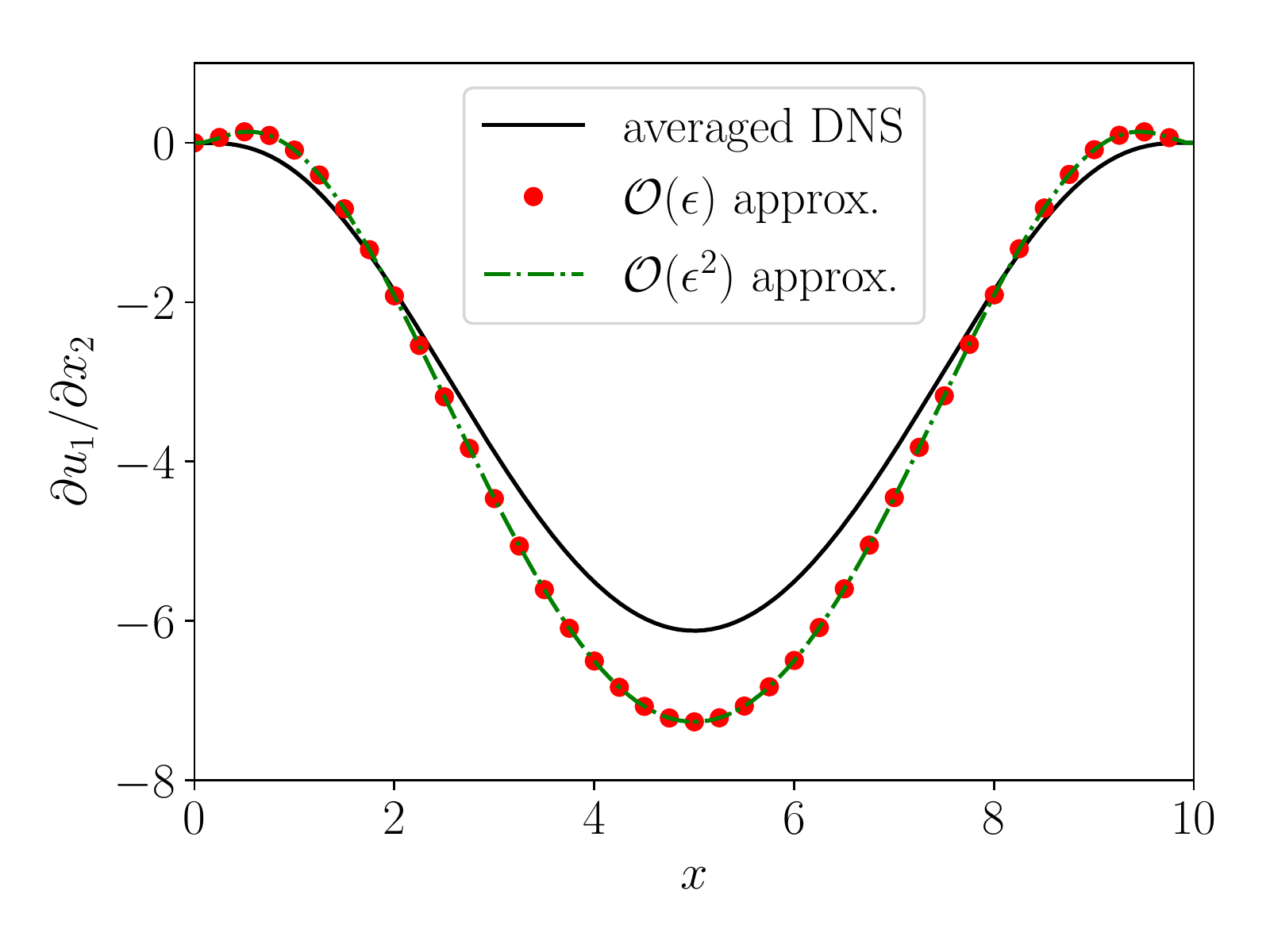}}\\
\subfloat[]{\includegraphics[trim = 0cm 0cm 0cm 0cm, clip, height=5cm]{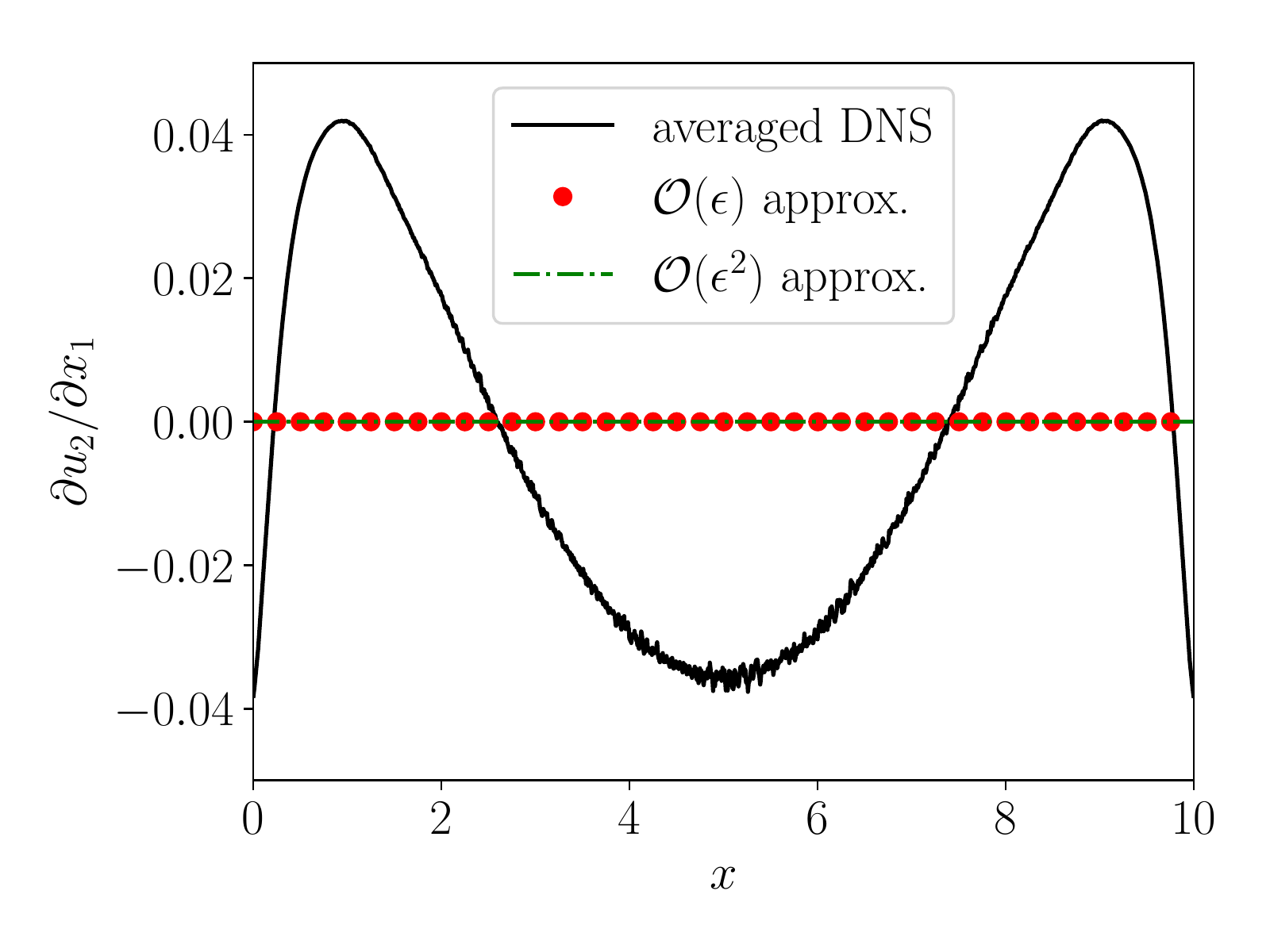}}
\subfloat[]{\includegraphics[trim = 0cm 0cm 0cm 0cm, clip, height=5cm]{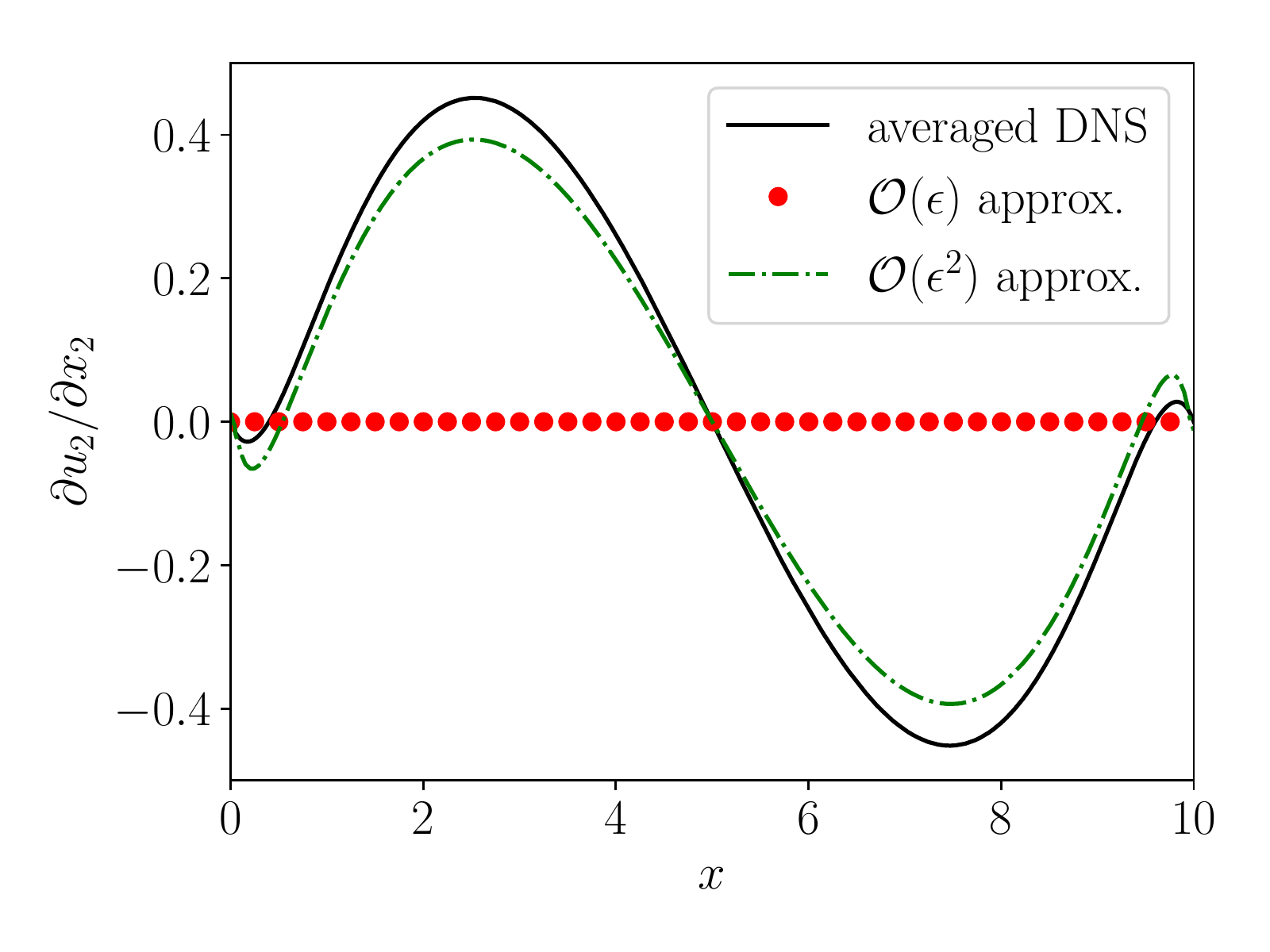}}\\
\caption{Approximation of velocity gradients with leading order $\mathcal{O}(\epsilon)$ and higher order $\mathcal{O}(\epsilon^2)$ approximation. Averaged DNS implies $\left< u_i\right> _{,j}$, $\mathcal{O}(\epsilon)$ and $\mathcal{O}(\epsilon^2)$ approximation imply right hand side of equations~\eqref{eqn:gradlead} and \eqref{eqn:gradhigh} respectively.}
\label{fig:gradvalid}
\end{figure}

We carry out this validation on the example of isotropic circular inclusions, and the results are presented in figure~\ref{fig:gradvalid}. From the figure it is evident that there is a hierarchy of amplitudes presented in all the derivatives at the interface. The velocity derivative $\partial u_1/\partial x_2$ corresponding to shear has the largest magnitude, followed by the derivatives $\partial u_1/\partial x_1$  and $\partial u_2/\partial x_2$ , which are order of magnitude smaller, and finally the derivative $\partial u_2/\partial x_1$ is two orders of magnitude smaller compared to the $\partial u_1/\partial x_2$ . We see from figure~\ref{fig:gradvalid} that $\mathcal{O}(\epsilon)$  approximation provides only the largest derivative $\partial u_1/\partial x_2$, while the higher order $\mathcal{O}(\epsilon^1)$ approximation provides a reasonable approximation to $\partial u_1/\partial x_1$  and $\partial u_2/\partial x_2$  also. However $\partial u_2/\partial x_1$  is still out of reach. This serves to demonstrate the better accuracy of the higher order approximation to gradients of velocity.
 
 \section*{Appendix 2: Relation between porous and rough wall interface conditions}
 The velocity boundary condition for the porous wall, given in equation~\eqref{eqn:epsnondim}, reads
\begin{equation}
 \left<u_i^+\right>=\mathcal{L}_{ij}s_j+\left(\overline{\mathcal{M}}_{ijk}+\mathcal{K}_{ik}\left<B_j^-\right>\right)d_{jk}-\mathcal{K}_{ij}p^-_{,j}+\mathcal{O}(\epsilon^2),
 \label{eqn:app-start}
 \end{equation}
 where
 \begin{equation}
 \overline{\mathcal{M}}_{ijk}=\mathcal{M}_{ijk}
-\mathcal{L}_{ik}\mathcal{L}_{mj}n_m-\mathcal{L}_{im}\mathcal{L}_{mj}n_k
-\mathcal{L}_{il}\left<M^+_{ljk}\right>_{,m}n_m
-\mathcal{L}_{il}\left<M^+_{mjk}\right>_{,l}n_m
 \end{equation}

In order to use this condition for a rough wall, we need write quantities defined in porous domain ($\left<p^-\right>_{,j}$ and $\left<B_j^-\right>$) in terms of their equivalent free-fluid quantities. This is because in the effective simulations of rough wall  flows, we cut off the domain below the interface, and hence these quantities are undefined.

In order to so, we consider the pressure jump condition given in equation~\eqref{eqn:epsnondim},
\begin{equation}
\llbracket p \rrbracket =\mathcal{B}_ks_k+\mathcal{O}(\epsilon).
\end{equation}
Taking derivative along the interface tangential direction of the above expression, we obtain
\begin{equation}
\left(P+\left<p^+\right>\right)_{,j}-\left<p^-\right>_{,j}=\mathcal{B}_kd_{kj}+\mathcal{O}(\epsilon^2).
\label{eqn:tangpgrad}
\end{equation}
 Substituting the above expression in equation~\eqref{eqn:app-start}, we get the desired result,
\begin{equation}
  \left<u_i^+\right>=\mathcal{L}_{ij}s_j+\left(\overline{\mathcal{M}}_{ijk}+\mathcal{K}_{ik}\left<B_j^+\right>\right)d_{jk}-\mathcal{K}_{ij}\left<p\right>_{,j}+\mathcal{O}(\epsilon^2),
 \end{equation}
where $\left<p\right>=P+\left<p^+\right>$, which is the quantity defined in the free-fluid region. The above expression is used to arrive at the boundary conditions for rough walls given in equation~\eqref{eqn:simplerough}.

We stress that the derivative of pressure, defined in equation~\eqref{eqn:tangpgrad} is valid only along the direction tangent to the interface. Derivative of $\mathcal{B}$ along the interface normal direction is undefined, which means it is not possible to relate  wall-normal pressure gradient across either sides of the interface using such an expression. 

 \section*{Appendix 3: Performance of TR model for porous media}
\begin{figure}
\centering
\subfloat[Tangential velocity]{\includegraphics[trim = 0cm 0cm 0cm 0cm, clip, height=5cm]{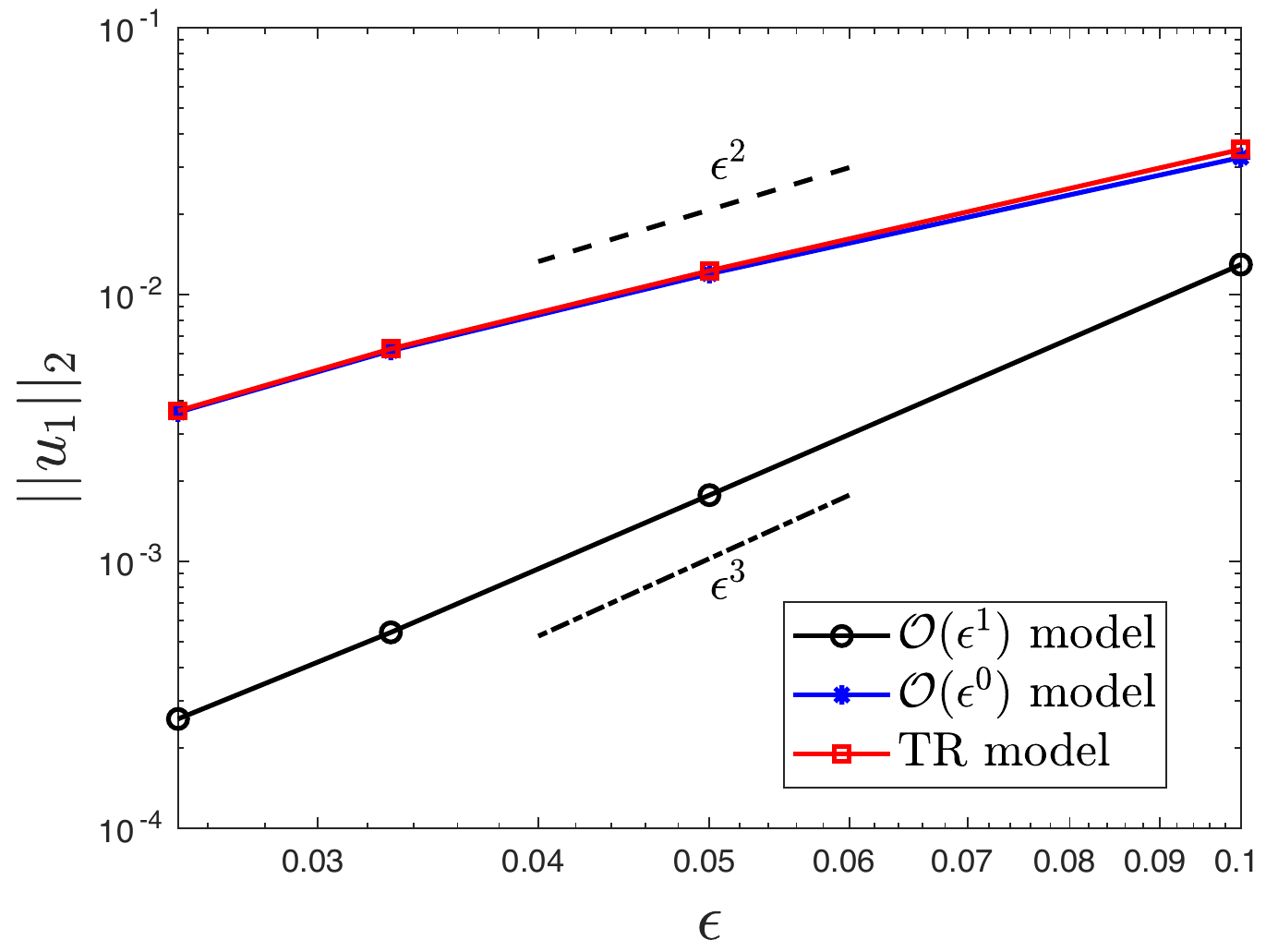}}
\subfloat[Transpiration velocity]{\includegraphics[trim = 0cm 0cm 0cm 0cm, clip, height=5cm]{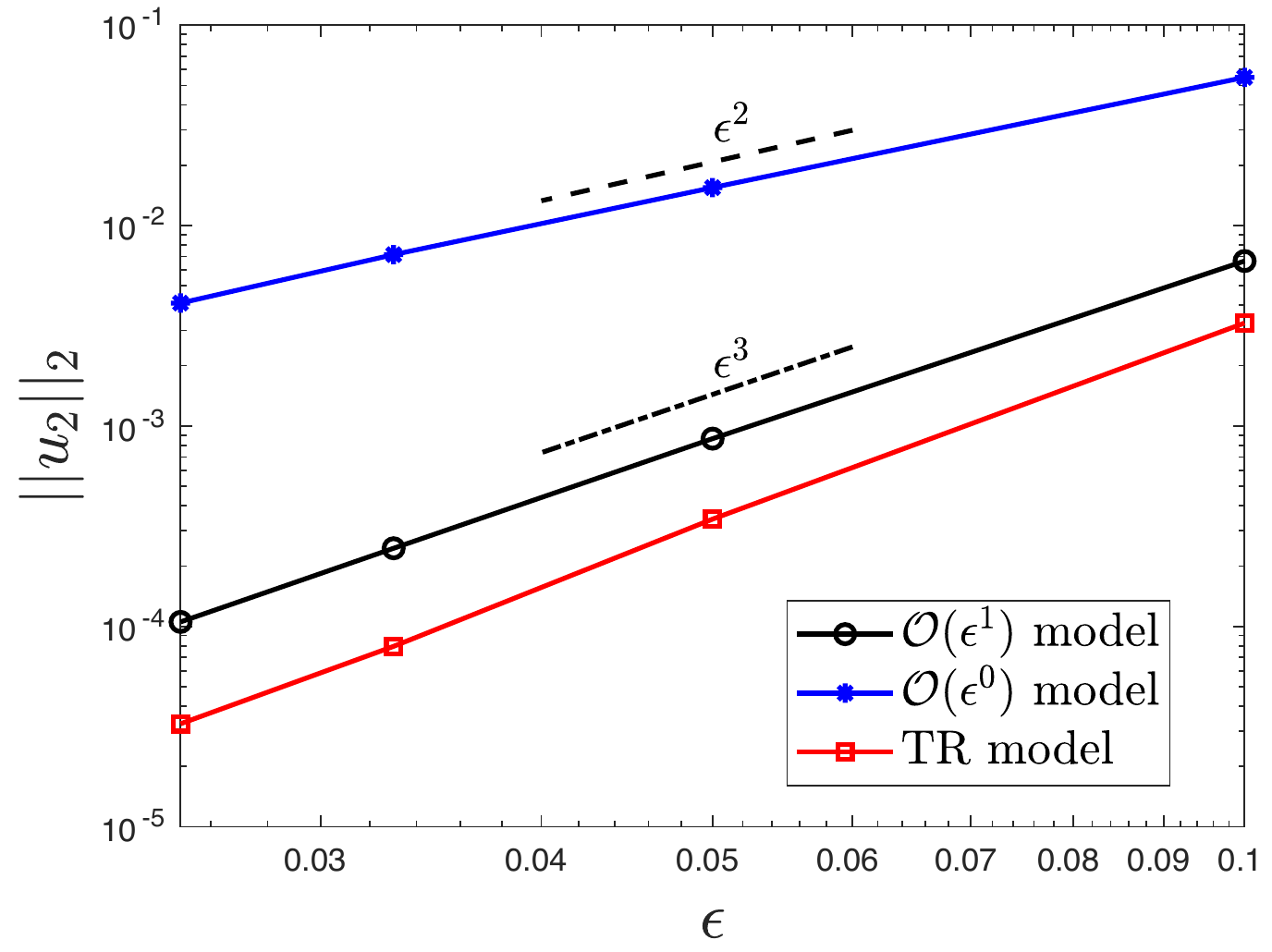}}
\caption{Convergence plots for porous medium with isotropic solid inclusions.}
\label{fig:poro-conv-tr}
\end{figure}

In the order of convergence study for isotropic porous medium presented in section~\ref{sec:results-poro-conv}, we replace BJ model by TR model, and the results are presented in figure ~\ref{fig:poro-conv-tr}. The tangential velocity formulations of TR model is same as that of $\mathcal{O}(\epsilon^0)-$model. Hence the convergence rate as well as errors produced by TR model is approximately same as that of $\mathcal{O}(\epsilon^0)-$model. However, for transpiration velocity TR model exhibits a surprising behavior; it produces more accurate results than the higher order model while maintaining the same order of convergence as can be seen from figure~\ref{fig:poro-conv-tr}b. This might be due to a more accurate pressure jump representation by the TR model than $\mathcal{O}(\epsilon^1)-$model (results not shown here). The reason for this superior behavior is unclear, and will be probed in future. %The same behavior is observed for both anisotropic and layered porous geometries described in figure~\ref{fig:poro-pres}.

%\begin{table}[h!]
%  \begin{center}
%    \caption{Stokes flow over cavity with anisotropic inclusions.}
%    \label{tab:por-lay-loc}
%    \begin{tabular}{l c c c c c c c}
%    \noalign{\smallskip}\hline
%      &\multicolumn{4}{c}{Error in $u_1^{min}$ (\%)} &   \multicolumn{3}{c}{Error in $u_2^{max}$  (\%)} \\ \cline{2-5} \cline{6-8}
%      $\frac{x_2^{\Gamma}}{l}$& $\mathcal{O}(\epsilon^0)$ model & BJ model & TR model & $\mathcal{O}(\epsilon^1)$ model  & BJ & TR model & $\mathcal{O}(\epsilon^1)$ model  \\ \hline
%%      0.0 & $ -1.36305\times 10^{-3}$&  $ 3.753\times 10^{-5} $ & 7.61$\pm$+0.1  & 7.63$\pm$+0.09 & 4.35$\pm$+0.1 & 21.13$\pm$0.39 & 17.04$\pm$0.42 \\
%      0.1 & 4.85 &  1.57 & 5.00 (5.00) & 5.00 (5.00) & 45.98 & 6.47 (9.66) & 3.84 (4.07) \\
%      0.2 & 5.45 &  3.39 & 5.69 (5.69) & 5.69 (5.69) & 66.37 & 1.87 (4.38) & 6.27 (6.41) \\
%      0.3 & 5.93 &  4.45 & 6.31 (6.31) & 6.30 (6.30) & 77.39 & 6.47 (0.65) & 7.96 (8.06) \\
%      0.4 & 6.19 &  5.04 & 6.74 (6.73) & 6.73 (6.73) & 83.66 & 9.15 (1.34) & 8.57 (8.60) \\
%      0.5 & 6.23 &  5.29 & 6.97 (6.96) & 6.96 (6.96) & 87.51 & 10.76 (3.60) & 7.29 (7.63) \\
%      \noalign{\smallskip}\hline
%    \end{tabular}
%  \end{center}
%\end{table}

%\begin{acknowledgements}
%\end{acknowledgements}

\bibliographystyle{plainnat}
\bibliography{porous}

\end{document}